\documentclass[]{loi}

\newcommand{\AmS}{{\protect\the\textfont2
  A\kern-.1667em\lower.5ex\hbox{M}\kern-.125emS}}

\newcommand{\units}{\,\mathrm}

\newcommand{\gevtwo}{\units{GeV^2}}

\begin{document}
\setcounter{page}{-1}
%
\vskip 3.cm
\title{A Detector for Forward Physics at eRHIC \\ ~ \\ Feasibility Study \\~\\}
\vskip 4.cm
\author{I.~Abt, A.~Caldwell, X.~Liu, J.~Sutiak \\ \\
        Max-Planck-Institut f\"ur Physik \\
        (Werner-Heisenberg-Institut) \\}
\date{July 20, 2004}
\maketitle
\vspace{-5.0cm}
\begin{figure}[hbpt]
\begin{center}
\epsfig{file=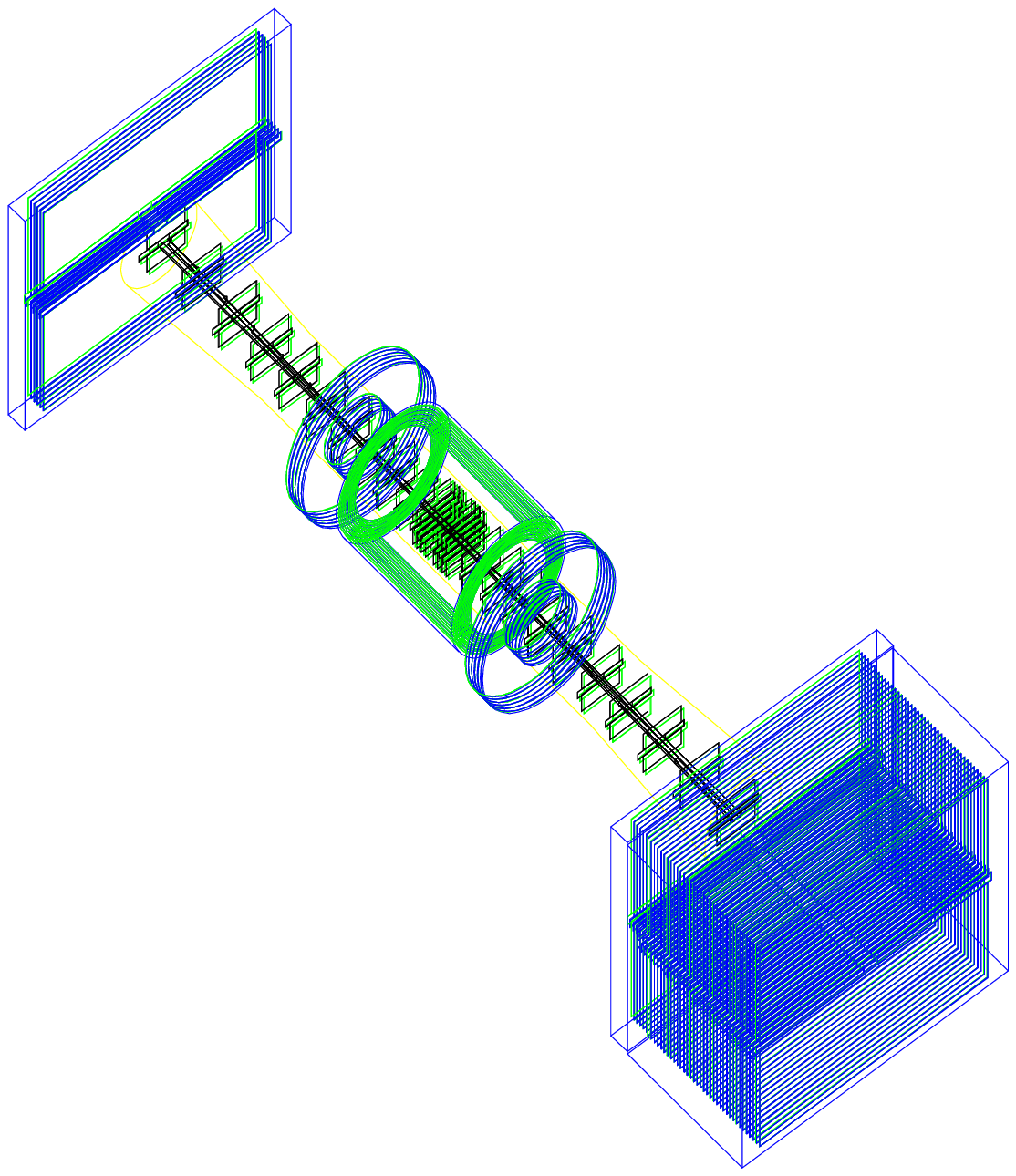,width=14cm}
\end{center}
\end{figure} 
\thispagestyle{empty}
\newpage
\hskip 1cm 
\thispagestyle{empty}
\newpage
\tableofcontents
\newpage

\newpage
\section{Introduction}
Quantum Chromodynamics is now widely acknowledged to
be the correct theory to describe strong hadronic interactions.  
Knowing the correct Lagrangian
for the theory does not, however, imply that we know everything we want to know
about the strong interactions.  Experimental observations such as the 
confinement of color or the energy dependence of scattering cross sections
cannot today be predicted from the QCD Lagrangian.
These are striking results which should be calculable from first principles
if we are to claim to understand our universe.

It has been proposed to build an electron accelerator at the Brookhaven
National Laboratory~\cite{ref:eRHIC} to allow electron-proton and electron-ion
collisions.  The electron beam would intersect the existing RHIC accelerator
at one or more locations.  The research program which would be made possible
with such an eRHIC facility is of great scientific interest, and it would
certainly push our understanding of strong interactions to a completely new
level.

The world of
small-x physics has been opened up by the HERA accelerator in Hamburg, Germany,
and it is clearly fascinating and full of surprises.  We now know that 
understanding electron-proton collisions is dependent on 
an understanding of the gluon sea. Unexpected
effects such as large diffractive cross sections were observed, which can only
be understood as the correlated exchange of multiple gluons.  With eRHIC,
it will be possible to study the role of gluons in electron-ion collisions, 
and to understand whether hadronic matter and its interactions at high
energies is fundamentally described by a gluon condensate (the Color Glass
Condensate~\cite{ref:CGC}).  The CGC is expected to provide a universal 
description of hadronic matter at small distances, and its discovery would 
revolutionize particle physics.

The large-x measurements will open up completely new territory.  
In this regime
the electron scatters off the valence quarks of the proton.  The distribution
of valence quarks has not been measured above x=0.75 at large $Q^2$ and 
new surprises may await us.  

The fundamental
measurements mentioned here require the extraction of inclusive cross
sections (the details of the hadronic final state are not analyzed), i.e., 
structure functions.  This document describes studies of a detector optimized
for structure function measurements in the small-x and large-x regimes.
The same detector would obviously be capable of making very precise
measurements of semi-exclusive and exclusive reactions, but these processes
have not been studied in detail.  Rather, we refer to a Letter of Intent
submitted to the DESY PRC for further experiments with the HERA accelerator
where these topics are covered in more detail~\cite{h3loi}.

The outline of this document is as follows: 
we first review the structure function measurements which have motivated the
new detector design and follow this with
a description of the resulting detector.
Detailed simulations of the
performance of the detector are presented, and, finally, we show 
the range and precision of structure 
function measurements which could be
extracted with moderate amounts of luminosity.  The studies were all performed
for electron-proton scattering, but the results carry over to electron-ion
collisions as well.

\newpage
\section{Precision structure function  measurements}
\vskip -1cm \hskip 10.2cm \footnotemark
\vskip 1cm
\footnotetext{We would like to thank Halina Abramowicz 
for providing most of the text in this section.}
\label{sec:precsf}
The measurements of the $F_2$, $F_L$ and $xF_3$ structure functions of
the nucleon were instrumental in discovering the point-like nature of
hadron constituents, quarks and gluons, in establishing QCD as the
theory of strong interactions and in determining the value of the
strong coupling constant.
\begin{figure}[hbt]
\begin{center}
\includegraphics[height=0.4\vsize]{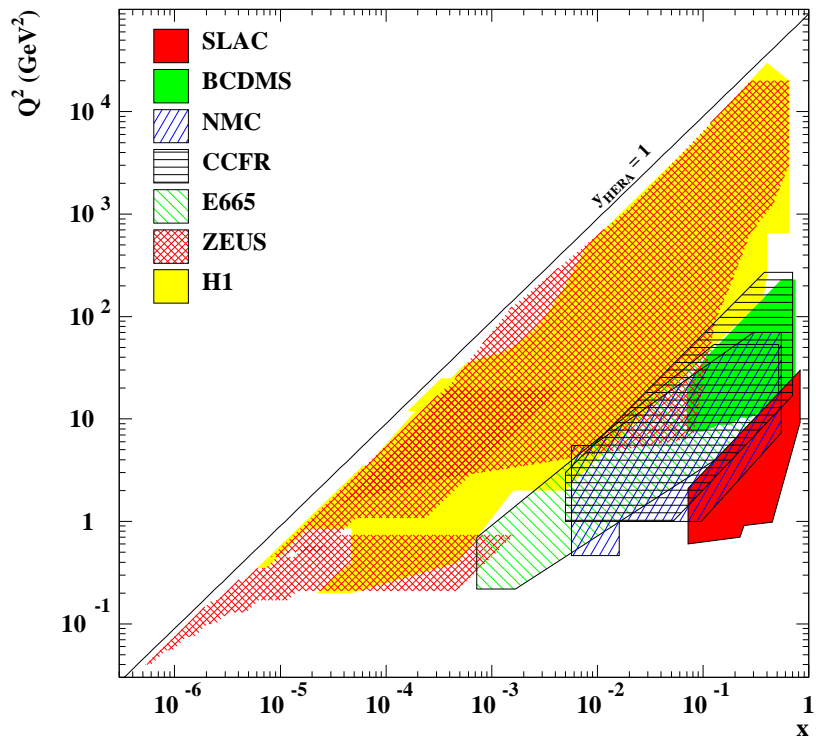}
\caption{$x$ and $Q^2$ plane covered by measurements of $F_2$ in charged 
lepton-nucleon scattering. }
\label{fig:kinematic-plane}
\end{center}
\end{figure}
Today, the measurements of $F_2$ cover a large phase space in Bjorken
$x$ and $Q^2$, as shown in Fig.~\ref{fig:kinematic-plane}, and are
used to determine a universal set of parton densities to be used for
predicting and calculating processes in phase space regions within and
beyond the probed space. The conventional approach is to postulate the
$x$ dependence of parton densities at some fixed $Q^2_0$ and to fit
the shape parameters, so that the parton densities evolved through the
DGLAP evolution equations reproduce best the $F_2$ data. There are
many unknowns in this procedure. In particular
\begin{itemize}
\item the shape of the parton densities cannot be derived from first
principles,
\item the scale at which the perturbative evolution can be safely
applied is not known,
\item there is no prior knowledge when to stop the perturbative expansion
of the splitting functions and the corresponding coefficient functions
entering the DGLAP formalism.
\end{itemize}

Using the NLO DGLAP evolution equations, the shape parameters
can be fitted very well to the
$F_2$ data, down to $Q^2$ values as low as $1
\gevtwo$~\cite{mrst2001,zeus-nlo,H1-nlo}. However, at low
$Q^2$ values, the solution~\cite{mrst2001,zeus-nlo}
requires a negative gluon density at low
$x$.  While it may be disputed whether unobservable gluon densities
can be negative, $F_L$, as a physical observable, has to be
positive. For DGLAP fits which include measurements of high
transverse momentum jets at the Tevatron~\cite{d0-jets,cdf-jets}, and
therefore require larger gluon content at high $x$, negative $F_L$
values for $x<2\cdot 10^{-4}$ appear already at $Q^2=2
\gevtwo$.  This puts in doubt the validity of the DGLAP formalism in this
low $x$ and low $Q^2$ region. Moreover, it becomes apparent that a
good fit to the $F_2$ data does not guarantee the applicability of the
DGLAP evolution equations, at least in NLO. This is especially true in
the low $x$ region, where the lever arm in $Q^2$ is small.

One striking experimental result from HERA
is the observation of a transition in the behavior
of $F_2$ at $Q^2 \approx 0.5$~GeV$^2$.  The $x$ dependence of $F_2$ at
fixed $Q^2$ can be directly related to the energy dependence of the 
photon-proton cross section.  Below $Q^2 \approx 0.5$~GeV$^2$
the virtual photon-proton
cross section has a very similar energy dependence to that seen in
hadron-hadron scattering.  Above this $Q^2$, the cross sections show
a steeper energy dependence.
The location of this transition is
particularly interesting, as it is near the perturbative regime, and far
from the scale set by the dimensions of the proton. 

At the other end of the $x$ scale, $x>0.5$, higher twist effects are
known to play an important role. At $x>0.6$ higher twist contamination
plays a role up to $Q^2$ of about $100 \gevtwo$ and is simply
parameterized in the fit procedure. The large $x$
region is important in the momentum balance of the parton densities,
and, via sum rules, affects the densities extracted at small 
$x$. This interplay
affects the determination of $\alpha_S$ from the evolution of $F_2$.
The region of large $x$ is important in the calculation of moments of
parton distributions which can be directly compared to lattice
calculations.

\paragraph{Importance of {\bf $F_L$}}

At low $x$ and relatively low $Q^2$, there are various dynamical
arguments why the NLO DGLAP conventional approach may fail. At low $x$,
higher order perturbative corrections may become important and
possibly the fixed-order perturbative expansion in the strong coupling
constant $\alpha_S$ may become inadequate~\cite{catani,thorne}. 
Present $F_2$ measurements alone are not sufficient to address these
issues. One of the main reasons is that in the $F_2$ analysis there
are only two experimental inputs, $F_2(x,Q^2)$ and $\partial
F_2(x,Q^2)/\partial \ln Q^2$. They allow to determine quark and gluon
densities, provided the theoretical framework is fixed. The latter can
only be checked if other observables are available. Such an observable
is the longitudinal structure function. 

\begin{figure}\begin{center}
\includegraphics[width=0.45\hsize]{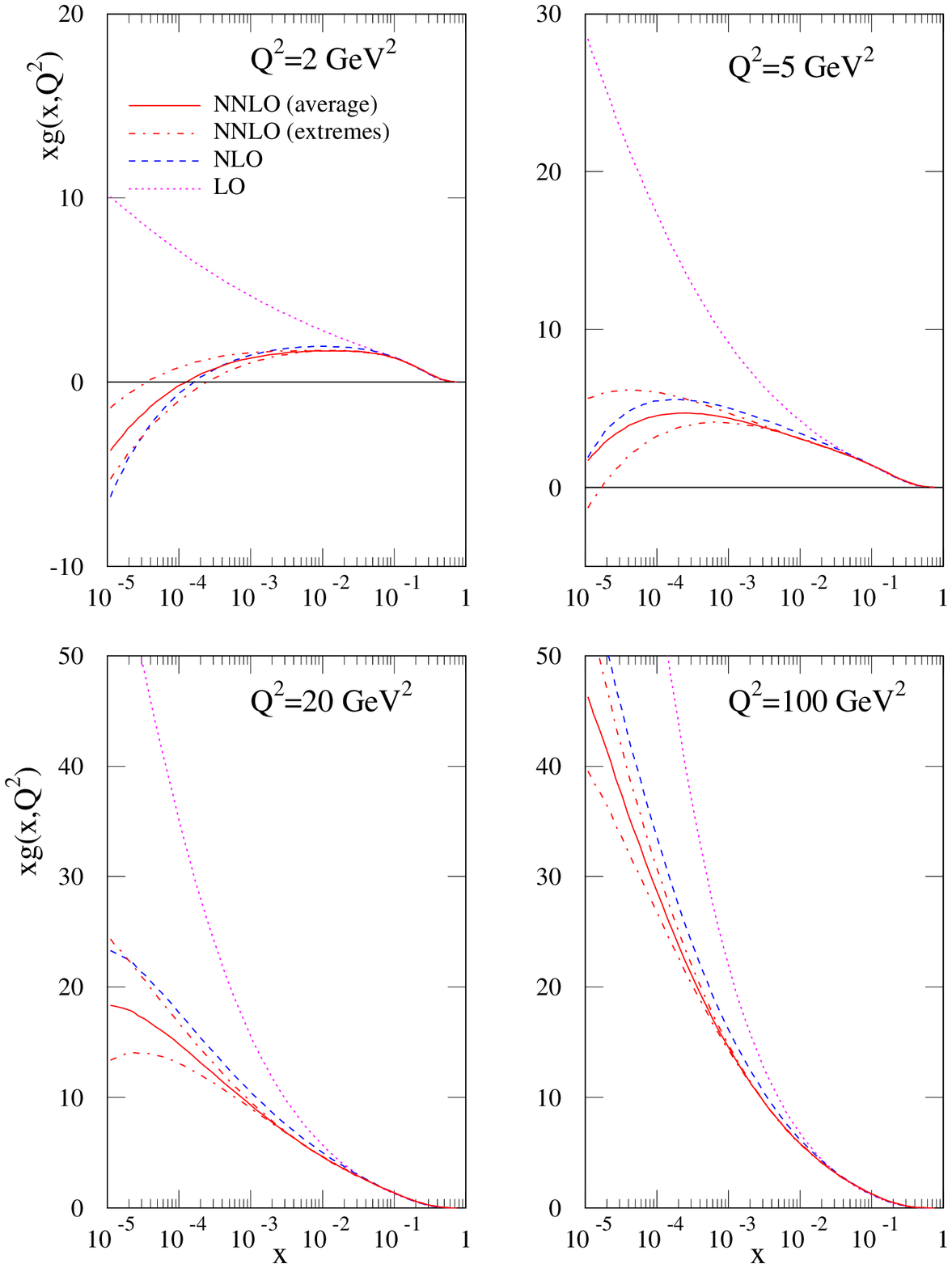}
\includegraphics[width=0.45\hsize]{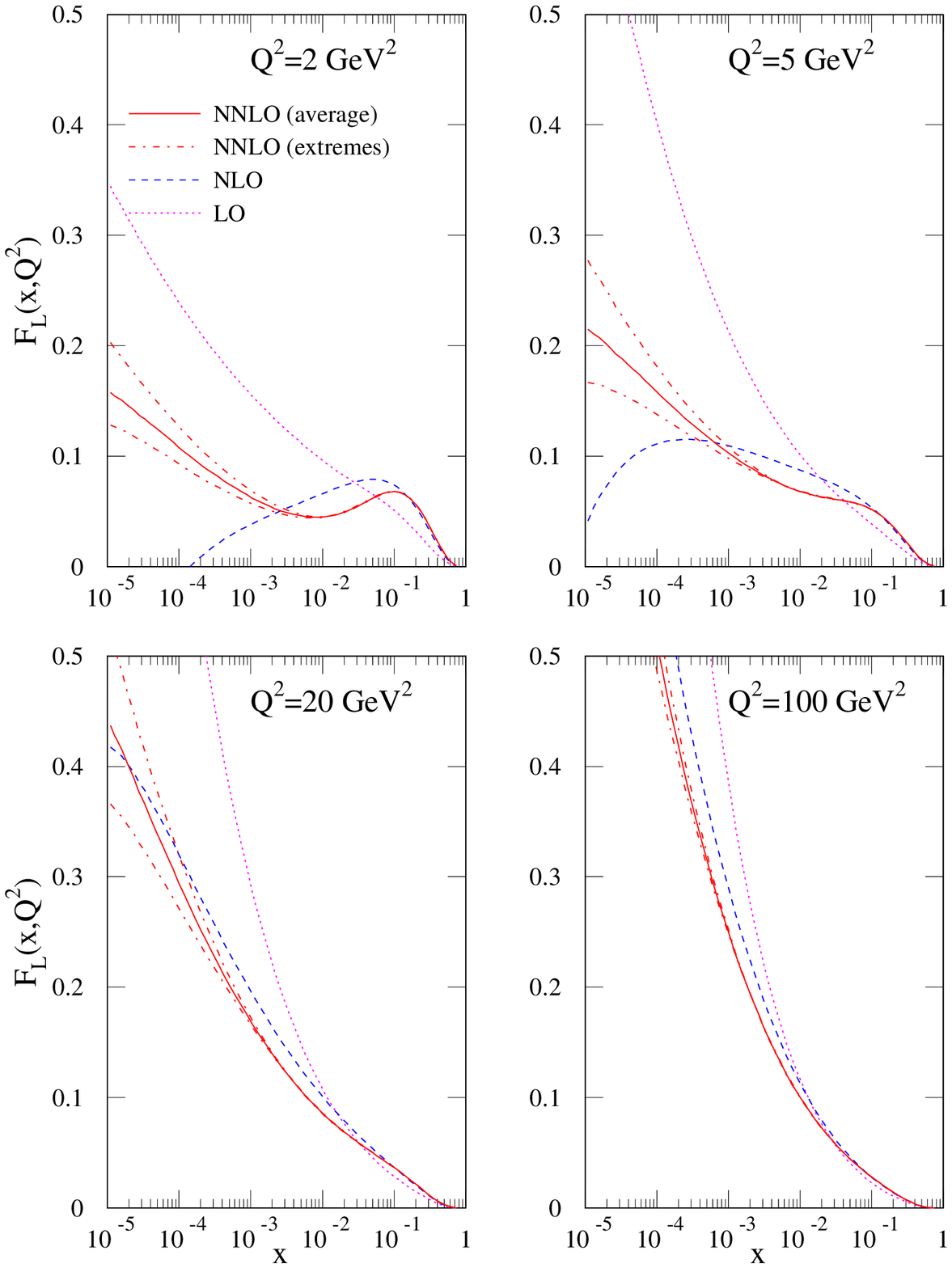}
\caption{Solution of the DGLAP evolution equation for gluons (left) at
  different orders of perturbative expansion and the corresponding
  expectations for $F_L$ (right) as a function of $x$ for fixed $Q^2$
  values, from ~\cite{thorne}. }
\label{fig:fl-thorne}
\end{center}
\end{figure}

The effect of using the DGLAP evolution equation at LO, NLO and NNLO for
extraction of the gluon density and for the calculation of the
corresponding $F_L$~\cite{thorne} is shown
in Fig.~\ref{fig:fl-thorne}.
The effects are quite dramatic especially at low $Q^2$.

\begin{figure}\begin{center}
\includegraphics[width=0.35\hsize]{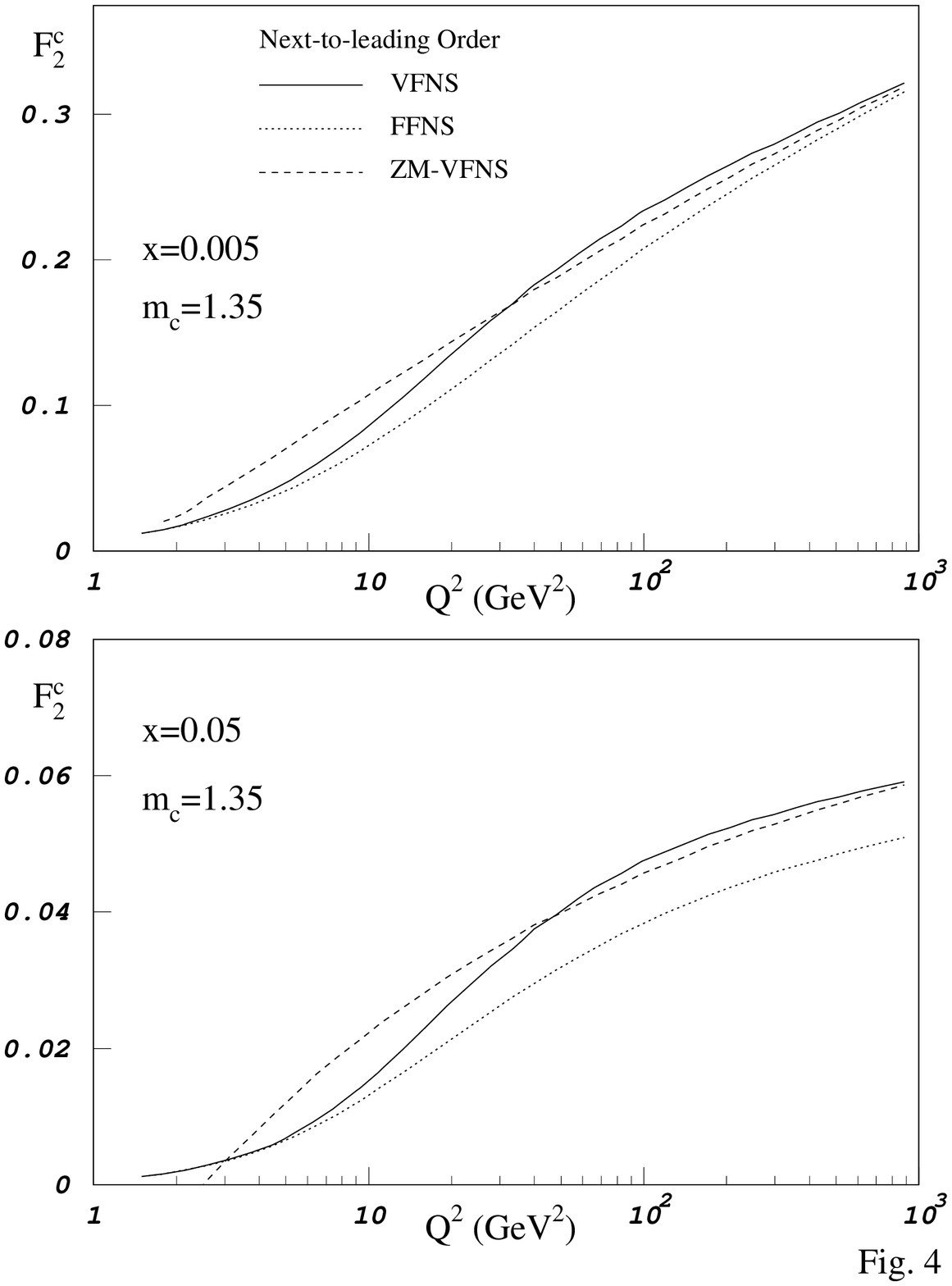}
\includegraphics[width=0.35\hsize]{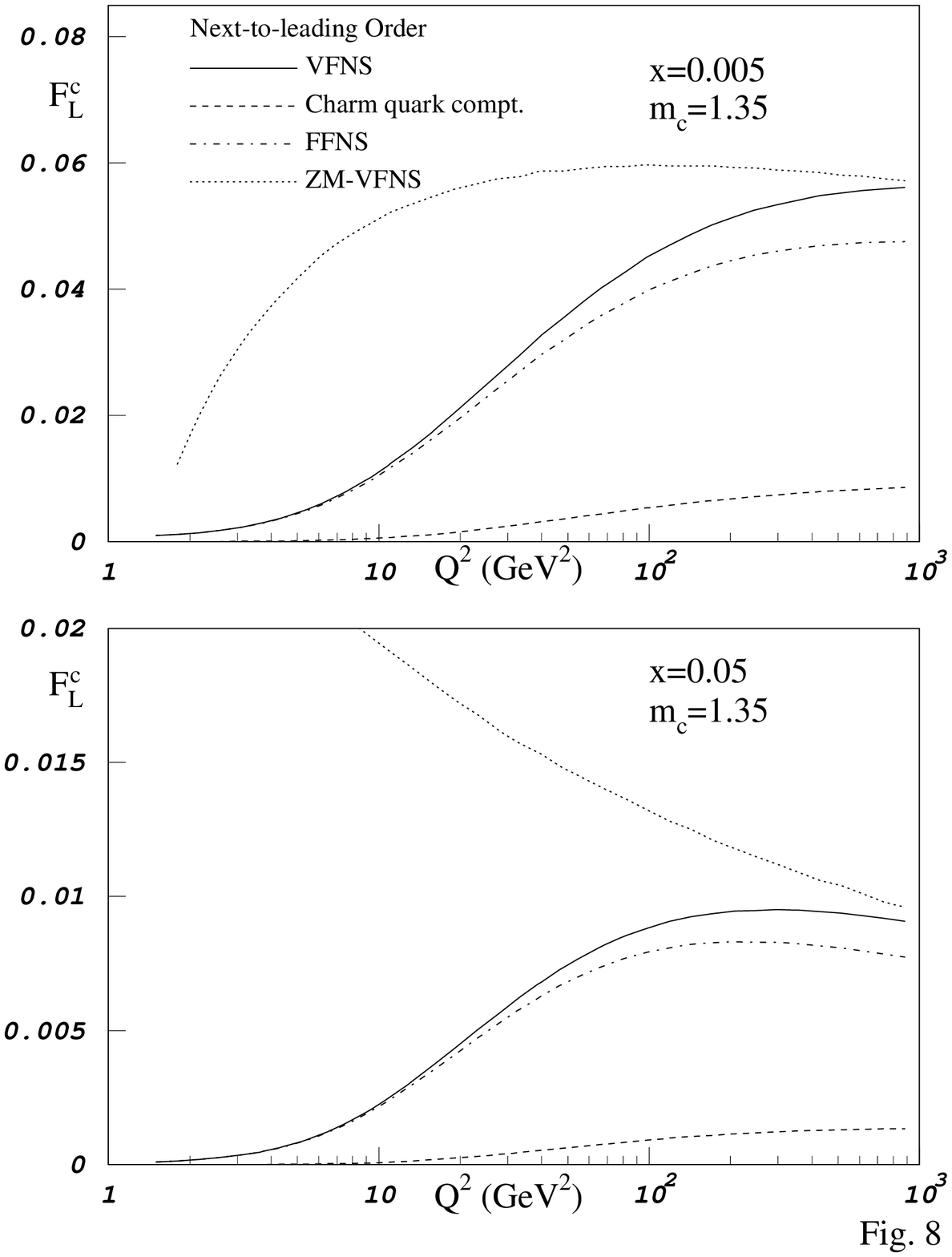}
\caption{Expectations from the 
DGLAP evolution equations for charm contribution 
  to $F_2$ (left) and $F_L$ (right) for various heavy-flavor schemes,
  as denoted in the figure (VFNS - variable flavor number scheme;
FFNS - fixed flavor number scheme; ZM-VFNS zero mass variable flavor
number scheme).}
\label{fig:fl-charm}
\end{center}
\end{figure}

The proper treatment of heavy flavor production in the low $x$ regime
is very important~\cite{grv-hf}. Various schemes have been proposed
and their predictions for the charm contribution to $F_2$, $F_2^c$, agree to
within few percent. However, the variations for the expected $F_L^c$
are very large, as shown in Fig.~\ref{fig:fl-charm}.

\begin{figure}
\begin{center}
\includegraphics[width=0.45\hsize]{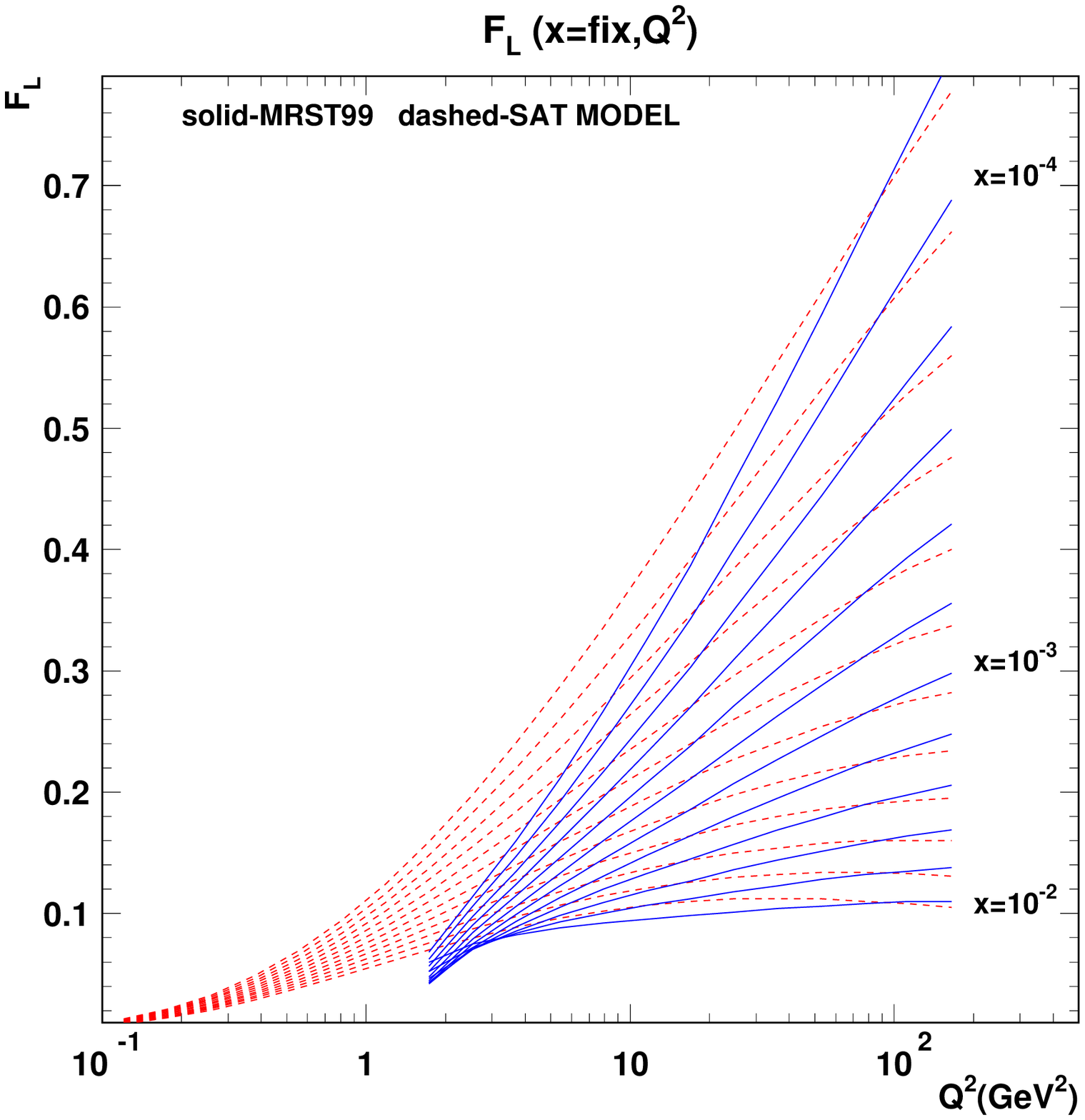}
\includegraphics[width=0.45\hsize]{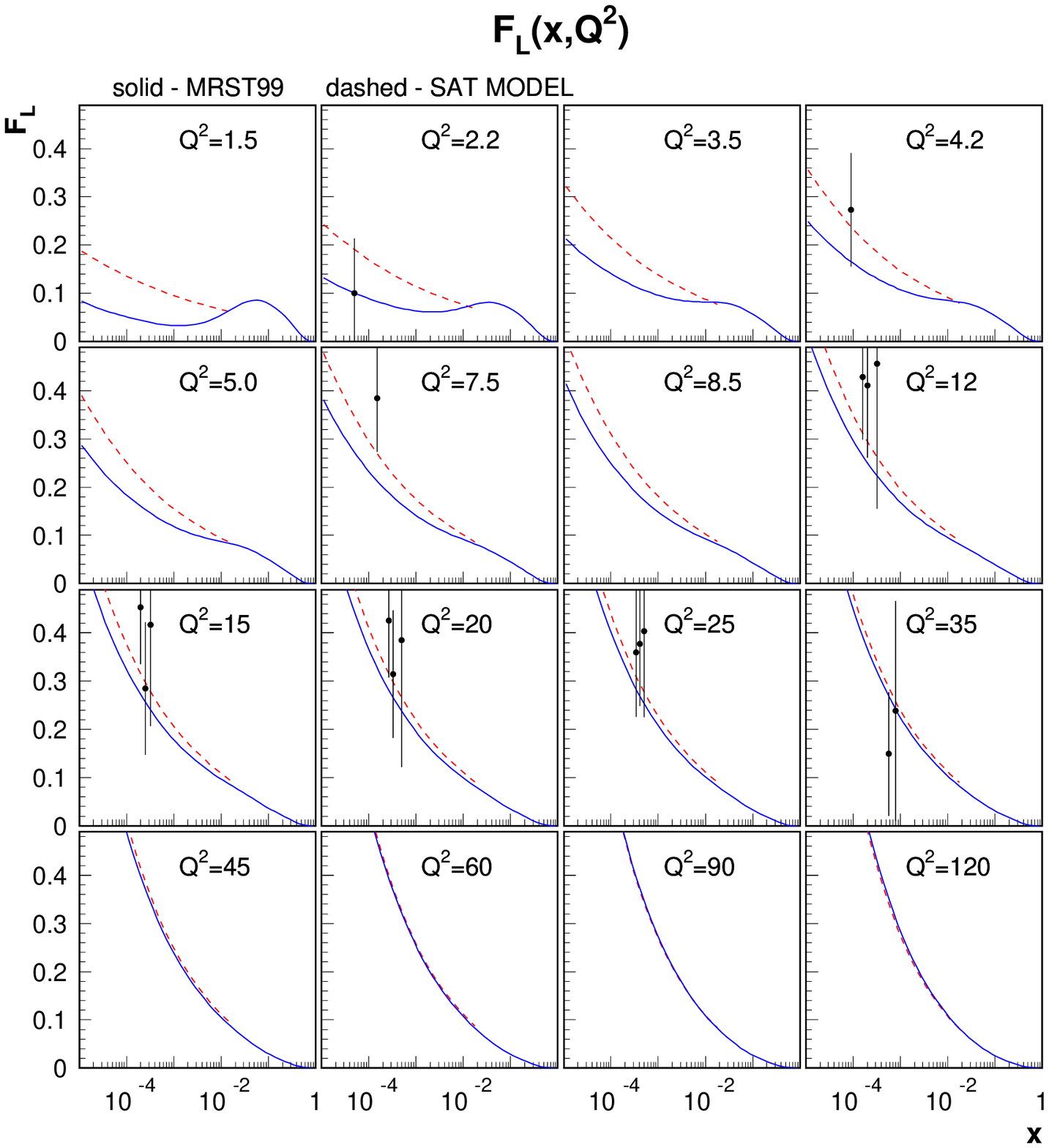} 
\caption{Comparison of the expectations for the $Q^2$ dependence at fixed $x$ 
  (left) and $x$ dependence at fixed $Q^2$ (right) of $F_L$ in the DGLAP
  approach (MRST99) and the saturation dipole model. The right-hand plot
also includes data from H1~\cite{h1fl}.}
\label{fl:dipole}
\end{center} 
\end{figure}

In the region where $F_2$ rises steeply with decreasing $x$, the
parton densities are large and one might expect coherent effects, such
as multi-parton interactions, leading to saturation (shadowing)
effects in the interaction cross section of the virtual photon with
the proton. The observed copious presence of diffractive events in the
DIS regime could be one of the manifestations~\cite{levin-dd}. There
are theoretical arguments for the presence of higher twist effects,
positive for $F_L$ and negative for $F_T$ ($F_2=F_L+F_T$), that would
approximately cancel for $F_2$~\cite{bartels-ht}, invalidating however
the leading twist DGLAP approach. The dipole model of
DIS~\cite{amirim}, in which the photon fluctuates into a $q\bar{q}$
pair before interacting with the proton, in a realization proposed by
Golec-Biernat and Wuesthoff~\cite{gbw}, seems to have all these
dynamical properties and will be used as a reference. The comparison
between the NLO calculations of $F_L$ and the calculation of $F_L$ in
the dipole model is shown in Fig.~\ref{fl:dipole}. Also shown are
the present measurements of $F_L$ by the H1 experiment~\cite{h1fl}.
The differences are indicative of the contribution of positive
higher twists in $F_L$. The most striking difference is in the $Q^2$
dependence of $F_L$ at fixed $x$, where with adequate measurement
precision, it should be possible to distinguish even between the NNLO
and the dipole model expectations.

\paragraph{$F_2$ at small and large $x$}

\begin{figure}[hbt]
\begin{center}
\includegraphics[height=8.cm]{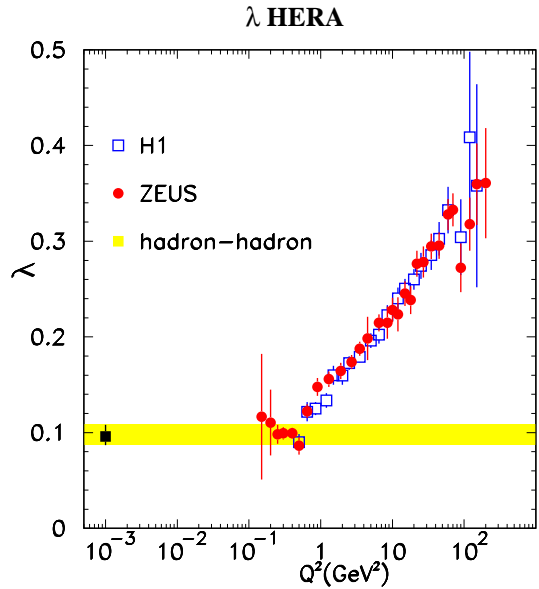}
\caption{ 
The energy dependence of the $\gamma^*p$ cross section measured at HERA
compared with the results from hadron-hadron scattering (yellow band).
The yellow band corresponds to the results of a fit to hadron-hadron
data~\cite{Cudell}.  The HERA data are from ZEUS and H1 and only include data
with $x<0.01$.}
\label{fig:lamda}
\end{center}
\end{figure}

One of the important parameters in the study of high energy processes
is the logarithmic derivative $\lambda$ of $F_2$, defined as
\[ \lambda = - \partial ln F_2/\partial \ln x  \,. \] 
Its value is related to the contribution of soft and hard dynamics in
the (virtual) photon-proton interactions. The value of $\lambda$ 
extracted from the small $x$ ($x<0.01$) HERA data is shown in 
Fig.~\ref{fig:lamda}.   The present data  point to a 
transition  around $Q^2\approx 0.5 \gevtwo$, where, for smaller values of
$Q^2$, the data are compatible with the equivalent quantity measured in
hadron-hadron scattering. The energy dependence of hadron-hadron scattering
is dominated by non-perturbative physics and has never been satisfactorily
explained.  On the other hand, the high $Q^2$ HERA data follows the 
expectations from perturbative QCD calculations using the DGLAP formalism.
The HERA data therefore span the kinematic region where a transition from a
partonic description to a hadronic description of photon-proton collisions
is observed. A detailed study of this transition region could lead us to an
understanding of fundamental non-perturbative QCD phenomena at work in
hadrons.

 The quality of the
existing data in the transition region is not as high as in
other regions, because this corresponds to a geometrical location for
the scattered electron
where the H1 and ZEUS detectors have very limited acceptance.
A detector dedicated to this region could make much higher 
precision measurements in an extended kinematic range.

No measurements of $F_2$ are available for $x>0.75$ in the large $Q^2$
region, where higher twist effects are likely to disappear.  The
non-perturbative corrections to $F_2$ at large $x$ are expected to
scale like $1/Q^2(1-x)$~\cite{dokshitzer}. If this claim can be
verified by performing measurements up to the highest $x$ values, this
would give a very important handle on the parton densities at large
$x$.  The current
H1 and ZEUS detectors are only able to measure large-x events at very high
$Q^2$, i.e., with small statistics.  The main limitation is the
acceptance of the detector to very forward jets.  A new detector, designed
with this physics in mind, would be capable of extending the large-x
measurements to much lower $Q^2$, thus allowing for significant event
rates and the ability to extract precision cross sections.

\paragraph{Structure Functions and the Color Glass Condensate}

The Color Glass Condensate~\cite{ref:CGC}, CGC, represents a new way
to look at hadronic matter.  It is known that
QCD predicts the gluon density to increase
steeply at small-x as $Q^2$ increases.  However, it is not possible
to calculate the gluon density in a hadron from the QCD Lagrangian.
The CGC approach starts from the concept that hadrons are represented
by collections of gluons in a saturated (maximal number density)
state.  The number density defines a distance scale, and thereby a
saturation scale, $Q_S$.  The variation of this scale with $W$ is
predicted by the CGC, and the current excitement lies in the belief
that the saturation scale may have already been seen in the HERA 
data. The HERA results are consistent
with a CGC with $Q_S \approx 1$~GeV for $x=10^{-4}$, but the existence
of the CGC is not proven~\cite{ref:DIS04}.  What is needed are data
beyond that which HERA has delivered, either in an extended $x$ range or
with different targets.  Of greatest importance are 
inclusive cross section data for ep and eA scattering.  The
CGC makes predictions on the behavior of the structure functions as A 
varies, and these could be tested with an electron-ion collider.  
Other important tests of the presence of the CGC would come
from semi-inclusive and exclusive measurements.

An example of a semi-inclusive measurement is the study of hadron
production and correlations among hadrons as a function of momentum and
rapidity for fixed electron kinematics ($Q^2,x$).  The presence of the
CGC would yield different patterns than expected from standard fragmentation
approaches.  These studies require large rapidity acceptance and good
momentum resolution for the hadronic final state.

Exclusive measurements, such as $ep(A) \rightarrow ep(A) V$ or
$ep(A) \rightarrow ep(A) \gamma$, where $V$ represents a vector
meson, allow to test the radial profile of hadronically interacting matter.
A full three-dimensional reconstruction of the nucleon (nucleus) is
then possible.  Extracting the scattering amplitude for virtual
photons on nuclear targets would directly test whether the saturation
regime has been reached.  These types of measurements require the
maximum rapidity coverage for the outgoing vector meson or photon, good
momentum and energy resolution, and the means to verify that the
scattered proton remains intact.

\paragraph{Required precision}

To validate the universality of parton densities derived from the HERA
data, $F_L$ has to be measured in the region of $1<Q^2<20 \gevtwo$
with a precision of about $5 \%$. This would allow to distinguish
between the various theoretical approaches at a $4\sigma$ level.

The precision achieved in the measurements of $F_2$ at HERA has
reached about $2\%$. This leads to an error of $\lambda$ of about $5
\%$ for $\lambda = 0.2$. In the transition region of interest, $\lambda
\simeq 0.1$. Therefore, all efforts should be made to measure $F_2$
at low $Q^2$ with a precision better than $2\%$.

At large $x$, the goal is to push the measurement beyond $x=0.75$ for
$Q^2$ values in the DIS regime.  The accuracy should be better than
$10$~\% to be sensitive to novel effects.

For more detailed tests of our understanding of QCD and the possible
presence of the CGC, it is important to measure the hadronic final state
over the largest possible rapidity range.  We set as a goal to have
momentum analysis of particles in the range $|\eta|<5$.

\newpage
\section{Summary of measurements and requirements}
\subsection{Summary of measurements}
In summary, the following measurements are seen as the highlights of the
proposed program:
\begin{itemize}
\item The high precision measurement of $F_2$ at low x from $Q^2=0.05$~GeV$^2$
to $Q^2 = 5$~GeV$^2$ to better understand the observed
transition of the cross sections from hadronic to partonic behavior.
\item The measurement of the longitudinal structure function, $F_L$,
particularly at $Q^2$ values below $10$~GeV$^2$, where present theoretical
and experimental uncertainties are very large.
\item The measurement of forward jets and forward particle production up
to pseudorapidities of at least 
$\eta=4$ to test in a direct way our understanding
of parton branching in strong interactions and to see the onset of
collective phenomena.  Acceptance for 
forward jets will also allow the measurement of $F_2$ to $x=1$ at
moderate $Q^2$.
\item The measurement of diffractive and exclusive reactions (VM production
and DVCS) over the full $W$ range, and to values of $|t| \le 1.5$~GeV$^2$,
with no proton dissociation background, to perform a three dimensional
mapping of the proton and perform first extractions of generalized
parton distributions.
\end{itemize}
All of the above measurements should be performed with protons and 
with at least two nuclear targets to search for the gluon condensate,
and understand nuclear effects in parton distributions.

\subsection{Accelerator requirements}
The luminosity requirements for the e~p~program
are set by exclusive cross section measurements
at high $t$, $F_2$ measurements near $x=1$, and by the $F_L$ measurement. 
For the latter we
anticipate requiring data sets with at least two different
proton energy settings (e.g., $E_P=100, 200$~GeV), and preferably with three 
or four. A luminosity requirement of
approximately $100$~pb$^{-1}$ at each energy is anticipated.
The $F_2$ measurements near $x=1$ will be statistics limited for luminosities
below 100~$fb^{-1}$.  First interesting measurements in a new kinematic
domain will already be possible with $100$~pb$^{-1}$.
High-$t$ exclusive processes have not been studied in detail with this
detector design.  However, using a parametrization of exclusive $\rho$
meson production at HERA~\cite{Arik}, it was determined that 
$1$~fb$^{-1}$ would yield $10^5$ elastic rho events in the 
interesting kinematic
range $30<W<90$~GeV, $Q^2>2$~GeV$^2$ and $|t|>1$~GeV$^2$.  This would be a
minimum requirement
for an extraction of a three dimensional cross section (in $W,Q^2,t$)
at high $t$, which corresponds to small proton (nucleus) impact parameters.
Summarizing these requirements, we find that a luminosity at the level of
$2\cdot 10^{32}$~cm$^{-2}$s$^{-1}$ with an active time of $10^{7}$~s/year
would satisfy the minimum requirements for all the measurements above 
except for structure function measurements at the very highest $x$.

The luminosity required for the e~A 
program has been estimated at $2$~pb$^{-1}$ per 
nucleon~\cite{ref:heraws96_krasny}.  We have not investigated eA measurements
in this study, and so take this figure at face value.
Finally, we do not anticipate needing positron
beams - electron beams will suffice - since we are concerned primarily with
the $Q^2$ region where photon exchange dominates.  

This study does not focus on beam polarization. However,  
it would of course be very desirable and allow for
a broadened physics program.

\subsection{Detector requirements}

The main focus of the detector design is large
angle coverage. Emphasis is put on forward/backward physics, which
distinguishes the design from typical HEP experiments such as the 
existing H1 and ZEUS detectors.
The measurement of F$_L$ and F$_2$ at small Q$^2$ ($0.05 < Q^2 <5$~GeV$^2$)
requires the tracking and identification of scattered electrons 
with energies between 1~GeV and 10~GeV
up to high rapidities.

The rapidity range is extended
by a dipole field that separates lower energy
scattered electrons from the beam. 
In the electron hemisphere tracking and identification require
\begin{itemize}
\item high precision tracking with 
 \begin{itemize}
 \item $\Delta$p/p of $\approx$2~\%
 \item angular coverage down to a pseudo-rapidity of $\approx$-6 over the
       full energy range 
 \end{itemize}
\item em calorimetry with 
\begin{itemize}
\item an energy resolution of better than $\approx$20~\%~$\sqrt{E}$
\item e-$\pi$ separation with a pion rejection factor of at least 20
      at 90~\%  efficiency over the electron energy range from 1~GeV to
      the beam energy of 10~GeV.
\end{itemize}
\end{itemize}

Another focus of the program is the study of forward jets, high-$x$
events, and exclusive processes. 
This leads to the following additional requirements for the
proton hemisphere:
\begin{itemize}
\item high precision tracking with  
 \begin{itemize}
 \item similar precision in the proton hemisphere as
       in the electron hemisphere
 \item similar angular coverage up to a pseudo-rapidity of $\approx$6 
 \end{itemize}
\item em and hadronic calorimetry in the proton hemisphere with 
\begin{itemize}
\item an em energy resolution of  better than $\approx$20~\%~$\sqrt{E}$
\item a hadron energy resolution of  better than $\approx$50~\%~$\sqrt{E}$
\item e-$\pi$ separation with a pion rejection factor of at least 10
      at 90~\% efficiency above $1$~GeV energy.
\end{itemize}
\end{itemize}

\newpage
\section{A detector designed for forward/backward physics at eRHIC}
\subsection{Detector concept} 
The main idea is to build a compact detector with 
tracking and central em calorimetry inside a magnetic dipole field
and
calorimetric end-walls outside. To keep the magnetic volume of reasonable size,
the design limits the detector radius inside the magnet[s]
to a radius of 80~cm.
The coordinate system has the z-axis parallel to the proton beam,
the x-axis horizontal and the y-axis vertical. The electrons thus
point towards negative z.

The tracking focuses on forward and backward tracks. The calorimetry
is to show the best performance in the central region where
momentum measurements are 
intrinsically less precise due to the field configuration
and thus e-$\pi$ separation is more difficult.
Tracking for $|\eta|<0.5$ is currently not foreseen.

The detector presented here is 
an adaption and optimization of the detector which was proposed for
an extension of the HERA program~\cite{h3loi}. 

\subsection{Interaction region}
\label{sec:ia}

\begin{figure}[hbpt] 
\begin{center}
\epsfig{file=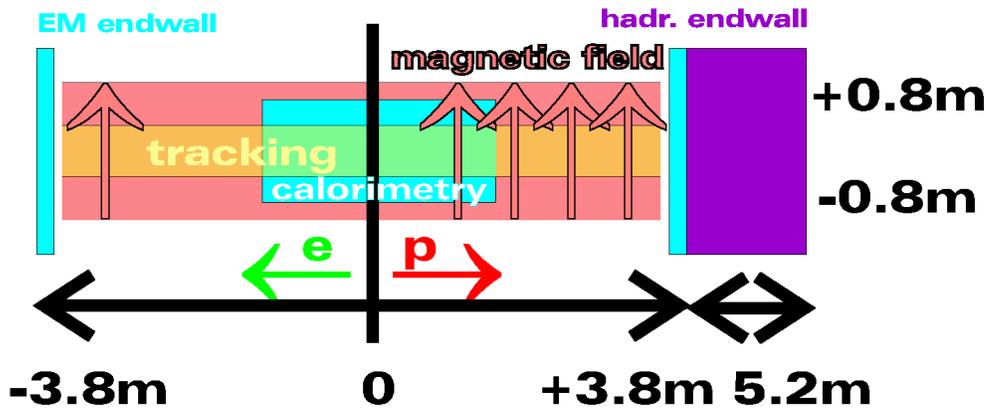,width=13.0cm}
\caption{Conceptual layout of the detector with
         a 7m long dipole field and an interaction region
         without machine elements extending from -3.8~m
         to +5.2~m}
  \label{det:fig:bl}
\end{center}
\end{figure} 

The interaction region is characterized by the presence of the
detector dipole field which has to become part
of the
machine lattice. 
Other machine elements cannot be incorporated inside the detector,
because they would reduce the acceptance for scattered particles.
Fig.~\ref{det:fig:bl} shows a block diagram of the interaction zone.
The first machine elements are placed outside the area from -3.8~m~
to~+5.2~m.
Longitudinal space for cabling and support is included in each block.
In x and y there are a priori no restrictions, so that the endwalls
can have their complete infrastructure outside.

The dipole magnet causes the electron beam to create a strong
synchrotron radiation fan. This is reduced by a longer magnet
with a weaker field. However, 
the distance between the interaction point and the first quadrupole
cannot be arbitrarily large, as the quadrupole aperture has to contain
the synchrotron fan.
The current machine studies for eRHIC envision an electron ring
or an electron linear accelerator as the source of leptons. For the
ring option also luminosity considerations limit the distance
to the first quadrupoles.

The 7~m long field indicated in Fig.~\ref{det:fig:bl} requires
a large aperture quadrupole in the electron direction.
The width of the radiation fan could be significantly reduced
by a split field. The magnetic field orientation would be opposite in
the electron and proton hemispheres. A small area around 
the interaction point would be ``field free".
This solution was chosen for the proposed detector at HERA, because
the higher electron energy at HERA required a substantially longer
magnetic field of 9~m.
For eRHIC the single field option would, however, be easier to integrate
into the machine lattice. A single dipole field is also favorable from
the analysis point of view, because a split field magnet would add 
significant complications to the reconstruction process.

The integration of the detector and its field into the machine 
lattice will be a crucial point in future studies.
The synchrotron radiation fan will require excellent masking.
In addition detailed studies of beam-gas backgrounds will be needed.

\subsection{Beam-Pipe}

\begin{figure}[h]
\begin{center}
\epsfig{file=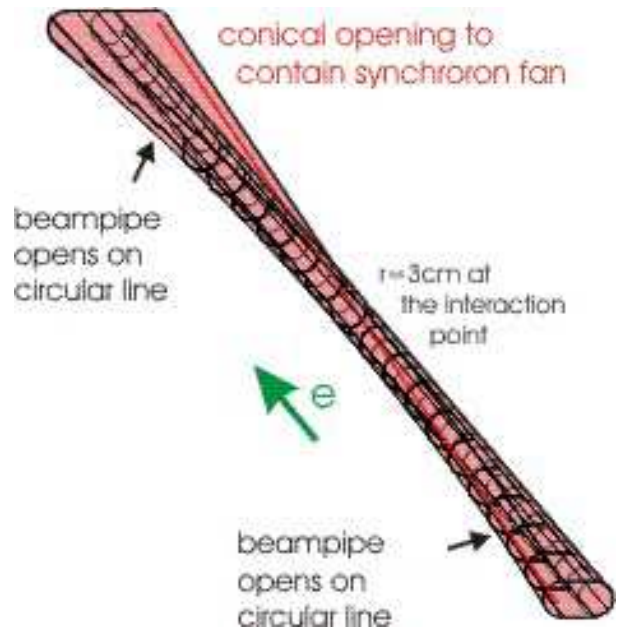,width=8.0cm}
\caption{Conceptual beam-pipe design for a long dipole field around the
    interaction point.
    } 
  \label{det:fig:bp1}
\end{center}
\end{figure} 

The beam-pipe currently envisioned is depicted in Fig.~\ref{det:fig:bp1}.
Vertically the radius is fixed to 3~cm.
Horizontally it follows the circular path of the electron beam on
one side and opens up conically on the other.
The electron beam follows an orbit with a radius of 110~m inside the
dipole field. At a distance of 350~cm from the interaction point
this requires and additional opening of the beam-pipe by 5.6~cm, 
adding to a total radius
of 8.6~cm. 

\begin{figure}[h] 
\begin{center}
\epsfig{file=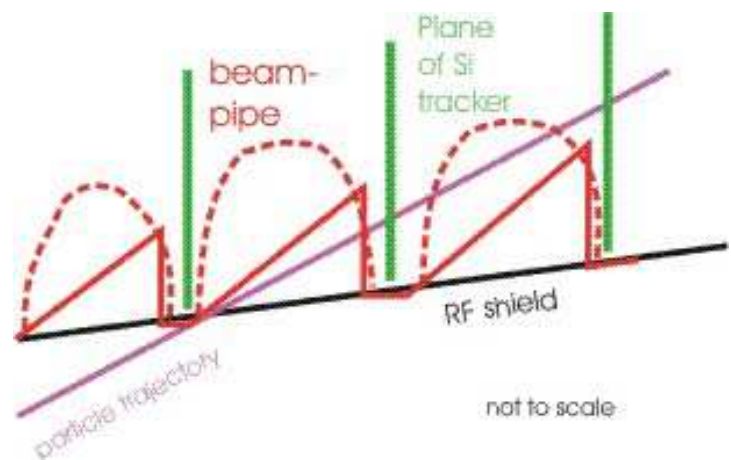,width=9.5cm}
\caption{Possible schematic beam-pipe designs to minimize the impact of
         the material of the beam-pipe on the particle trajectories.
         The solid red line depicts an opening of the beam-pipe in
         front of each plane; the dashed indicates another option
         considered.   
    } 
  \label{det:fig:bp2}
\end{center}
\end{figure} 

At the moment the material assumed is aluminum with a wall thickness of
~0.5~mm. 
This is not quite realistic for such a large beam-pipe, but
it serves as a reasonable start for the optimization of the design
and is used in performance studies.
The central part of the beam-pipe could be manufactured in
beryllium to reduce material.
The forward parts 
where particles traverse the pipe at shallow angles can
undergo geometric optimization.
Two possible  design options are depicted in
Fig.~\ref{det:fig:bp2}.
These types of designs introduce 
the additional complication of RF-shielding.
It is also possible to integrate a full Roman pot system 
like the one used in
HERA-B~\cite{hbvds}. This would, however, make the 
construction significantly more complicated and expensive.

As beam-gas events will be a severe background, the quality of the vacuum
will be of utmost importance. Therefore we envision to integrate
ion-getter-pumps into the beam-pipe. 
In case of a split field magnet configuration the beam-pipe 
would not need an extra conical opening for the synchrotron fan.
The smaller volume would reduce the pumping capacity needed.

\begin{figure} 
\begin{center}
\epsfig{file=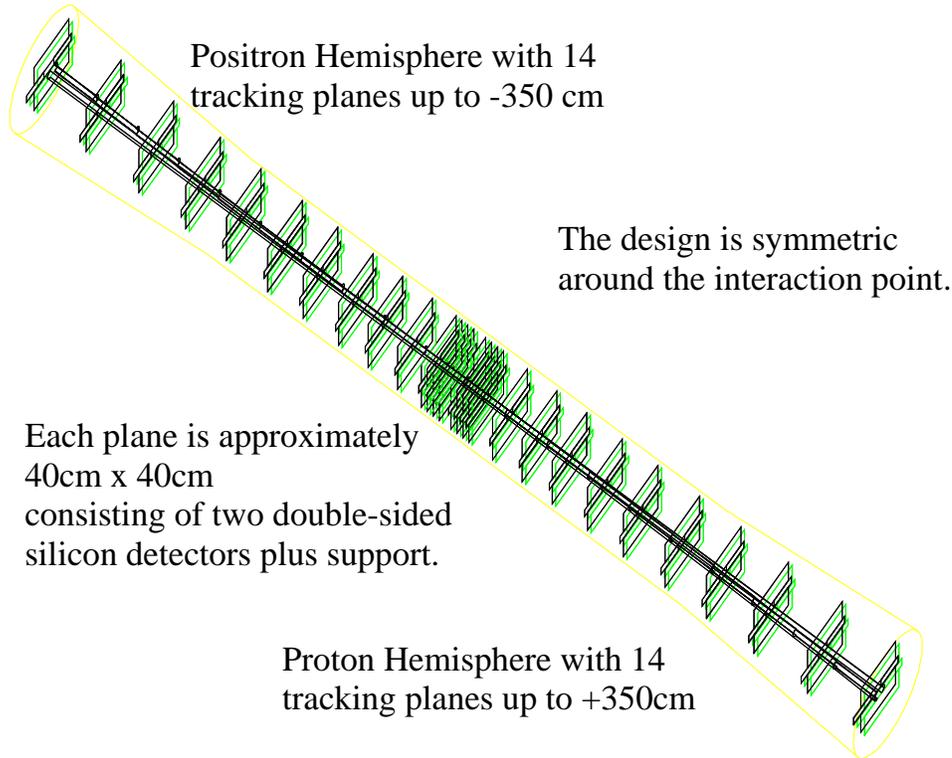,width=12.5cm}
\caption{Overview over the silicon tracker. In the central
         region the system is confined
         to cylinder with a radius of 30~cm in order to leave space for
         the calorimetry inside the magnet.
    } 
  \label{det:fig:siliov}
\end{center}
\end{figure}

\subsection{Tracking System}

The precision tracking required translates into hit resolutions
of less than about 50~$\mu$m.
This makes silicon strip detectors an obvious choice of technology.
As the material budget will be important, double-sided detectors
are desirable. A read-out-pitch between 50~$\mu$m and 100~$\mu$m
will be adequate.
Two double-sided detectors with appropriate stereo angle design
can yield unambiguous space-points.
For the baseline design we assume silicon tracking stations 
with two double-sided, 300~$\mu$m thick silicon strip detectors 
and support structures with material equivalent to 1.2~mm of carbon fiber.
This results in stations with a material budget equivalent to
$\approx$1~\% of a radiation length X$_0$. 

Figure~\ref{det:fig:siliov} gives an overview over the design.
The complete tracker is composed of planes oriented perpendicularly
to the beam. 
The planes are centered around the proton beam line and
measure approximately 40$\times$40~cm$^2$. They have a central cut-out
that follows the beam-pipe design.
Each plane is composed of a top and a bottom half and two
horizontal plugs that are adjusted to the required cut-out.
The positioning of the planes was optimized for acceptance
and momentum resolution. Each hemisphere features 14~silicon planes.
The silicon plane furthest from the interaction point starts at
z~=~3.5~m. 
Close to the interaction point the planes are
relatively densely packed; they are used to track low momentum
tracks with a large curvature and tracks with pseudo-rapidities less than
$\approx$1. Further out the planes have larger distances to enhance
the lever arm for tracking particles with higher momenta.

\begin{figure} 
\begin{center}
\epsfig{file=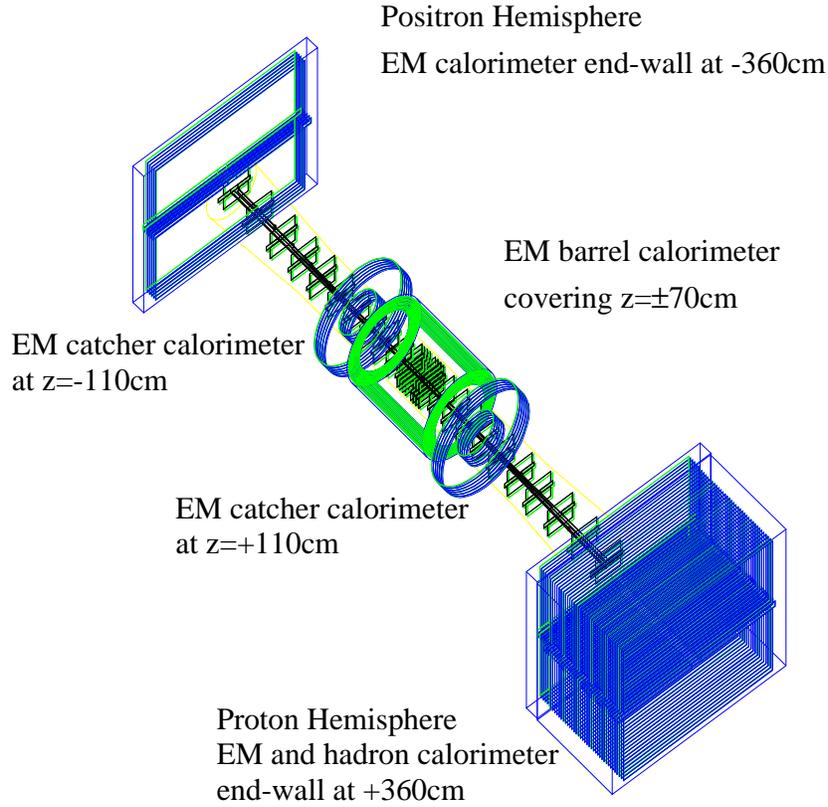,width=11.0cm}

\caption{ Schematic overview over the detector components
          within $\approx\pm$5~m of the interaction point.
          The silicon planes are visible inside the 
          yellow tracking volume.
          The calorimeter system consisting of
          a central barrel, a catcher ring on each side
          and end-walls is depicted in blue and green.
          } 
  \label{det:fig:detov}
\end{center}
\end{figure}

\begin{figure} 
\begin{center}
\epsfig{file=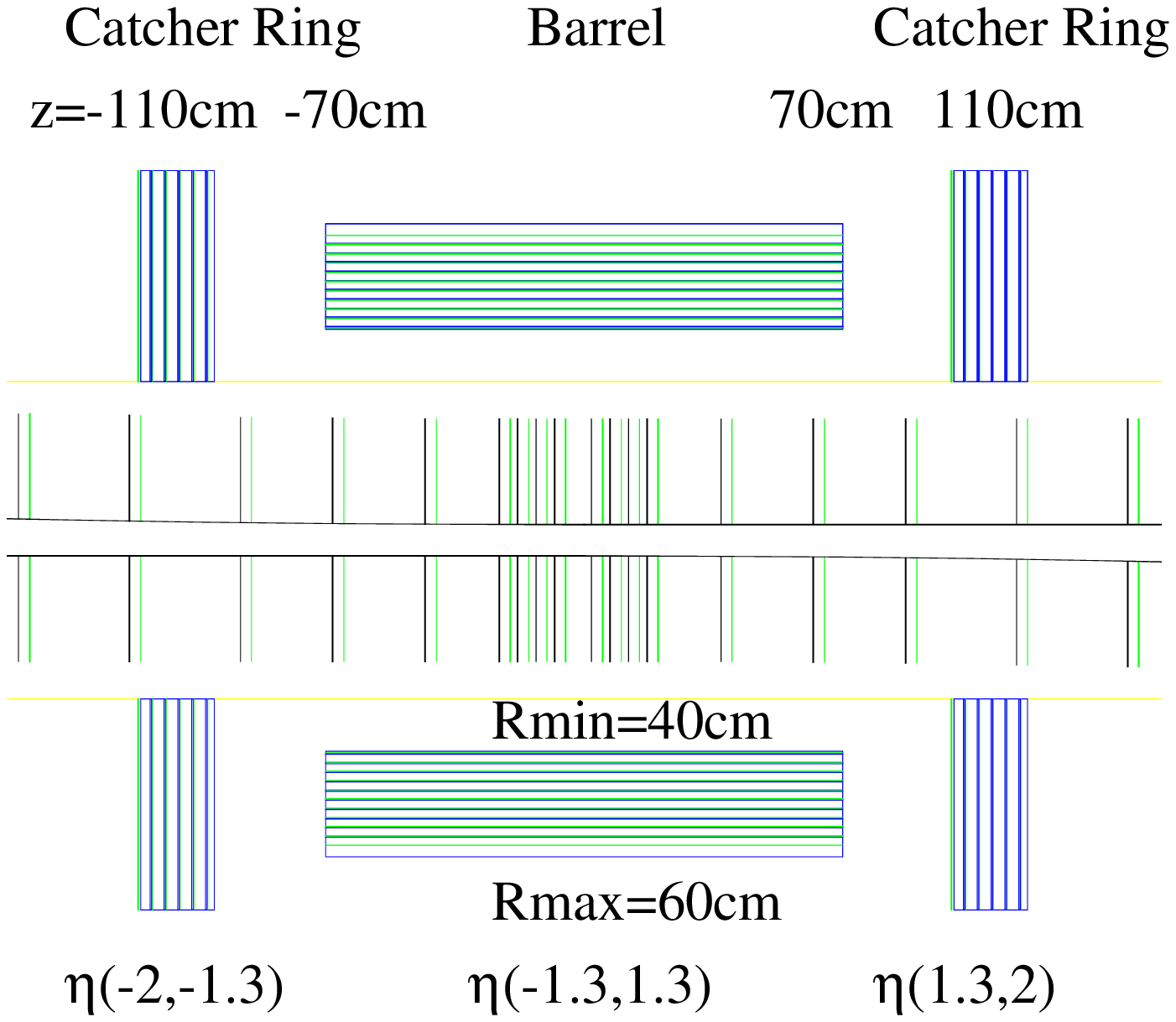,width=12.5cm}

\caption{Overview over the central part of the EM calorimeter system. 
         The central barrel has a
         radius of 60~cm. The 2 catcher rings extend from r~=~30~cm
         to r~=~70~cm.
         Also visible are the active silicon planes[green] and support
         structures [black].   
    } 
  \label{det:fig:calocent}
\end{center}
\end{figure} 

\subsection{Calorimetry}

Over most of 
4$\pi$ only electromagnetic calorimetry is required.
The goal of a compact detector
leads to silicon-tungsten as the choice of technology.
It is assumed that the magnet starts at r~=~80~cm and 
in order to leave sufficient space for support and cabling 
any barrel structure inside is confined
to a tube with a radius of 60~cm. 
In order to cover 4$\pi$ in an elongated design,
multiple structures have to be adapted.  
Figure~\ref{det:fig:detov} shows an overview over the full
detector, Fig.~\ref{det:fig:calocent} depicts the side
view of the central region.

The central region in pseudo-rapidity [$|\eta|<1.3$] is covered
by a barrel structure with an inner radius of 40~cm.
It extends to $\pm$70~cm in z.
For a simple, non pointing layer geometry this results in
a doubling of the path-length through a layer for a particle
coming from the interaction point and hitting the end of the barrel,
as compared to a particle impacting the calorimeter at 90$^o$.
The tungsten layers have a thickness of 1.75~mm, which is equivalent
to half a radiation length [X$_0$] per layer. 
For particles hitting the end of the barrel the effective layer
thickness is 1 X$_0$.
There are 50 layers resulting in an overall thickness of 25-50 X$_0$.
The active medium consists of 500 $\mu$m thick silicon pads.
The pad~[cell] size will be 
1~cm$\times$1~cm and
there will be three longitudinal sections.
The segmentation will be discussed in the context of e-$\pi$~separation.

High pseudo-rapidities, $|\eta|>2.$, are covered by end-walls.
The aperture of the dipole magnet limits rapidity coverage of the end-walls.
Therefore, the intermediate range in rapidity is covered
by ``catcher rings''.
Their absorber plates are 
perpendicular to the beam-line
and cover radii from 
30~cm to 70~cm. 
A catcher is placed on each side of the interaction point.
They start at z=$\pm$110~cm 
and cover
the pseudo-rapidity ranges from 1.3 to 2.0 with some reasonable
additional overlap of about 10~\% of a unit in pseudo-rapidity.
Both catchers and the 
end-walls have tungsten absorber
plates of 3.5~mm, i.e.1X$_0$, thickness. They have 25 layers and
again silicon pads of 500~$\mu$m thickness as active elements. 

In the
proton hemisphere an additional uranium-scintillator calorimeter
is deployed to provide hadronic calorimetry. The design follows
the hadronic part of the existing ZEUS calorimeter
with 160 uranium absorber plates, each 3.3~mm thick, providing
5.3~interaction length. 
Each uranium plate has 0.4~mm steel cladding which provides
and additional 0.4~interaction length.
The lateral granularity is 10~cm~$\times$~10~cm.
The complete assembly is 1.4~m~thick with 5.7~interaction length
absorber strength. 
The uranium-scintillator option is chosen because of its compact design
and its proven compensating nature. The combined use of a silicon-tungsten 
electromagnetic section and a uranium-scintillator hadronic calorimeter
has not been studied in detail and may not be optimal.  For the purposes
of this study, the performance of the ZEUS calorimeter has been assumed.

\newpage
\subsection{Forward and backward detectors}
\label{sec:fb}

Three devices along the electron and
proton~[ion] beam-line are foreseen:
\begin{itemize}
\item
a photon calorimeter along the electron beam direction,
\item
a proton remnant tagger,
\item
a zero degree neutron detector.
\end{itemize}

As the machine lattice for a possible eRHIC collider is not yet
defined, it is impossible to precisely place the instruments or study
their performance.
In the letter of intent for an extension of the HERA program \cite{h3loi}
more details can be found. Here only the basic requirements and usages
are outlined.

\vskip 1cm

The {\bf photon detector} 
at zero angle with respect 
to the lepton beam direction  
will cover about 1~mrad of the polar angle.
The purpose of the detector is: 
\begin{itemize}
\item measurement of the luminosity through the observation of 
      bremsstrahlung photons,
\item lepton beam diagnostic,
\item tagging and measurement of events with initial state radiation (ISR).
\end{itemize}
The device will be an electromagnetic calorimeter
placed well beyond the separation of the electron and
proton~[ion] beams. It will need a shield against the low energy
Bremsstrahlung photons due to the dipole field and other bending
magnets.
A quartz-fiber calorimeter~\cite{fibercalo} is one 
possible solution for the high-radiation environment the device has to operate
in.

\vskip 1cm

The {\bf proton remnant tagger} 
will cover the range between approximately 1~mrad and 10~mrad.
The purpose of the device is:
\begin{itemize}
\item tag events in e~p collisions where the proton stays intact
      in order to measure cross-sections of diffractive and
      exclusive processes,
\item study proton dissociation.
\end{itemize}
The device will consist of a
high density hadron calorimeter for energy measurements and
a [silicon] tracker to study multiplicities and provide spatial information.
It will be located beyond the electron and proton~[ion] beam separation.
The measured energy and multiplicity will be used to
distinguish between proton-dissociative and elastic reactions 
at large $|t|$.
It is expected that for $-t > 1$~GeV$^2$, the elastically scattered proton
will enter the detector.
The measurement of the impact point of the scattered
proton will help in the reconstruction of the variable $t$ for DVCS
and Vector Meson production with non-negligible values of $Q^2$.

\vskip 1cm

The {\bf zero degree neutron detector}
compliments the proton remnant tagger by identifying neutrons
below approximately 0.5~mrad.
The purpose of the device is:
\begin{itemize}
\item study proton dissociation.
\item tag spectator nucleons in e~d collisions. 
\item measure the centrality of interactions for eA collisions.
\end{itemize}
The device will consist of a hadron calorimeter placed
at the location where the neutrons will leave the beam-pipe.
This will most likely be further away from the interaction region
than the location of the proton remnant tagger.
It is foreseen to complement the neutron detection with a device to
identify protons of low energy on the other side of the beam-line.

\newpage
\section{Monte Carlo}
\label{sec:MCR}
\vskip 0.5cm

\noindent
{\bf Detector simulation:}

\vskip 0.5cm
A full GEANT3
simulation of the detector described in the previous
section was set up to study
the detector performance 
and the physics reach of the experiment.
Interactive GEANT was used embedded 
in ``atlsim''~\cite{bib:det:atl}, a tool developed for ATLAS
\footnote{Special thanks to Denis Salihagic and Pavel Nevski for
their help setting up the simulation.}.
For historical reasons the simulation uses a 
positron beam and a split magnetic field
without any field free region in the center. 
This is an unphysical field
which is, however equivalent to the homogenous long field as decribed in
section~\ref{sec:ia}. The beam-pipe simulated does not have
a conical opening, but follows the positron beam on both sides of the
interaction point.

Three material budgets are used in the detector simulation.
They are listed in tab.~\ref{tab:mc:mat}.
These assumptions do not correspond to real technical designs; they
are merely used to study the influence of material to guide later
technical specifications.
If not otherwise specified, the standard detector is used.

\begin{table}[htb]
\caption {Material budgets used in the detector simulation.
          600~$\mu$m silicon represent 2 silicon wafers, 1.2~mm CF
          assume a uniform support structure of 1.2~mm carbon fiber
          and the beam-pipe is assume to have a thickness of 500$\mu$~m
          aluminium.}
\begin{center}
\begin{tabular}{|l||r|r|r|} \hline
  Scenario      &   silicon     &  support   &  beam-pipe  \\ \hline  
  standard      &   600~$\mu$m  &  1.2~mm CF   &  500$\mu$~m Al   \\ \hline  
  light         &   600~$\mu$m  &  1.2~mm CF   &  none   \\ \hline  
  extra-light   &   600~$\mu$m  &       none   &  none   \\ \hline  
\end{tabular}
\end{center}
\label{tab:mc:mat}
\end{table}

\vskip 1cm
\noindent
{\bf Event generation:}
\vskip 0.5cm

Two kinds of events were used to study the performance.
Either positrons or pions were ``injected" at the interaction point or full
neutral current~[NC] events were generated.

The following samples of injected particles were generated:
\begin{itemize}
\item {\bf I1}
 1.000.000 positrons
 in a flat random distribution 
 with energies from 0.5~GeV to 15~GeV, $\phi \in$~[0,2$\pi$], 
 and rapidities from 0 to -8.
 Only positrons kinematically possible in a collision between a 10~GeV
 positron and a 200~GeV proton were kept.
\item {\bf I2}
 1.000.000 positive pions
 in a flat random distribution 
 with energies from 0.5~GeV to 100~GeV, $\phi \in$~[0,2$\pi$],
 and rapidities from 0 to +8.
\item {\bf I3}
 Sets of 500000~electrons and pions at fixed $\eta$~=~0,~-1~-1.7~and~-3 and 
 flat in energy between 0.5 and 15~GeV.
\end{itemize}
The particles were processed through the full detector simulation
for all material budgets.

Two sets of NC events were generated for a proton energy of 200~GeV
\begin{itemize}
\item {\bf NC1}
1.6 million events with Herwig, version~5.9. 
\item {\bf NC2}
1.8 million events with DjangoH version 1.1.  
\end{itemize}
\noindent
For a proton energy of 100~GeV a MC data set {\bf NC1L} was produced
similar to {\bf NC1}.

The {\bf NC1} and {\bf NC1a} events 
have a  Q$^2$ between 0.01~GeV$^2$ and 100~GeV$^2$
and a y between 0.001 and 0.99.
The events were reweighted according to the ALLM parton distribution
functions~\cite{ref:ALLM}.
\newpage
\noindent
The total luminosities generated were
\begin{itemize}
\item {\bf NC1:}
\item  ~9~pb$^{-1}$ for Q$^2$ $\in$ [0.01,0.10] 
\item ~8~pb$^{-1}$ for Q$^2$ $\in$ [0.10,1.00] 
\item ~13~pb$^{-1}$ for Q$^2$ $\in$ [1.00,10.0] 
\item ~93~pb$^{-1}$ for Q$^2$ $\in$ [10.0,100.] \\
\item {\bf NC1a:}
\item ~10~pb$^{-1}$ for Q$^2$ $\in$ [0.01,0.10]
\item ~9~pb$^{-1}$ for Q$^2$ $\in$ [0.10,1.00]
\item ~14~pb$^{-1}$ for Q$^2$ $\in$ [1.00,10.0]
\item ~230~pb$^{-1}$ for Q$^2$ $\in$ [10.0,100.]
\end{itemize}
The events were processed through the full detector simulation
for all material budgets.

\vskip 0.7cm

\noindent
The {\bf NC2} events have a Q$^2$ between 4~GeV$^2$ and
1000~GeV$^2$,  and an x between 0.1 and 1.0.
\noindent
The total luminosities generated were 
\begin{itemize}
\item ~150~pb$^{-1}$ for Q$^2$ $\in$ [4,10]
\item ~4.8$\times$ 10$^6$ pb$^{-1}$ for Q$^2$ $\in$ [10,100]
\item ~1.5$\times$ 10$^8$ pb$^{-1}$ for Q$^2$ $\in$ [100,1000]
\end{itemize}
As described in section~\ref{sec:perf:cal} 
the full GEANT simulation of the forward hadronic detector 
suffers from large uncertainties.
Therefore, the jet response of the detector was
simulated by assuming perfect pattern recognition
and fixed resolutions. 
The energy of all charged particles within the tracker acceptance
is smeared by 3.5~\%. The energy of all remaining particles is 
smeared by 35\%/$\sqrt{E}$.

\newpage
\section{Detector performance}

The detector performance was studied using 
the MC~events described in the previous section, i.e.
either positrons or pions
injected at the center or full neutral current events.

\subsection{Acceptance}

The calorimeter has full acceptance for tracks 
with $\mid\eta\mid~\leq$~5.0. For higher $\eta$ it follows the
acceptance of the tracker. 

The geometrical acceptance of the tracker was 
studied using MC sample I1, i.e.
positrons injected at the
interaction point.
A minimum of three space-points is needed to reconstruct a track.
Tracks which penetrate three silicon planes are thus called accepted.

\begin{figure}[hbpt] 
\begin{center}
\epsfig{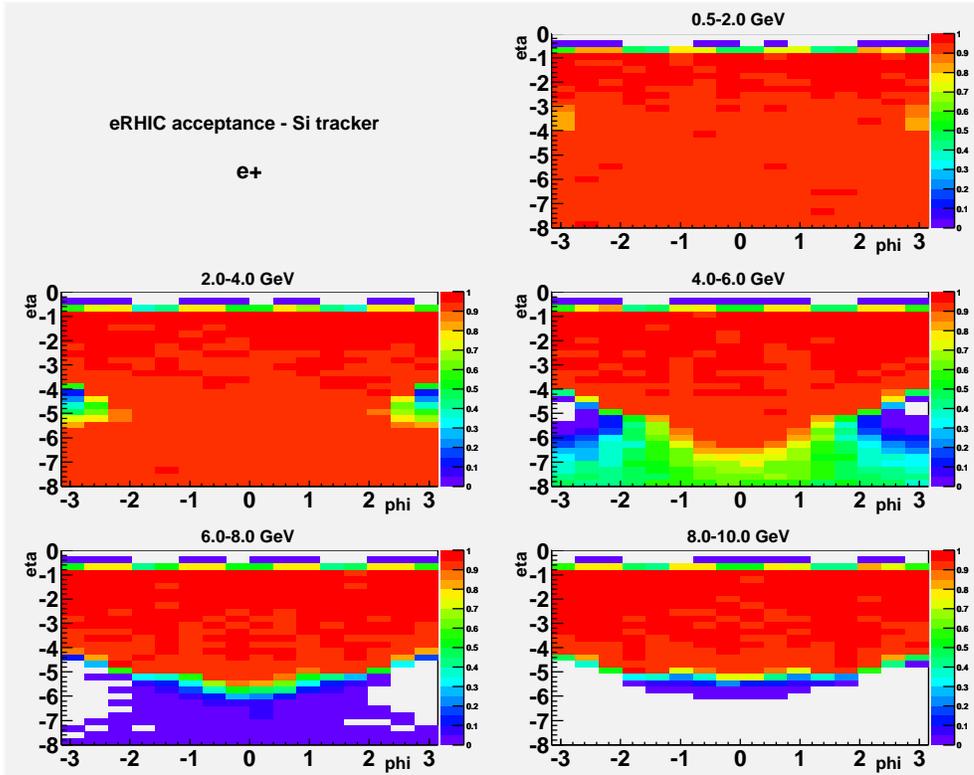}
\caption{Acceptance of the tracking system in different energy bins
         vs. azimuthal angle and pseudo-rapidity. A track is counted as accepted
         if there are at least three space-points reconstructable
         in the tracking system.
    } 
  \label{det:fig:acc3h}
\end{center}
\end{figure}

Figure~\ref{det:fig:acc3h} 
shows the acceptance binned in $\eta$ and $\phi$  for several energy bins.
Although three space-points are sufficient it is desirable to
have redundancy. For $\eta \in [-6,0]$  and E~$\in [0.5,10]$~GeV
90\% of the accepted tracks have 4 or more space points.

\begin{figure} 
\begin{center}
\epsfig{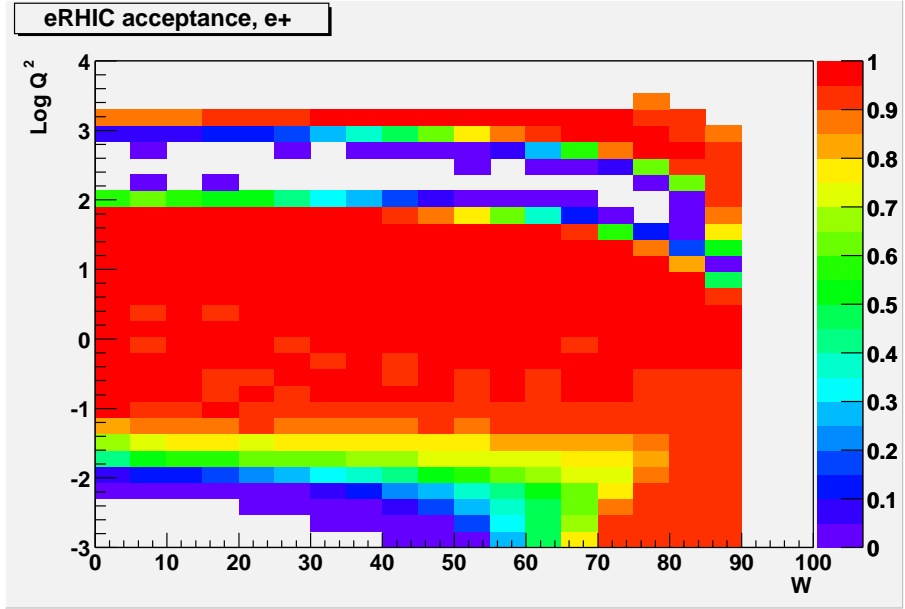}
\caption{Acceptance of the tracking system 
         vs. W and $Q^2$ assuming the phase space of e~p neutral
         current interactions with $E_e$=10~GeV
         $E_p$=200~GeV. 
         A track is counted as accepted,
         if there are at least three space-points reconstructable
         in the tracker.
    } 
  \label{det:fig:acc}
\end{center}
\end{figure}

Up to an energy of about 6~GeV 
good acceptance is obtained up to the maximum pseudo-rapidity studied,
$\eta= -8$.
For larger energies and $\eta < -5$,
the magnetic field is not able to separate the
positrons sufficiently from the beam to produce enough hits
in the tracking stations.

The system is not designed for independent tracking in the central
region, $\mid\eta\mid < 0.5$. Full acceptance is achieved 
for $\mid\eta\mid > 0.75$. 
For $\mid\eta\mid \in [0.5,0.75]$ the acceptance is 57~\%. 
However, in this range 100~\% of the tracks have at least
2 hits; in  $\mid\eta\mid < 0.5$ this is still 50~\%.
This information can be combined
with calorimeter information which is best in this region. 
We do not pursue this further at this stage.

The $\phi$~dependence of the acceptance is purely geometrical
due to the field configuration and the shape of the beam-pipe.
The acceptance 
in the +z direction (proton hemisphere)
is similar.

The acceptance of the tracker  in
the physics variables $W$ and $Q^2$ was studied with 
electrons of the NC~event
sample NC1.
It is shown in 
Fig.~\ref{det:fig:acc} 
for the requirement of at least three reconstructable space-points.
There is an acceptance gap for events with $Q^2$ between 100~GeV$^2$ 
and 1000~GeV$^2$ due to the lack of tracking in the central region. 
This gap can be closed by using calorimetry in this region.
This study focuses on the low~$Q^2$ region where
good acceptance is provided down to $Q^2= 0.01$~GeV$^2$ 
over the complete W range. 

\subsection{Momentum Resolution}
\label{sec:perf:mom}

The momentum resolution is studied using 
the MC event sets I1~[electrons] and I2~[pions] 
for momenta up to 15~GeV and 30~GeV respectively.
The tracks are simulated for
the three material budgets described in
sect.~\ref{sec:MCR}.
In addition a hit resolution of 20~$\mu$m is applied.
Two physics processes directly influence the resolution.
One is multiple scattering, the other one Bremsstrahlung.
While multiple scattering changes the track in a random way,
Bremsstrahlung always causes an energy loss and thus a bias.
Some Bremsstrahlung photons convert and cause extra hits.
In addition, some of the primary electrons/pions interact and create a shower.

In the following a perfect pattern recognition is assumed; i.e.,
all hits caused by the primary particle as well as those from secondary 
processes are handed to the reconstruction program for a particular track.
As the magnetic field is parallel to the y-axis, a high-momentum 
track is a straight line in the y~z-plane.
For each track
a $\pm$2~cm wide corridor is defined using the y-position 
of the first hit in~z 
and the origin.  
In order to reject hits from secondary processes only hits 
falling into this corridor 
are
handed to the fitting routine
\footnote{Special thanks to Volker Blobel who provided the routine and
to Christian Kiesling for implementing it.}.
This procedure fails to provide enough points for fitting for approximately
$\approx$0.3~\% of the otherwise accepted electron
tracks in the standard detector for $|\eta| \in$~[1,5].
For higher $|\eta|$ this number increases to the several percent level
due to showering in the beam-pipe.
The momentum fit is first performed in the x~z-plane which is perpendicular
to the magnetic field.
The components perpendicular and parallel are then recombined.

\begin{figure} [h]
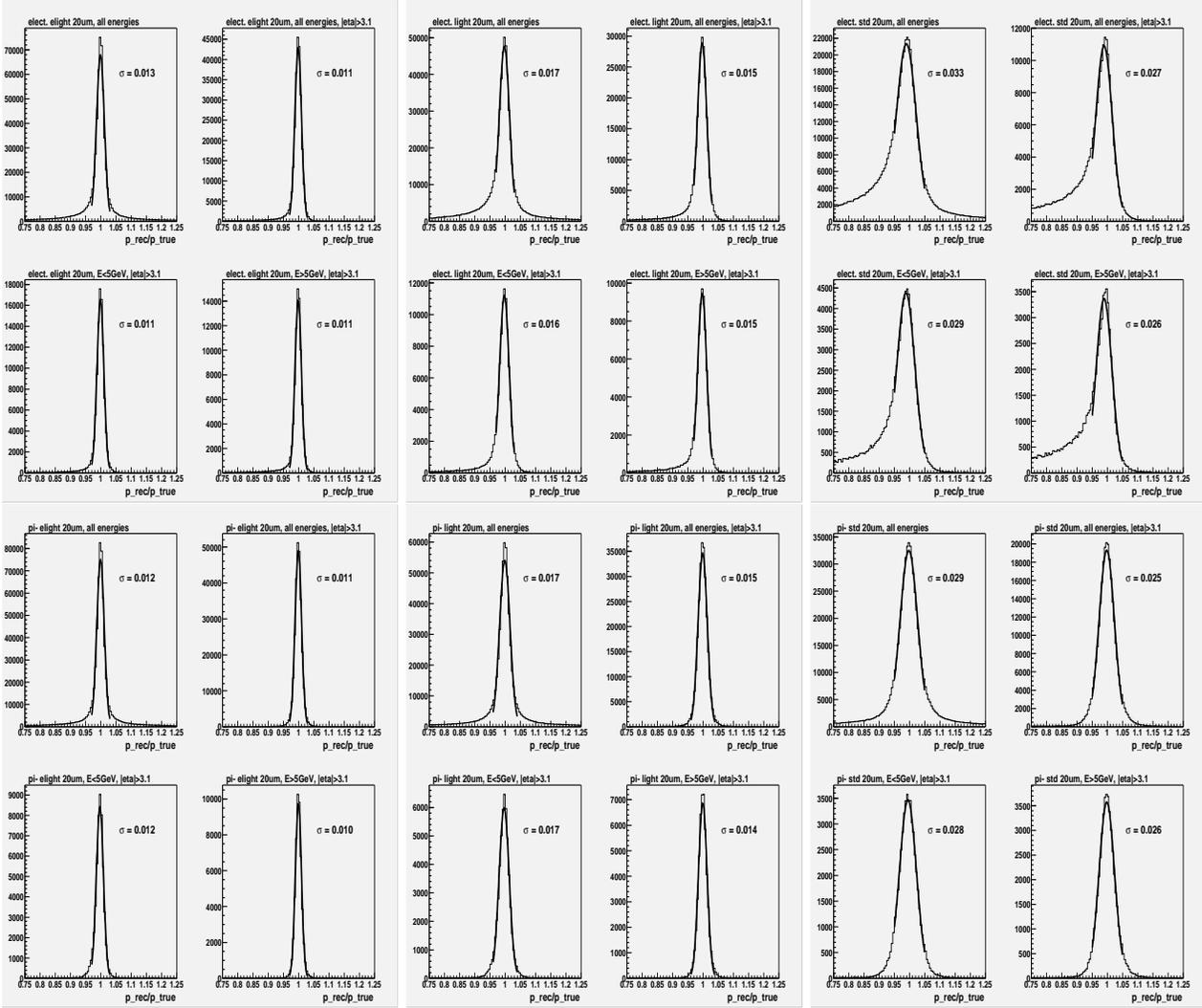

\begin{center}
\hbox to \linewidth{\hss
\put(0.,200.){\epsfig{file=electron.elight.20.res_page3.epsi,width=5.5cm,
                                                              height=7cm}}
\put(160.,200.){\epsfig{file=electron.light.20.res_page3.epsi,width=5.5cm,
                                                              height=7cm}}
\put(320.,200.){\epsfig{file=electron.std.20.res_page3.epsi,width=5.5cm,
                                                              height=7cm}}
\put(0.,0.){\epsfig{file=piminus.elight.20.res_page3.epsi,width=5.5cm,
                                                              height=7cm}}
\put(160.,0.){\epsfig{file=piminus.light.20.res_page3.epsi,width=5.5cm,
                                                              height=7cm}}
\put(320.,0.){\epsfig{file=piminus.std.20.res_page3.epsi,width=5.5cm,
                                                              height=7cm}}

\hss}
\caption{
         Distribution of the ratio of the
         reconstructed 
         and 
         generated  momentum 
         for  electrons~[top blocks] and pions~[bottom blocks] for the
         the extra-light~[left],
         the light~[center], 
         and the standard~[right] detector. For material budgets see
         Table~\ref{tab:mc:mat} in sect.~\ref{sec:MCR}. 
         In each block
         the overall distribution~[top left],
         the distribution for forward tracks~[top right],
         for forward tracks with momentum below~5~GeV~[bottom left] 
         and above~5~GeV~[bottom right] 
         are depicted. 
         A hit resolution of 20~$\mu$m is assumed.
    } 
  \label{fig:perf:momres-overview}
\end{center}
\end{figure}

\begin{table} [h]
\caption {$\delta$p/p for electrons in different energy bins  for all or only forward tracks for all three
material budget. 
"In peak" refers to the percentage of tracks described by the Gaussian fit.
A hit resolution of 20~$\mu$m is assumed.}
\begin{center}
\begin{tabular}{|rcr|c||c|r||c|r||c|r|} \hline
\multicolumn{4}{|c||} {} & \multicolumn{2}{|c||}{extra-light}  & \multicolumn{2}{|c||}{light} & \multicolumn{2}{|c|}{standard}   
                                                                                                                    \\ \hline 
\multicolumn{3}{|c|}{Energy} & $|\eta|$      & $\delta$p/p    & in peak            & $\delta$p/p    & in peak & $\delta$p/p  & in peak            \\\hline 
\multicolumn{3}{|c|}{all}  &  all      
                      & 1.30 \%   &  68  \%      &   1.87 \%   &  65  \%             &   3.28 \%   &  54   \%            \\
\multicolumn{3}{|c|}{all}       & $>$3.1 
                      & 1.11  \%  &  90  \%     &   1.61 \%   &  84       \%        &   2.69 \%   &  56   \%            \\
\multicolumn{3}{|c|}{$<$5GeV}   & $>$3.1
                      & 1.14 \%   &  90  \%     &   1.66 \%    &  86  \%              &   2.95  \%     &  63    \%    \\
\multicolumn{3}{|c|}{$>$5GeV}   & $>$3.1
                      & 1.07 \%   &  89  \%      &   1.56 \%   &  84   \%              &   2.59  \%    &  52   \%         \\
0.5 & - & 2.0GeV & all 
                      & 1.42 \%   &  77  \%      &   2.06 \%   &  75    \%             &   3.45  \%    &  66   \%         \\
2.0 & - & 4.0GeV & all 
                      & 1.16 \%   &  75  \%     &   1.66 \%   &  72  \%               &   3.30  \%    &  59   \%      \\
4.0 & - & 6.0GeV & all 
                      & 1.15 \%   &  70  \%    &   1.63 \%   &  67   \%              &   3.39  \%    &  54   \%              \\
6.0 & - & 8.0GeV & all 
                      & 1.36 \%   &  64  \%     &   1.96 \%   &  61  \%               &   3.29  \%    &  51   \%              \\
8.0 & - & 10.0GeV & all 
                      & 1.40 \%   &  63  \%      &   2.02 \%   &  60  \%               &   3.11  \%    &  49   \%              \\
10.0& - & 15.0GeV & all  
                      & 1.40 \%   &  61  \%      &   2.03 \%   &  58   \%              &   3.15  \%    &  48  \%               \\
0.5 & - & 2.0GeV   &   $>$3.1
                      & 1.31 \%   &  91  \%        &   1.92 \%  &  88   \%               &   3.21  \%    &  71    \%             \\
2.0 & - & 4.0GeV   &    $>$3.1 
                      & 1.07 \%   &  90   \%     &   1.54 \%   &  85    \%             &   2.81 \%     &  60    \%             \\
4.0 & - & 6.0GeV   &    $>$3.1
                      & 1.04 \%   &  90   \%      &   1.51 \%   &  85    \%            &   2.73  \%    &  54     \%            \\
6.0 & - & 8.0GeV   &    $>$3.1
                      & 1.07  \%  &  89   \%      &   1.55 \%   &  84    \%            &   2.63  \%    &  52    \%             \\
8.0 & - & 10.0GeV   &   $>$3.1
                      & 1.09 \%   &  89   \%    &   1.59 \%   &  84    \%            &   2.48  \%    &  51    \%             \\
10.0& - &15.0GeV  &   $>$3.1
                      & 1.10 \%   &  89   \%    &   1.60 \%   &  83    \%             &   2.45   \%   &  51    \%             \\\hline 

\end{tabular}
\end{center}
\label{tab:perf:momres-electrons}
\end{table}

\begin{table} [h]
\caption {$\delta$p/p for negative pions in different energy bins  
for all or only forward tracks for all three
material budget. 
"In peak" refers to the percentage of tracks described by the Gaussian fit.
A hit resolution of 20~$\mu$m is assumed.}
\begin{center}
\begin{tabular}{|rcr|c||c|r||c|r||c|r|} \hline
\multicolumn{4}{|c||} {} & \multicolumn{2}{|c||}{extra-light}  & \multicolumn{2}{|c||}{light} & \multicolumn{2}{|c|}{standard}   
                                                                                                                    \\ \hline 
\multicolumn{3}{|c|}{Energy} & $|\eta|$      & $\delta$p/p    & in peak            & $\delta$p/p    & in peak & $\delta$p/p  & in peak            \\\hline 
\multicolumn{3}{|c|}{all}  &  all      
                      & 1.24 \%   &  70  \%      &   1.73 \%   &  70  \%             &   2.92 \%   &  72   \%            \\
\multicolumn{3}{|c|}{all}       & $>$3.1 
                       & 1.08    \% &  96         \% &   1.51    \% &  95               \% &   2.53    \% &  90               \%  \\
\multicolumn{3}{|c|}{$<$5GeV}   & $>$3.1
                      & 1.18    \% &  93         \% &   1.66    \% &  93               \% &   2.80    \% &  93               \%  \\
\multicolumn{3}{|c|}{$>$5GeV}   & $>$3.1
                      & 1.02    \% &  96         \% &   1.42    \% &  95               \% &   2.56    \% &  92               \%  \\
0.5 & - & 2.0GeV & all 
                       & 1.55    \% &  78         \% &   2.14    \% &  79               \% &   3.39    \% &  83               \%  \\
2.0 & - & 4.0GeV & all 
                       & 1.18    \% &  80         \% &   1.67    \% &  79               \% &   2.81    \% &  82               \%  \\
4.0 & - & 6.0GeV & all 
                      & 1.10    \% &  79         \% &   1.55    \% &  78               \% &   2.94    \% &  81             \%  \\
6.0 & - & 8.0GeV & all 
                      & 1.06    \% &  77         \% &   1.46    \% &  76               \% &   3.00    \% &  79               \%  \\
8.0 & - & 10.0GeV & all 
                      & 1.06    \% &  74         \% &   1.48    \% &  74               \% &   3.05    \% &  76               \%  \\
10.0& - & 30.0GeV & all  
                      & 1.30    \% &  65         \% &   1.81    \% &  65               \% &   2.84    \% &  66               \%  \\
0.5 & - & 2.0GeV   &   $>$3.1
                      & 1.40    \% &  91         \% &   1.99    \% &  92               \% &   3.18    \% &  93               \%  \\
2.0 & - & 4.0GeV   &    $>$3.1 
                       & 1.08    \% &  96         \% &   1.54    \% &  95               \% &   2.63    \% &  94               \%  \\
4.0 & - & 6.0GeV   &    $>$3.1
                      & 1.03    \% &  96         \% &   1.44    \% &  95               \% &   2.62    \% &  94               \%  \\
6.0 & - & 8.0GeV   &    $>$3.1
                       & 1.02    \% &  96         \% &   1.41    \% &  95               \% &   2.55    \% &  92               \%  \\
8.0 & - & 10.0GeV   &   $>$3.1
                       & 1.02    \% &  96         \% &   1.43    \% &  95               \% &   2.54    \% &  91               \%  \\
10.0& - &30.0GeV  &   $>$3.1
                       & 1.07    \% &  97         \% &   1.48    \% &  95               \% &   2.44    \% &  89     \%  \\ \hline
\end{tabular}
\end{center}
\label{tab:perf:momres-pions}
\end{table}

\begin{figure} 
\begin{center}
\epsfig{file=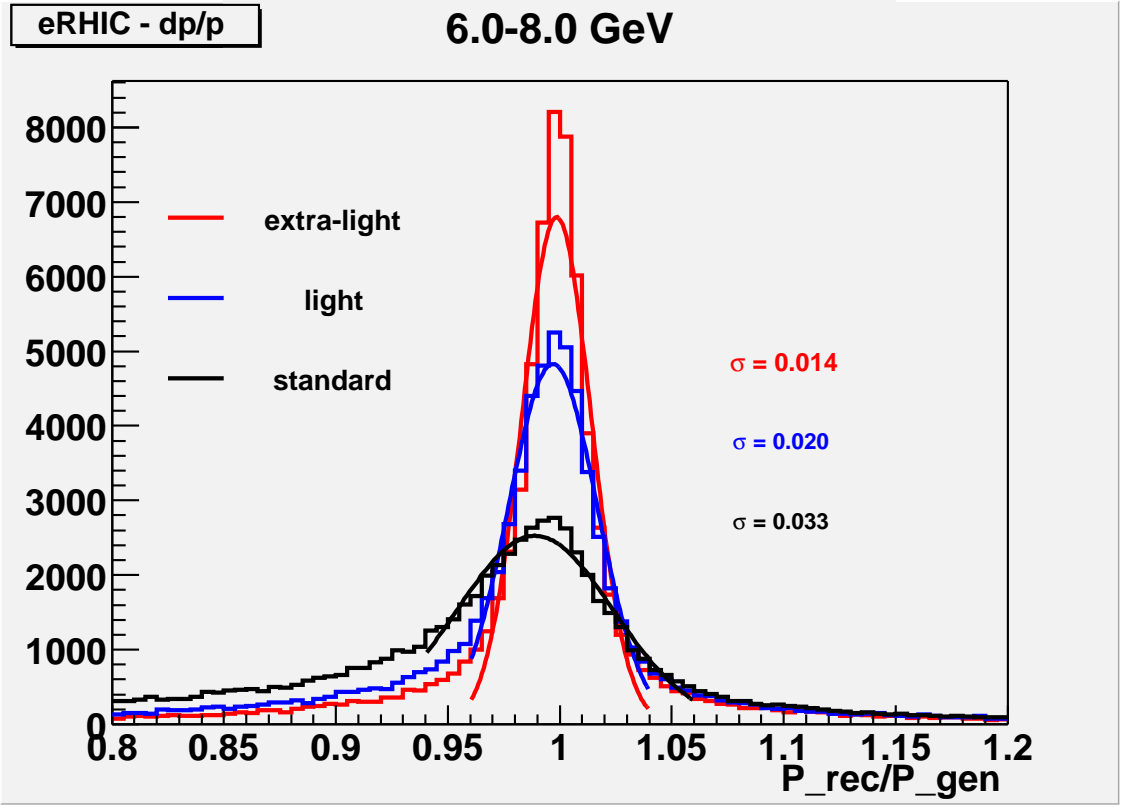,width=15cm}
\caption{The distributions of the ratios of
         reconstructed
         and 
         generated  momentum 
         are compared for the three different material
         budgets for electrons of all angles 
         with momenta between 6~GeV and 8~GeV.
         A hit resolution 20~$\mu$m is assumed.
    } 
  \label{fig:perf:momres-matter}
\end{center}
\end{figure} 

\begin{table}
\caption {$\delta$p/p for electrons for different hit
resolutions for forward~[$|\eta|>$3.1] tracks.
Values are given for the extra-light and the standrad material
and are also separated into two energy bins.
The percentages of tracks described by the Gaussian
fit are also listed.}
\begin{center}
\begin{tabular}{|l|l|r|r|r|r|r|r|r|} \hline

 & $\sigma_h$ [$\mu$m] & 0  & 20  & 50  & 100  & 150  & 200  & 500 \\  \hline\hline
{\bf extra-light}  &
   $\delta$p/p [\%]           & 1.10  & 1.11  & 1.14  & 1.24  & 1.35  & 1.47  & 2.17 \\
 all energies  
 & in peak [\%]  & 89.5  & 89.5  & 89.8  & 89.9  & 89.4  & 88.4  & 83.3  \\ \hline

 E$>$5GeV & $\delta$p/p[\%]           & 1.06  & 1.07  & 1.10  & 1.20  & 1.33  & 1.48  & 2.54 \\ 
 & in peak [\%]  & 89.1  & 89.2  & 89.4  & 89.8  & 89.8  & 89.6  & 88.2  \\ \hline

 E$<$5GeV & $\delta$p/p[\%]           & 1.14  & 1.14  & 1.15  & 1.17  & 1.21  & 1.25  & 1.65 \\ 
 & in peak [\%]  & 90.3  & 90.4  & 90.4  & 90.6  & 90.8  & 90.9  & 91.7  \\  \hline

{\bf standard}  &
$\delta$p/p[\%]           & 2.69  & 2.69  & 2.71  & 2.78  & 2.87  & 2.98  & 3.80 \\
all energies  
 & in peak [\%]  & 55.9  & 55.9  & 56.0  & 56.4  & 56.8  & 57.2  & 59.8 \\ \hline

 E$>$5GeV & $\delta$p/p[\%]           & 2.59  & 2.59  & 2.61  & 2.66  & 2.74  & 2.86  & 3.90 \\ 
 & in peak [\%]  & 52.0  & 52.0  & 52.1  & 52.3  & 52.7  & 53.2  & 57.6 \\  \hline

 E$<$5GeV & $\delta$p/p[\%]           & 2.94  & 2.95  & 2.95  & 2.96  & 2.98  & 3.00  & 3.33 \\ 
 & in peak [\%]  & 63.0  & 63.1  & 63.1  & 63.2  & 63.2  & 63.3  & 65.6 \\ \hline
\end{tabular}
\end{center}
\label{tab:perf:momres-vs-resolution}
\end{table}

\begin{figure} 
\begin{center}
\epsfig{file=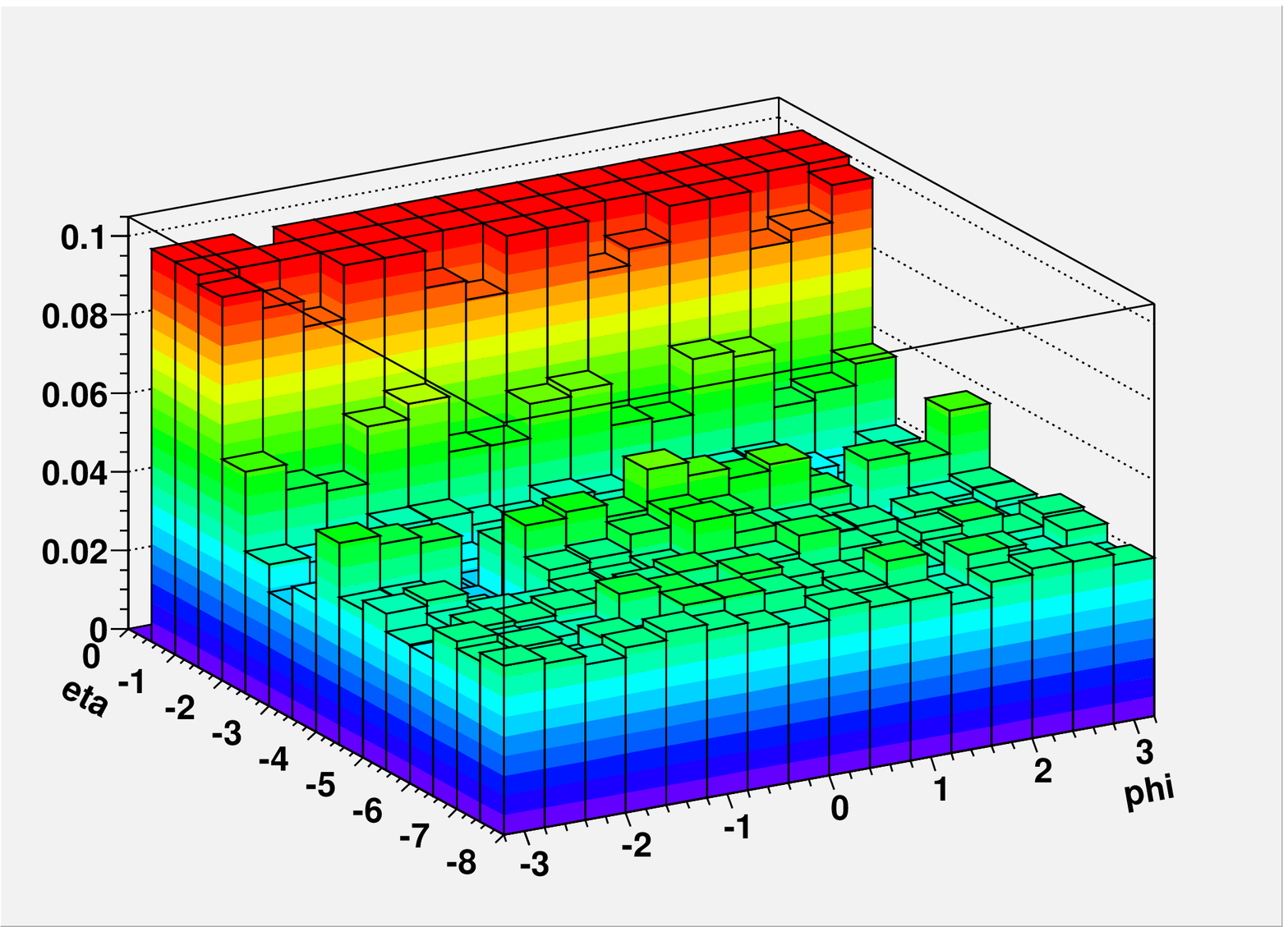,width=16cm}
\caption{$\delta$p/p vs. $\phi$ and $\eta$ integrated over all momenta
    for the standard material budget and 20~$\mu$m hit resolution.
    In the central region~[$|\eta|<1$] the plot is truncated at 10\%.}
  \label{fig:perf:momres-phieta}
\end{center}
\end{figure}

In Fig.~\ref{fig:perf:momres-overview}
the overall distributions
of the ratio  between the reconstructed and generated momenta
are shown for electrons and pions
for the three detector scenarios (see sect.~\ref{sec:MCR})
under consideration.
For each scenario the distribution for all energies and angles and for
all energies and $\eta > 3.1 $, i.e. $\theta>175^o$
is provided. For the forward tracks also a separation in
energies below and above 5~GeV is shown.
The central part of each distribution is fit with a Gaussian function.
The resulting $\sigma$ will always be
quoted as the resolution $\delta$p/p.
Obviously the material is important for the central part 
of the distributions which are dominated by multiple scattering.
However, for electrons it is even more important for the
tails due to energy loss caused
by Bremsstrahlung. Fig.~\ref{fig:perf:momres-matter} compares the
results for the three detector scenarios for electrons of 
all angles with energies between 6~GeV and 8~GeV.
The tail depicted for the standard
detector contains $\approx$50~\% of the tracks.
This number is calculated by integrating the Gaussian 
over~$\pm$~infinity and defining the tail as
the total number of fitted tracks minus this integral.
For pions, Bremsstrahlung does not present a problem and this tail
basically does not occur even for the standard detector.

In Tables~\ref{tab:perf:momres-electrons} and \ref{tab:perf:momres-pions}
the $\delta$p/p values are given for the three scenarios.
In addition the resolutions are
quoted for different energy bins.
The momentum resolution $\delta$p/p for pions is a little   
better than for electrons, as they are less affected by multiple scattering.
The difference is, however, not as striking as for the tails of the
distribution. 
This is reflected in the much larger percentages of tracks
contributing to the peaks of the distribution which are also listed in
Tables~\ref{tab:perf:momres-electrons} and \ref{tab:perf:momres-pions}.
The biggest difference is naturally seen for the standard detector.

Comparing the three scenarios it becomes evident that the extra support
material which distinguishes the light from the extra-light scenario
has a much smaller effect on the performance than the introduction of
a beam-pipe in the standard detector. This can be understood as the
precision of any tracking is massively influenced by the amount of material
in front of the first measurement. Thus, any technical development
aiming to reduce the material in each tracking station is only useful
if the beam-pipe is replaced with a Roman pot system.

Due to the geometrical layout of the detector the momentum resolution 
does not depend strongly
on the momentum itself.
Tracks with larger momenta have a higher probability to
have more than three space-points seen and usually have a longer
lever arm, thereby maintaining an approximately constant momentum
resolution.

The influence of the hit resolution $\sigma_h$ was also studied.
Table~\ref{tab:perf:momres-vs-resolution} lists the influence of $\sigma_h$ on
$\delta$p/p for the standard and extra-light detector in the 
forward phase-space region
previously used. 
The hit resolution is obviously not critical
below at least 50~$\mu$m when reasonable amounts of material for
a silicon tracker are assumed. 
This is understandable, as one 
tracking station with 1~\%~X$_0$ causes distortions in the next station
about 50~cm downstream
of about
70~$\mu$m from multiple scattering alone.
However, hit resolution is also important for pattern recognition issues
like double track
separation and space-point reconstruction
which are not addressed here.
Alternative technologies
like drift chambers, which could reduce the material in the tracker
beyond the extra-light scenario,
generally have hit resolutions above $\approx$100~$\mu$m and would thus
be limited by their resolution.
A silicon tracker is a more robust solution and a hit resolution
after alignment of 20~$\mu$m is realistic.

Due to the field configuration the momentum resolution  
depends on the angle $\phi$ and the pseudo-rapidity $\eta$.
This is demonstrated
in Fig.~\ref{fig:perf:momres-phieta}. The effect is most pronounced
at low $|\eta|$. The plot is truncated at 10~\% in the central region where
the momentum fit is intrinsically bad due to the field configuration and
orientation of the tracking planes.

\noindent
{\bf Summary:}

\noindent
The material budget seriously affects the momentum resolution. 
A beam-pipe even as thin as is assumed in the standard scenario
does not allow a momentum resolution of better than $\approx$3~\%
for electrons or pions. In addition
it causes tails in the $\delta$p/p distribution
for electrons which can contain up to 50~\% of the tracks.
It thus seems advisable to consider the possibility of
a Roman pot system for the electron hemisphere.  
This should allow for a momentum resolution around~2~\% and managable
tails in the distribution.

\subsection{Energy Resolution}
\label{sec:perf:cal}

\begin{figure} [h] 
\begin{center}
\epsfig{file=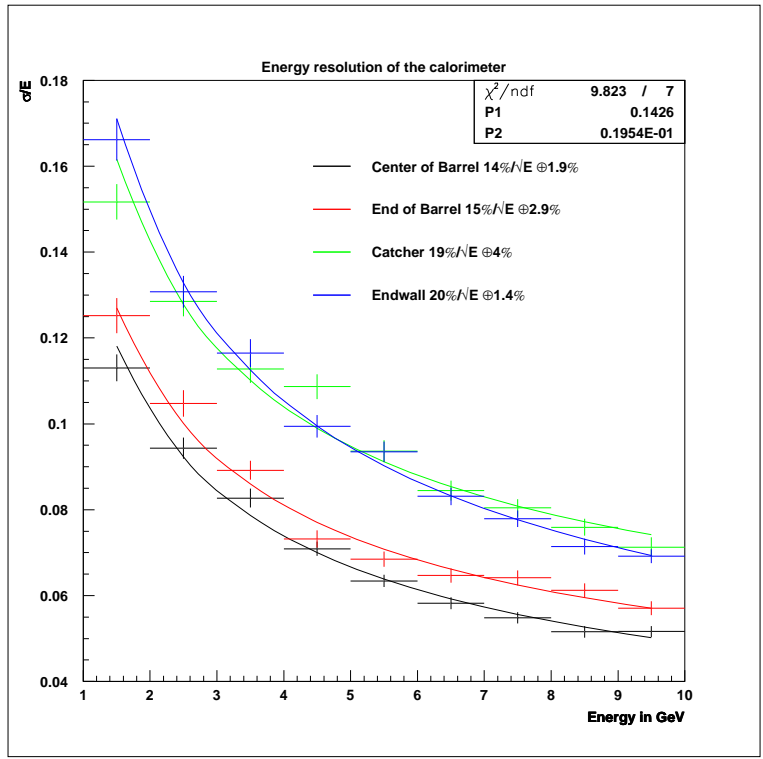,width=10cm}
\caption{Energy resolution of the silicon tungsten calorimeter
    for scattered positrons hitting at the center~[$\eta=0$] and
    the end~[$\eta=-1$], the catcher~[$\eta=-1.6$] and 
    the end-wall[$\eta=-3$].
    } 
  \label{det:fig:sigE}
\end{center}
\end{figure}

The energy resolution for electromagnetic showers
was studied using the Monte Carlo data sets I3
with the standard detector simulation.
The pattern recognition in the calorimeter 
is assumed to be perfect, i.e. all deposited energy is used,
and
no energy threshhold corresponding to  
the limitations of any
read-out electronics 
is taken into account. 
The results are shown in Fig.~\ref{det:fig:sigE}.

In the barrel the energy resolution depends on the 
pseudo-rapidity of the particle, i.e. on the impact angle. In Fig.~\ref{det:fig:sigE}
results are given for the center and the end of the barrel where
the effective absorber and sampling layer thickness doubles.
In the catcher and the endwall only the absorber thickness is doubled while
the silicon thickness remains the same. Thus the resolution is worse.
This is not a problem, as this phase space is 
completely covered by the tracker and thus is ``only'' needed 
for neutral particles and e-$\pi$~separation.

The energy resolution for electromagnetic showers in the barrel is

\[
 14\%\sqrt{E} \oplus 2\%  \qquad\mbox{and}\qquad 15\%\sqrt{E} \oplus 3\%
\]

\noindent for the central and end part 
respectively.
Here no correction is applied to adjust the calibration to the
change in sampling fraction. This is reflected in the larger
constant term for the end part. 

The energy resolution in the catcher and end-wall is
\[
 19\%\sqrt{E} \oplus 4\%  \qquad\mbox{and}\qquad 20\%\sqrt{E} \oplus 1\%
\]

The calibration is taken from the end-wall and also used for the catcher.
The larger constant term reflects the different angular distribution
of the showers and leakage in the catcher.

Hadronic calorimetry is only foreseen for
the proton hemisphere end-wall. 
A full GEANT study of the set-up, i.e.
the silicon-tungsten end-wall plus the
hadronic part of the ZEUS uranium-scintillator calorimeter, was
inconclusive.
The resolution observed depends strongly on the
hadronic shower simulation package and the cut-off energies used.
Resolutions between 40~\% $\sqrt{E}$ and 80~\% $\sqrt{E}$ were
``determined''. 
The ZEUS forward calorimeter 
has a measured hadronic energy resolution of 
35~\% $\sqrt{E}$. We assume that a similar resolution can be
achieved, and in the following the ZEUS calorimeter performance for
hadrons
was put in by hand.  The possible use of such a mixed technology for
the forward calorimeter will require detailed simulation studies as well
as test-beam data.

\subsection{e-$\pi$ Separation}
\label{sec:perf:epi}

The Monte Carlo event sets I3 with $\eta$~=~0,~-1~and~-3 were used.
The extralight detector was simulated to exclude radiative effects
from the material in the beam-pipe.
In order to define clusters
the transverse segmentation of the calorimeter in 1~cm~by~1~cm cells
is used to build  5~cm~by~5~cm towers.
If a tower shows energy, the neighboring towers are checked and
added to the cluster, if they too contain energy.
The cluster with the highest energy is analysed.
For electrons this procedure can result in the loss of energy if a radiated
photon creates a second cluster. For pions it is possible that 
non-contiguous 
parts of a shower are neglected.

As a first step
the energy response is studied.
For positrons the simulation confirms an energy independent response
over the range of 1~GeV to 15~GeV. 
The calibration constant needed to adjust the visible energy
to the generated energy changes by 5.5~\% 
over the length of the barrel calorimeter.
In contrast the calibration in the end-wall only 
differs by 0.2~\% from the central
barrel. This can be explained by the same
thickness of the silicon in the two devices.
All shifts in calibration constants are corrected for 
from now on.

\begin{figure}  
\begin{center}
\epsfig{file=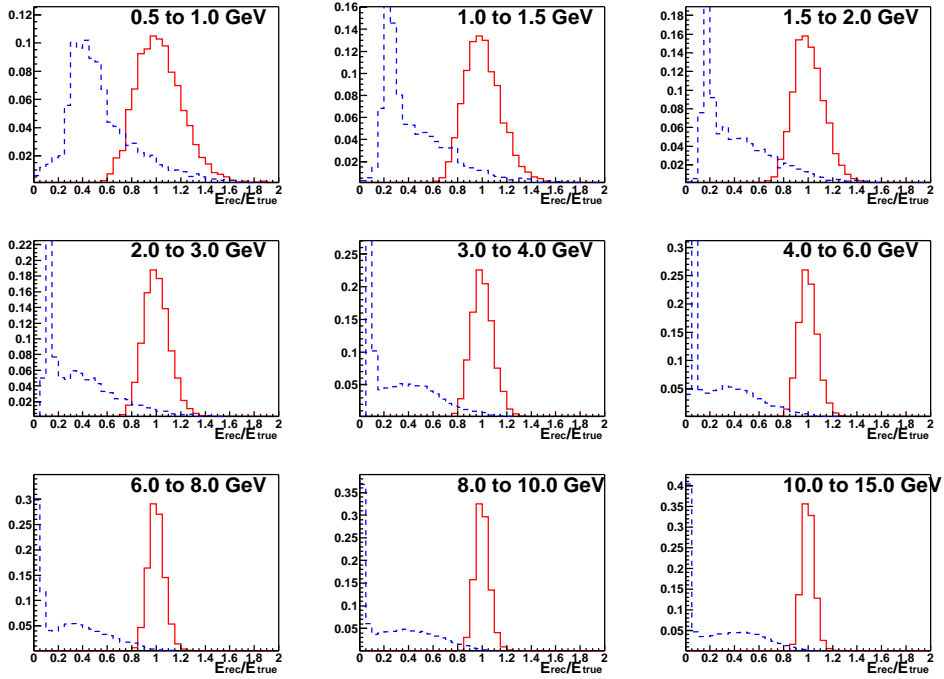,width=13cm}
\caption{Normalized visible energy 
         in the central part of the barrel calorimeter
         for positrons~[red] and pions~[blue] for energy bins
         from 0.5 to 15~GeV. 
         The normalized visible
         energy was calibrated to be 1 for electrons.
    } 
  \label{det:fig:eresp}
\end{center}
\end{figure}

The calibrated visible energy in the central part 
of the barrel calorimeter for positrons and positive pions is depicted
in Fig.~\ref{det:fig:eresp} for energies between 0.5 and 15~GeV.
The energy response for the pions clearly shows the peak at very low energies
associated with non interacting pions. For higher energies pion showers 
are not contained and thus the visible energy also for interacting pions
is lowered.

\begin{figure} 
\begin{center}
\epsfig{file=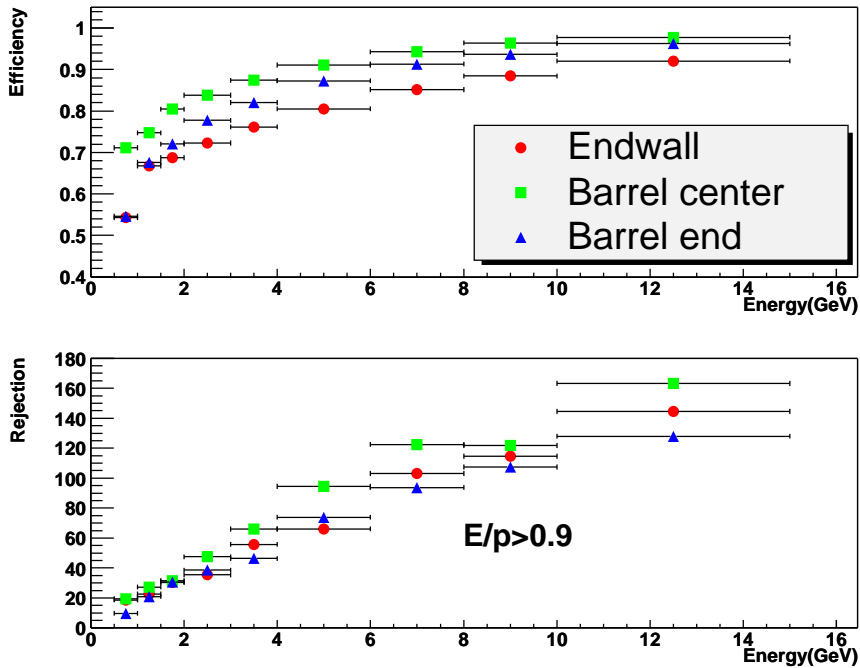,width=13cm}
\caption{Efficiency [top] and rejection power [bottom] 
         of a cut $E_{vis}/E_{true}>0.9$, where $E_{true}$ is to be 
         provided from a 
         measurement of the momentum p.
         Results are given for the central~[green squares] 
         and end~[blue triangles] part of the barrel and the
         end-wall~[red dots] 
         calorimeter.
    } 
  \label{det:fig:episeps}
\end{center}
\end{figure} 

Assuming knowledge of the particle momenta from the tracker
an e-$\pi$~separation solely
based on the energy response is possible. Figure~\ref{det:fig:episeps}
shows the result for a cut on the ratio $E_{vis}/E_{true}>0.9$
for the central~[$\eta=0$] and the
end part~[$\eta=-1$] of the barrel as well as the end-wall~[$\eta=-3$].  
The binning in energy is the same as in Fig.~\ref{det:fig:eresp}.

The separation is significantly worse for the end part of the barrel
than for the center.
At low energies this is caused by the cruder sampling due to the geometry.
At high energies the result just reflects the fact that pion showers are
better contained in the end part due to the larger effective depth.
The end-wall has the same effective absorber plate thickness 
as the end of the barrel,
but thinner silicon. This causes an even lower efficiency than in the
end of the barrel, but the rejection power is slightly better.
The end-wall has the same overall thickness as the central barrel
so the differences are only caused by the 
different sampling fraction.
However, the results from this simple approach show that
an efficient and powerful e-$\pi$ separation is rather easy at
energies above $\approx$3~GeV. 
Therefore the focus of further studies is on the
low energies.

Electromagnetic and hadronic showers are different in longitudinal
and in transverse development. Both aspects are studied.

\begin{figure} 
\begin{center}
\epsfig{file=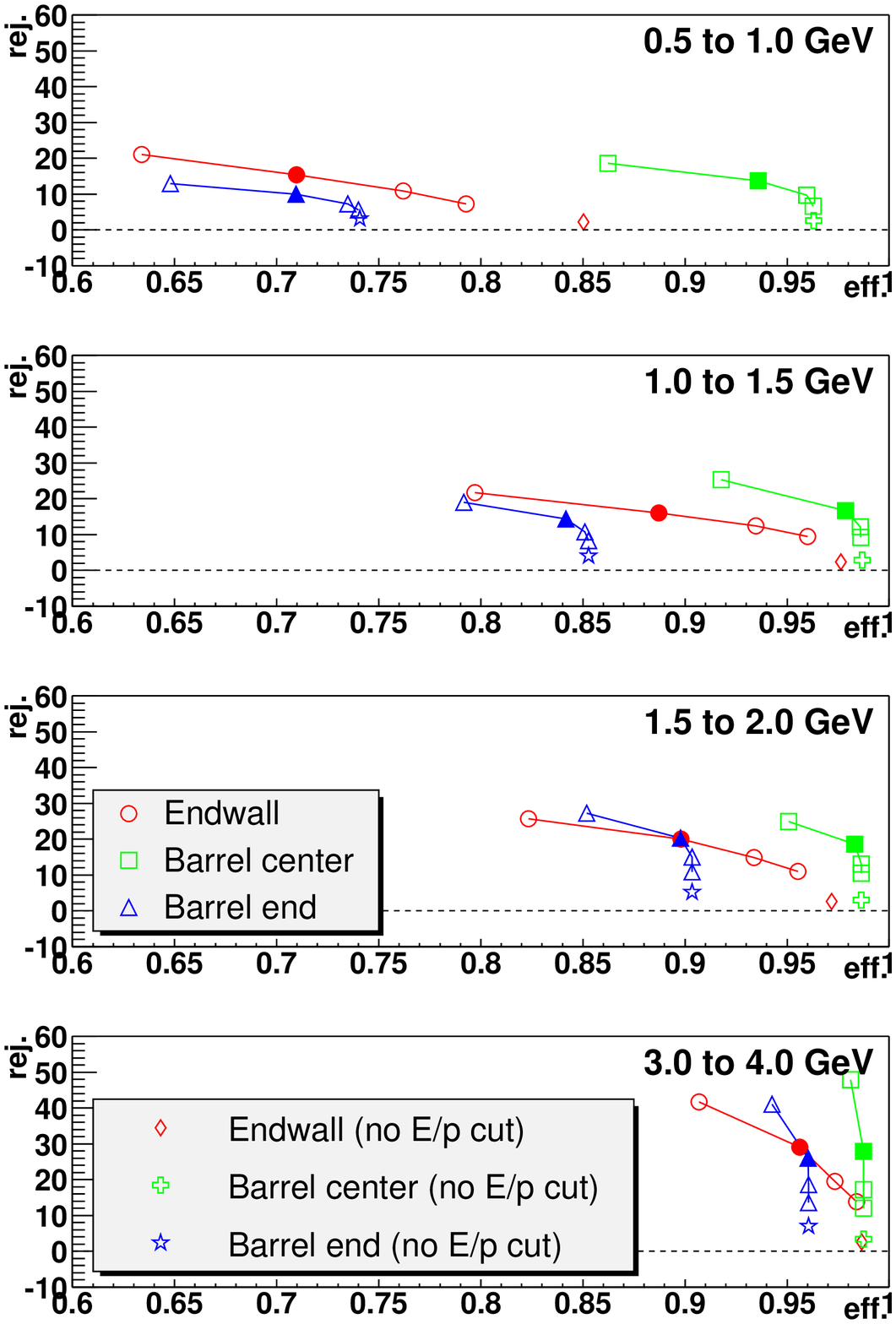,width=13cm}
\caption{Rejection power vs. efficiency for
         cuts on $E_{vis}/E_{true}>$~0.5,~0.6,~0.7,~0.8
         plus longitudinal segment cuts. 
         Results are given for the central~[green squares] 
         and end~[blue triangles] part of the barrel and the
         end-wall~[red dots] 
         calorimeter for 4 energy bins.
         Filled symbols represent our final choice of $E_{vis}/E_{true}>$0.7. 
         Special symbols indicate results using no cut 
         on $E_{vis}/E_{true}$~[E/p].
    } 
  \label{det:fig:episepl}
\end{center}
\end{figure} 

The barrel calorimeter has 50 layers, the end-wall has 25.
Both add up to 25 radiation length and in both cases
a grouping into three longitudinal segments is assumed.
The optimization of the longitudinal segments was done for
2~GeV showers, as this provides a good compromise
between optimization for low and high energy showers.
The segment borders were chosen to correspond to 50~\% and 150~\%
of the depth of the shower maximum.
In the barrel the grouping of layers changes with z in order to 
keep the conditions for e-$\pi$~separation similar.

The following cuts on the ratios of the longitudinal
segment energies $E_1$, $E_2$ and $E_3$ are used:
\[
 \frac{E_1}{E_2} < 1.0 \qquad\mbox{and}\qquad \frac{E_3}{E_2} < 4.0
\]
\noindent 
These cuts are combined with 
a varying cut on $E_{vis}/E_{true}$.
It starts at 0.5, yielding the maximum
electron finding efficiency, and is 
increased in steps of 0.1 up to 0.8.
Figure~\ref{det:fig:episepl} shows the results
for the three regions under study and 4 energy bins.

Above 1.5~GeV
a rejection power of 20
is reached with efficiencies above 90~\%.
Above 3~GeV the efficiency increases to 95~\%.
It is clear that below 1~GeV a separation between electrons and pions
becomes very difficult. 
For reasonable rejection power  very low efficiencies are achieved.
Only the central barrel with its fine granularity might support
e-$\pi$ separation at such low energies.
The preferred cut is $E_{vis}/E_{true}>0.7$. 
It is used for all energies. The corresponding points
are printed as filled symbols in Fig.~\ref{det:fig:episepl}.

We also determined the e-$\pi$ separation
without  a cut on $E_{vis}/E_{true}$.  
This reflects the situation where a track
cannot be found. 
In this case 
events with $E_{vis}<0.5$~GeV
are identified
as non-interacting pions. 
This is justified, as in subsequent analysis no attempt is made to
identify particles with an energy of less
then 1~GeV. The result is also shown in Fig.~\ref{det:fig:episepl}.
The longitudinal cuts are not optimized for this case. They would have 
to be made dependent on the visible energy to be more effective.
This is not done for this study.

The transverse cell size of the calorimeter was chosen
to be
1~cm~$\times$~1~cm, i.e. close to the Moliere radius.
For the moment
a non-pointing geometry
with a stable cell size 
is assumed 
over the whole barrel and also 
the end-wall has the same cell size.
The development of the transverse size of the shower in the three
longitudinal sections reveals that after 
cluster finding the transverse sizes in the three longitudinal compartments
look very similar. Thus, the overall transverse size $T$ is defined
as
\[
 T = \frac{ \sum_{i} E_{i} \sqrt{(\eta_i-\bar{\eta})^2 +
                                 (\phi_i-\bar{\phi})^2 } }
          { \sum_{i} E_{i} },
\]
where i runs over all calorimeter cells. $E_{i}$ is the energy in the cell,
$\eta_i$ and $\phi_i$ define its center. The baricenter of the
cluster is denoted by $\bar{\eta}$ and $\bar{\phi}$.

\begin{figure} 
\begin{center}
\epsfig{file=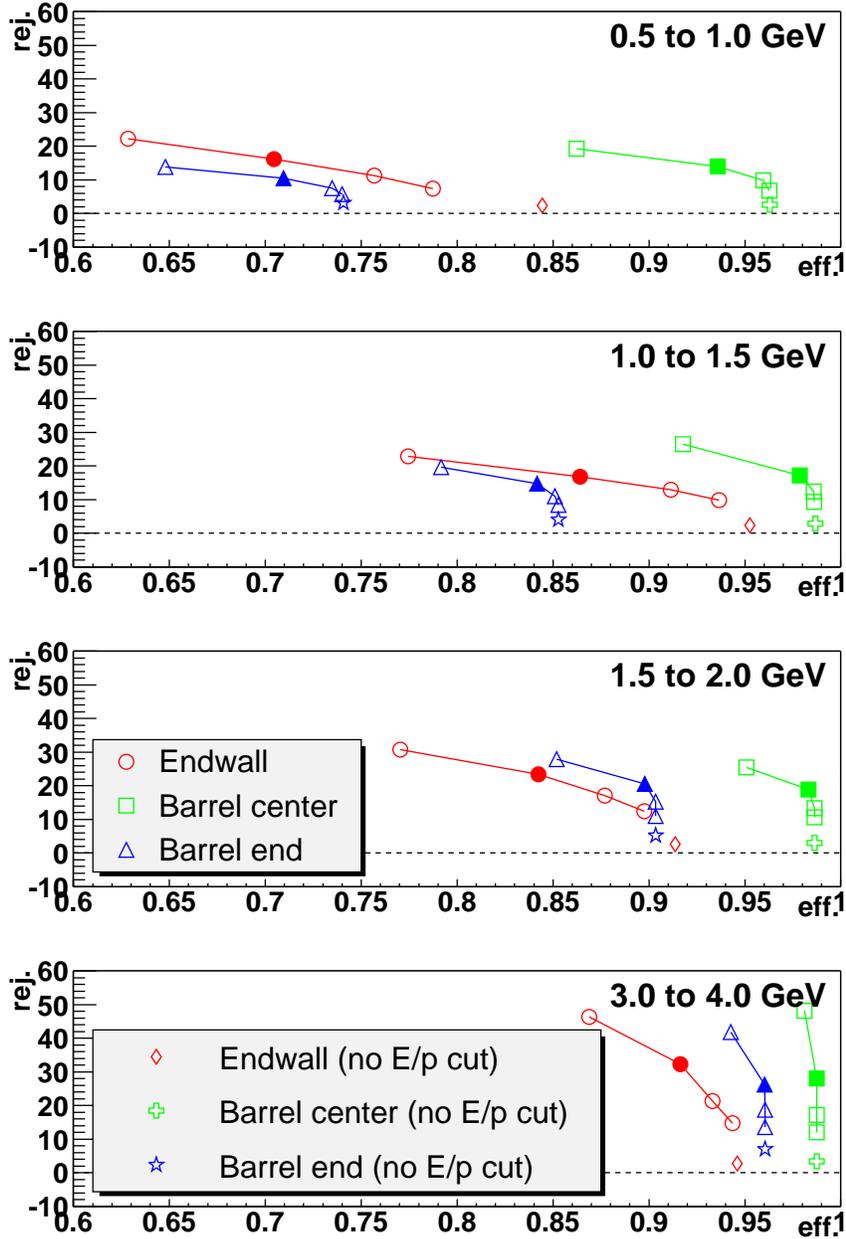,width=13cm}
\caption{Rejection power vs. efficiency for 
         cuts on $E_{vis}/E_{true}>$~0.5,~0.6,~0.7,~0.8
         plus longitudinal and transverse segment cuts. 
         Results are given for the central~[green squares] 
         and end~[blue triangles] part of the barrel and the
         end-wall~[red dots] 
         calorimeter for 4 energy bins.
         Filled symbols represent our final choice of $E_{vis}/E_{true}>$0.7. 
         Special symbols indicate results using no cut 
         on $E_{vis}/E_{true}$~[E/p].
    } 
  \label{det:fig:episeplt}
\end{center}
\end{figure}

The cut in $T$ to identify electrons is
\[
 T < 0.1 \frac{E_{true}}{1.5\mathrm{GeV}}. 
\]
Figure~\ref{det:fig:episeplt} depicts the results when this cut
is combined with the cuts used in Fig.~\ref{det:fig:episepl}.
The increase in rejection power is not convincing while the loss
in efficiency is quite significant, especially in the end-wall.
This is due to the widening of electromagnetic showers due to
the addition of a Bremsstrahlungs-photon. 
For all further analysis presented in this study the transverse
shower development is not used.

\begin{figure} 
\begin{center}
\epsfig{file=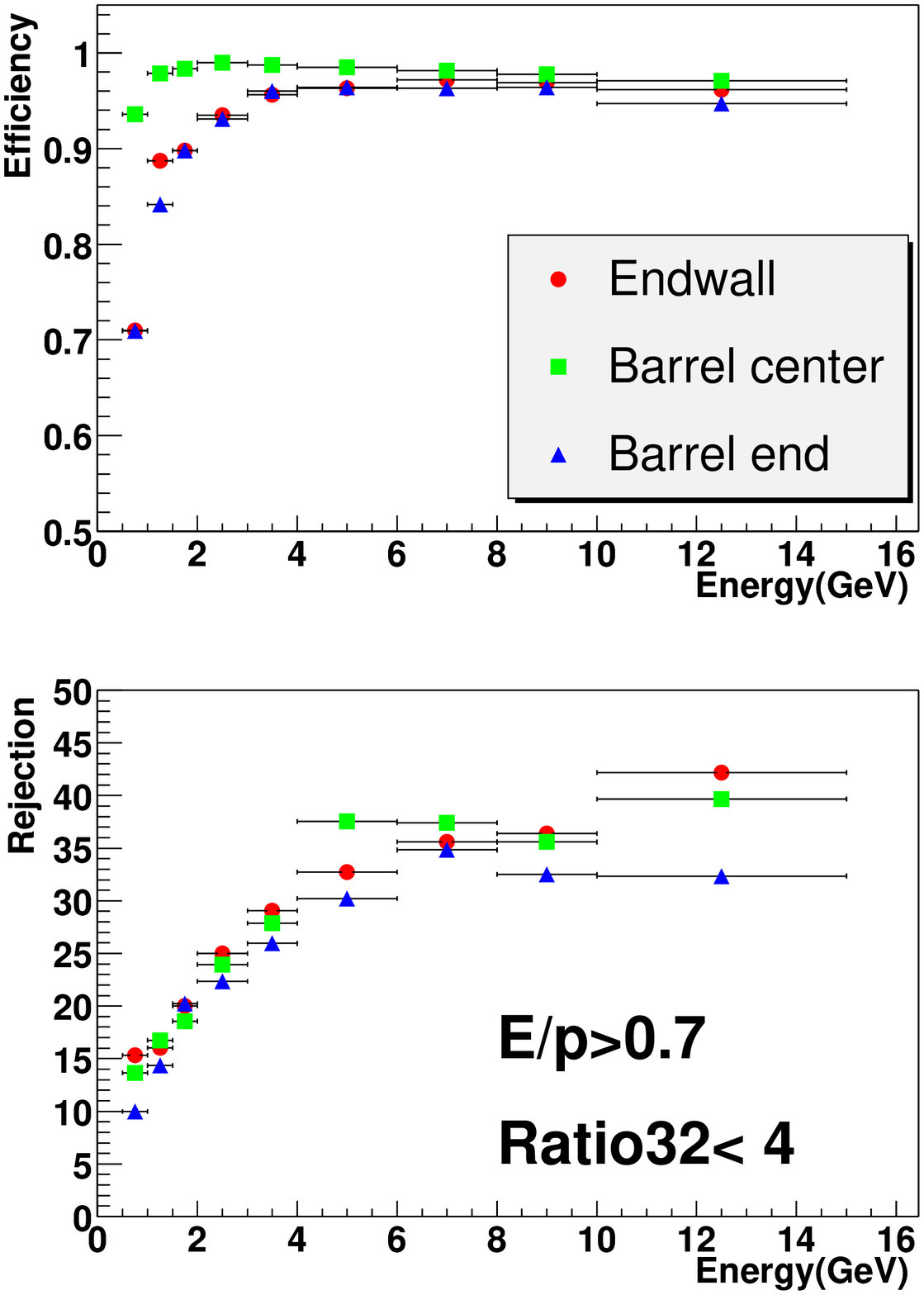,width=13cm}
\caption{Efficiency [top] and rejection power [bottom] 
         of a cut $E_{vis}/E_{true}>0.7$~[E/p] plus longitudinal cuts.
         Results are given for the central~[green squares] 
         and end~[blue triangles] part of the barrel and the
         end-wall~[red dots] 
         calorimeter.
    } 
  \label{fig:perf:epifinal}
\end{center}
\end{figure} 

The final result using a cut $E_{vis}/E_{true}>0.7$ and the
longitudinal energy cuts decribed above is given in 
Fig.~\ref{fig:perf:epifinal}.
Above 2~GeV a 90~\% efficiency to identify positrons 
with a rejection factor
above 20 for pions is  achieved.
Between 1~GeV and 2~GeV the efficiency stays above 85~\% while
the rejection factor is between 15 and 20.
Below 1~GeV only the good sampling of the central barrel
would support e-$\pi$~separation with good efficiency, if additional
tracking was added.

The question of transverse shower development should be
revisited
after further refinement of the clustering procedure. 
In addition the total number of
three dimensional cells with energy deposition could be considered. 
Also
a neural network analysis using full cell information
might be very useful.

More design optimization for the various parts of the calorimeter
is deferred to the actual design phase of the detector.
A projected geometry for the barrel and finer sampling
in the end-wall could support e-$\pi$~separation down to 0.5~GeV
everywhere.
The barrel calorimeter could also have a number of layers
varying with~z.
The effective depth would always be kept at 25~X$_0$. 
The 25~inner layers would extend to $\pm$70~cm and the 25~outer 
layers would 
gradually be reduced in length to $\approx \pm$20~cm.
This would reflect the change in longitudinal segmentation
described
earlier 
and not degrade the overall performance.
However, it would save
a substantial amount of silicon and tungsten and
the placement of read-out electronics and support mechanics
would become easier.   
Another possibility is to implement a digital instead of
an analogue readout.
This would allow for very fine longitudinal
and transverse segmentation. The small three dimensional cells
would reveal the shower structure. 
The number of holes within the shower
volume would be one of the observables to separate 
hadronic and electromagnetic 
showers.

Clearly more studies are advisable
to optimize e-$\pi$ separation  before a calorimeter is actually
constructed.
However, it seems feasible to meet the requirements 
to efficiently find and identify electrons down to an
energy of 1~GeV.

\clearpage
\newpage
\subsection{Background rejection}

\begin{figure} [h]
\begin{center}
\hbox to \linewidth{\hss
\put(0.,0.){\epsfig{file=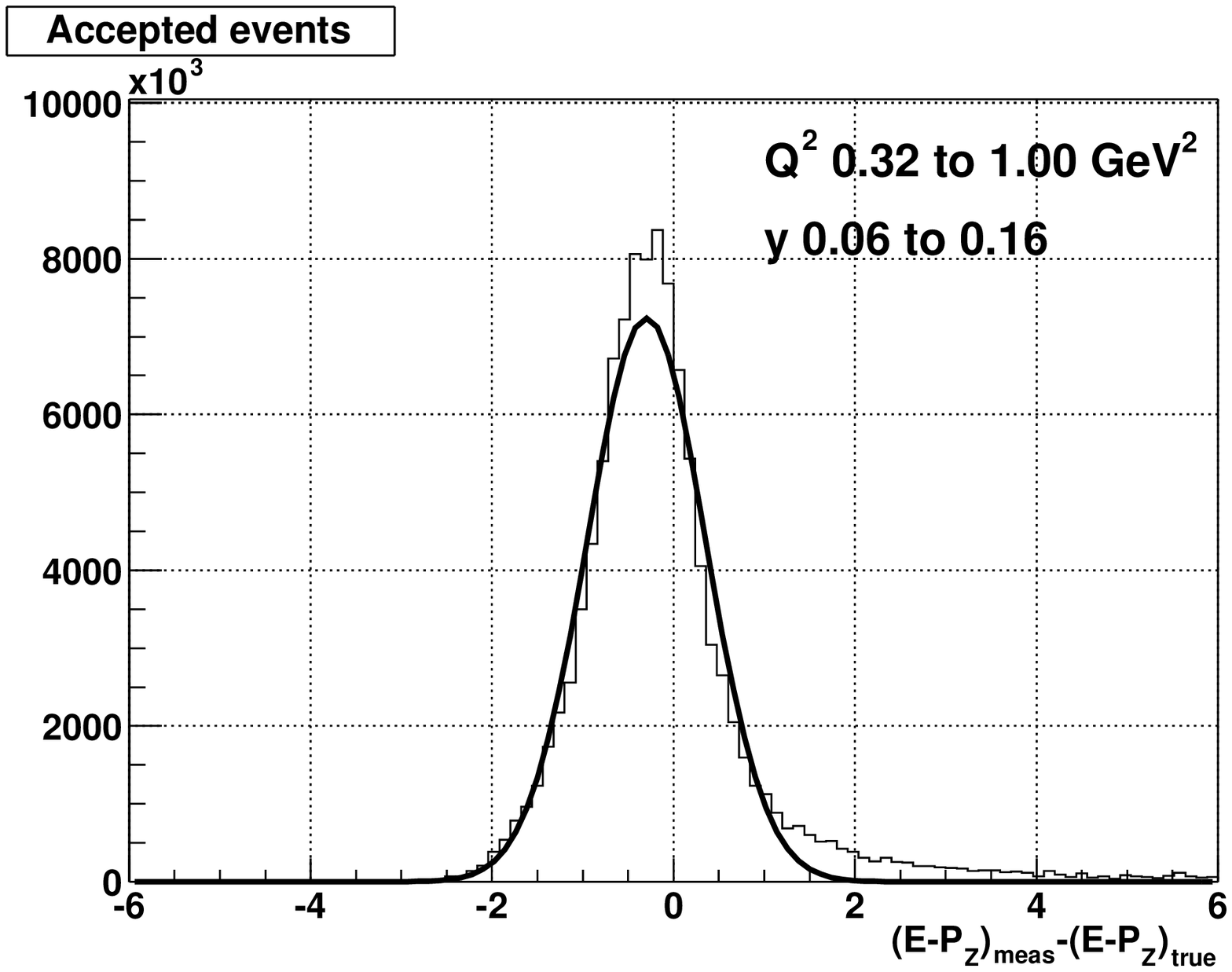,width=7.8cm}}
\put(250.,0.){\epsfig{file=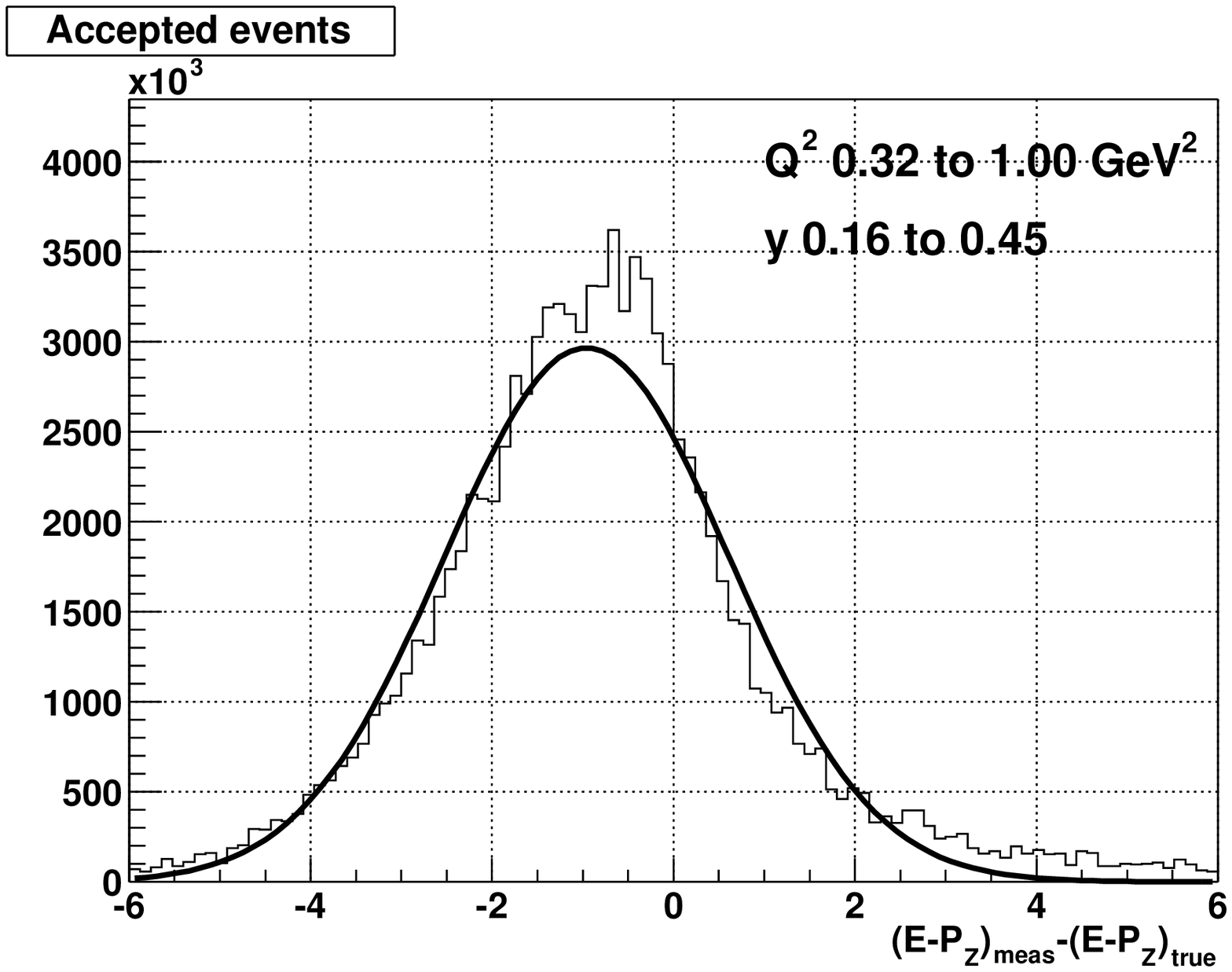,width=7.8cm}}
\put(205.,10.) {\small GeV}
\put(455.,10.) {\small GeV}
\hss}
\caption{The distribution of 
         $(E-P_z)_{\mathrm{meas}}-(E-P_z)_{\mathrm{true}}$
         as seen for NC events with an identified electron.
        } 
  \label{fig:perf:EmPzres}
\end{center}
\end{figure} 

There will be several types of backgrounds to deal with in the 
detector design discussed here:
\begin{itemize}
\item
Beam related background, such as synchrotron radiation, off-momentum electrons
and secondaries from upstream proton interactions.  We do not expect these
events to be mistaken for real events, as they can easily be removed with
timing and vertex constraints.  However, if the rate is too high, 
overlay or even radiation damage problems could occur.  A detailed study of
these backgrounds has not yet been performed, but would clearly be an
important issue in future studies of this type of detector design.

\item
Background from Bethe-Heitler overlays. There will be a high rate of
electrons in the detector from the e~p~$\rightarrow$~e~p$\gamma$ Bethe-Heitler
reaction.  The expected rate approaches 2~MHz  for a luminosity of \\
$1\cdot10^{32}$~cm$^{-2}$~s$^{-1}$.  These extra electrons will need 
to be properly treated  in the analysis procedure.  
As the electrons are produced under fixed zero degree angle,
their identification should be feasible.
This also
allows a high precision luminosity measurement with the main detector.

\item 
Photoproduction events ($Q^2 \sim 0$~GeV$^2$) mistakenly reconstructed as 
DIS events ($Q^2 > 1$~GeV$^2$) are a
source of background at HERA.  This is particulary true at high $W$, where the
scattered electron has low energy and a photon or $\pi^0$ in the hadronic
state can be misreconstructed as the scattered electron.
As the acceptance of this detector allows the reconstruction of the true
scattered electron down to very low Q$^2$ values, the
background from smaller $Q^2$ events is naturally reduced.  
Furthermore, at
high $W$, the scattered electron always enters the detector and will be
reconstructed. 

The low~W photoproduction background, where the scattered electron carries a 
large fraction of the beam energy and therefore remains in the beampipe, 
can be suppressed by a minimum requirement on the quantity $E-P_z$
observed in the detector.
The  $E-P_z$~resolution was studied with Monte Carlo events from set
NC1 which have identified electrons in the standard detector. 
Fig.~\ref{fig:perf:EmPzres}
gives the result for two $y$~bins 
and $Q^2$~$\in$~[0.32,1]~GeV$^2$.
The resolutions of 0.7~GeV~[left] and 1.6~GeV~[right] 
and shifts of 0.3~GeV~[left] and 1.0~GeV~[right]
should be compared to 
$E-P_z$~=~20~GeV. A cut of about 15~GeV to suppress photoproduction
is clearly feasible, especially for the low~W region which 
corresponds to the left picture.      

Given the ability to reject low~W photoproduction background and the 
containment of high~W events, photoproduction will not be a major source
of background with this detector.

\item
Events with initial state photon radiation.  These events can be controlled
both by measuring the radiated photon (see section ~\ref{sec:fb}) or by
measuring E-P$_z$ in the central detector.

\end{itemize}

\clearpage
\newpage
\section{Physics reach}
In this section, 
the kinematic limits of the set-up are discussed
and compared with the physics reach determined by full
MC studies.

\subsection{Structure function $F_2$}

Detailed studies of the low~$x$  and the high~$x$ regime 
are presented in later sections. Here we discuss the basic kinematic
limitations.

\begin{figure} [h]
\begin{center}
\epsfig{file=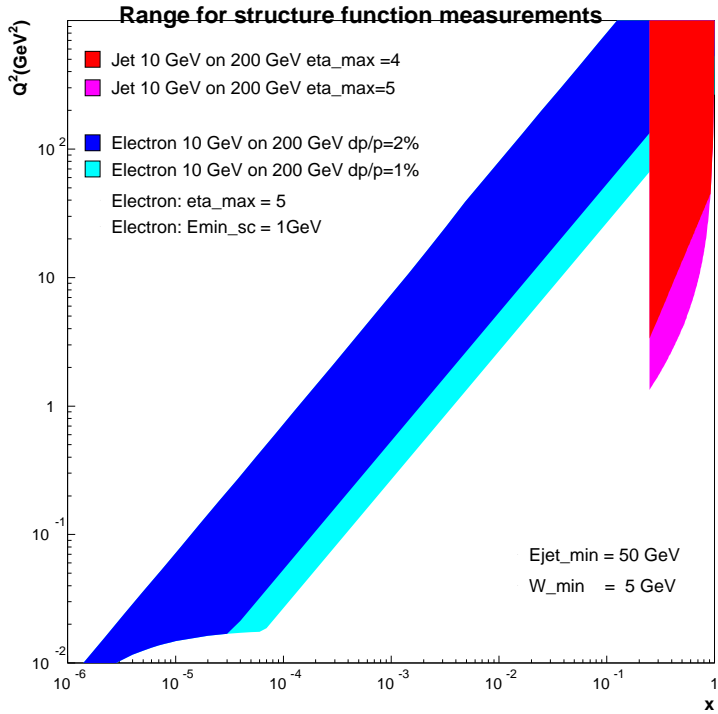,width=10cm}
\caption{The kinematic coverage with the electron method 
assuming a 2~\% or 1~\% electron momentum measurement
at small-$y$ (dark and light blue shaded areas) 
and with a combination of the electron and jet information, 
where the maximum $\eta$~in the acceptance is assumed to be
4~(red shaded area) or 5~(magenta shaded area).
The intermediate region requires a
measurement of $y$ from the hadronic system.
    } 
  \label{fig:F2kin}
\end{center}
\end{figure} 

In the low~$x$ regime only the scattered electrons are
used to reconstruct the kinematic
variables x and Q$^2$.
The acceptance for the electron in the tracking system
has already been shown in 
Fig.~\ref{det:fig:acc}. 
The scattered electron can be momentum analyzed
over the full $W$ (or $x$) range for 
$Q^2$ down to 0.05~GeV$^2$, since the electron scattering angle is
large enough for it to enter the detector independent of momentum. 
Below this $Q^2$, the acceptance is limited to the larger values of $W$
where the magnetic field is used to extract the scattered electron out
of the beampipe.  The detector has 100~\% acceptance for $W>80$~GeV
down to $Q^2=0$~GeV$^2$.

The $x$ resolution is given by
\begin{displaymath}
\frac{\delta x}{x} = \frac{1}{y}\frac{\delta p}{p} \oplus 
2 \frac{\delta \epsilon}{\epsilon}
\end{displaymath}
where $p,\epsilon$ are the scattered electron momentum and angle, respectively.
At the smallest values of $x$ for a given $Q^2$, the
electron energy is small and the kinematic quantities will be measured with
high precision.  However, as $x$ increases, $y$ decreases (for a fixed
$Q^2$) and the electron method rapidly loses resolution in the measurement
of $x$ (or $W$).  The kinematic range yielding 
$\frac{\delta x}{x} <$30~\% for the electron 
method assuming a momentum 
resolution of $1$~\% or 2~\% is shown in Fig~\ref{fig:F2kin}.  
The standard detector including a beam-pipe cannot quite reach that level, 
see sect.~\ref{sec:perf:mom}.
However, for a Roman pot system without a beam-pipe 
a 2~\% momentum resolution is realistic.
The low~$x$ edge is determined by the minimum energy required for the
electron. The cut of 1~GeV is a result of the choice of silicon-tungsten
calorimetry, as e/$\pi$~separation becomes impossible at lower energies.

In the high~$x$ regime the electron is used to
calculate $Q^2$ and the jet energy is used to calculate $x$.
Thus the minimum energy required for a jet determines the left edge of 
the accessible area.
The acceptance in $\eta$ limits the reach towards low~$Q^2$. 
In addition the minimum
$W$ required cuts into the $Q^2$-reach at extremely high~$x$.
The resulting kinematic coverage
using this mixed method is also shown in  Fig~\ref{fig:F2kin}.  

For intermediate $x$ values, where
neither the electron nor electron+jet methods work, other techniques such as
the double angle method must be employed.  
It is expected 
that these methods will 
fill the gap between low and high~$x$ seen in
Fig.~\ref{fig:F2kin}.  This region has not been studied in detail,
but experience at HERA indicates that it should be possible to perform
high precision measurements also in this kinematic range.

This study focuses on the low and high~$x$ regime, for which the
detector was optimized and which are the most interesting.

\subsection{Structure function $F_2$ at low~$x$}
\label{sec:phys:f2low}

In the last section the kinematic range in principle accessible
to the device was discussed including the key role of momentum
resolution. The experimental situation is, however, more complicated,
as not only the Gaussian part of the momentum resolution, but also
the tails as discussed in sect.~\ref{sec:perf:mom} are of importance.
In addition e/$\pi$ separation is crucial in identifying the 
scattered electron.

The study uses the events from MC~sample~NC1 as described in 
sect.~\ref{sec:MCR}. It should be noted that this sample 
contains only events with
$Q^2 <$~100~GeV$^2$.

\subsubsection{Event reconstruction}

The event reconstruction assumes perfect pattern recognition
in the tracker and uses the fit method described in sect.~\ref{sec:perf:mom}.
The calorimeter cell information is used to identify 
clusters which are subsequently matched to tracks.

The cluster finding algorithm does not use the longitudinal segmentation
of the calorimeter. 
The cells are grouped into 5 by 5 cell towers.
If a tower shows energy, the neighboring towers are checked and
added to the cluster, if they too contain energy.
The barycenter of the cluster is calculated using the full 
cell granularity of the calorimeter.

A track is matched to a cluster if its distance of closest approach
to the barycentre of the cluster is less than~10~cm.
If more than one cluster are within the cut the track is matched to the
closest cluster.

The electron is identified by ordering the tracks 
in the electron hemisphere and central tracker according to momentum.
Descending in momentum the corresponding cluster is analyzed
for its electron probability. The cuts described in sect.~\ref{sec:perf:epi},
with performance as shown in Fig.~\ref{fig:perf:epifinal}, are used. 
If no cluster could be assigned, the track is assumed to be a pion.
If no electron can be identified in the list of tracks, clusters
in the catcher and the barrel which do not have a track are analyzed.
As only events with $Q^2 <$~100~GeV$^2$ are considered, there is no need
to search for electrons in the proton hemisphere.

\subsubsection{Efficiency and purity}

\begin{figure} [ht]
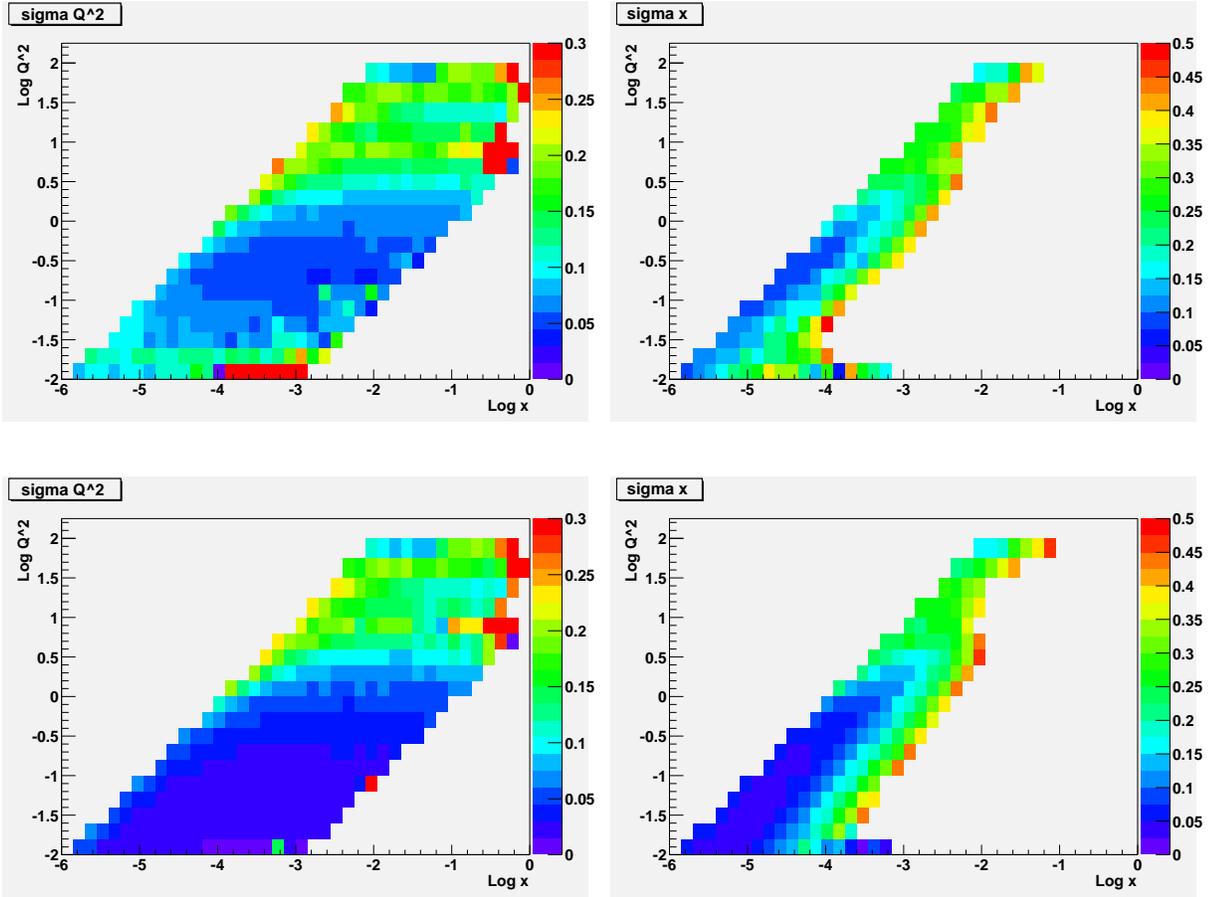

\begin{center}
\hbox to \linewidth{\hss
\put(0.,0.){\epsfig{file=elight200_h2d_sigma_q2.epsi,width=7.0cm,angle=-90}}
\put(230.,0.){\epsfig{file=elight200_h2d_sigma_x.epsi,width=7.0cm,angle=-90}}
\put(0.,180.){\epsfig{file=std200_h2d_sigma_q2.epsi,width=7.0cm,angle=-90}}
\put(230.,180.){\epsfig{file=std200_h2d_sigma_x.epsi,width=7.0cm,angle=-90}}
\hss}
\caption{
The relative resolutions in $Q^2$ and x are given for the standard~[top] and
extra-light~[bottom] detector as a function of $Q^2$ and x.
    } 
  \label{fig:phys:resol}
\end{center}
\end{figure}

The resolution in x and $Q^2$ was used to determine the binning
for the analysis. These resolutions are shown in Fig.~\ref{fig:phys:resol} for
both the standard material budget and the extra-light material budget.  As
is clear from the figure, the $Q^2$ resolution is very good for $Q^2<5$~GeV$^2$
and degrades as $Q^2$ increases.  This is because the electron scattering
angle increases, resulting in both a smaller $\int B\cdot dl$ and fewer 
planes crossed.  Also, at higher values of $Q^2$, no track is available and
only calorimeter information can be used, resulting in poorer resolution.

As described above, good resolution
in $x$ is only possible with the electron method
for the smallest values of $x$ for a fixed $Q^2$.  This
is clearly seen in Fig.~\ref{fig:phys:resol}, where the resolution in $x$
is given as a function of $Q^2$ and~$x$.  Only those areas are shaded where
the resolution is better than 50~\%. The resolution again degrades
at the larger values of $Q^2$ where the tracking is less effective or absent. 
These is a marked difference in the resolutions
determined for the standard and extra-light material budgets, adding emphasis
to the point that minimizing the material budget will be a crucial design
consideration.

The resolutions determined with the extra-light material budget were used
to define bins in which cross sections were evaluated.  
Efficiency $\epsilon$ and purity $P$  
for a given bin are defined as   

\begin{displaymath}
\epsilon = \frac{\#~events~generated~and~reconstructed~in~bin}{\#~events~generated~in~bin}
\end{displaymath}
and
\begin{displaymath}
P = \frac{\#~events~generated~and~reconstructed~in~bin}
     {\#~events~reconstructed~in~bin} .
\end{displaymath}
 
\begin{figure}
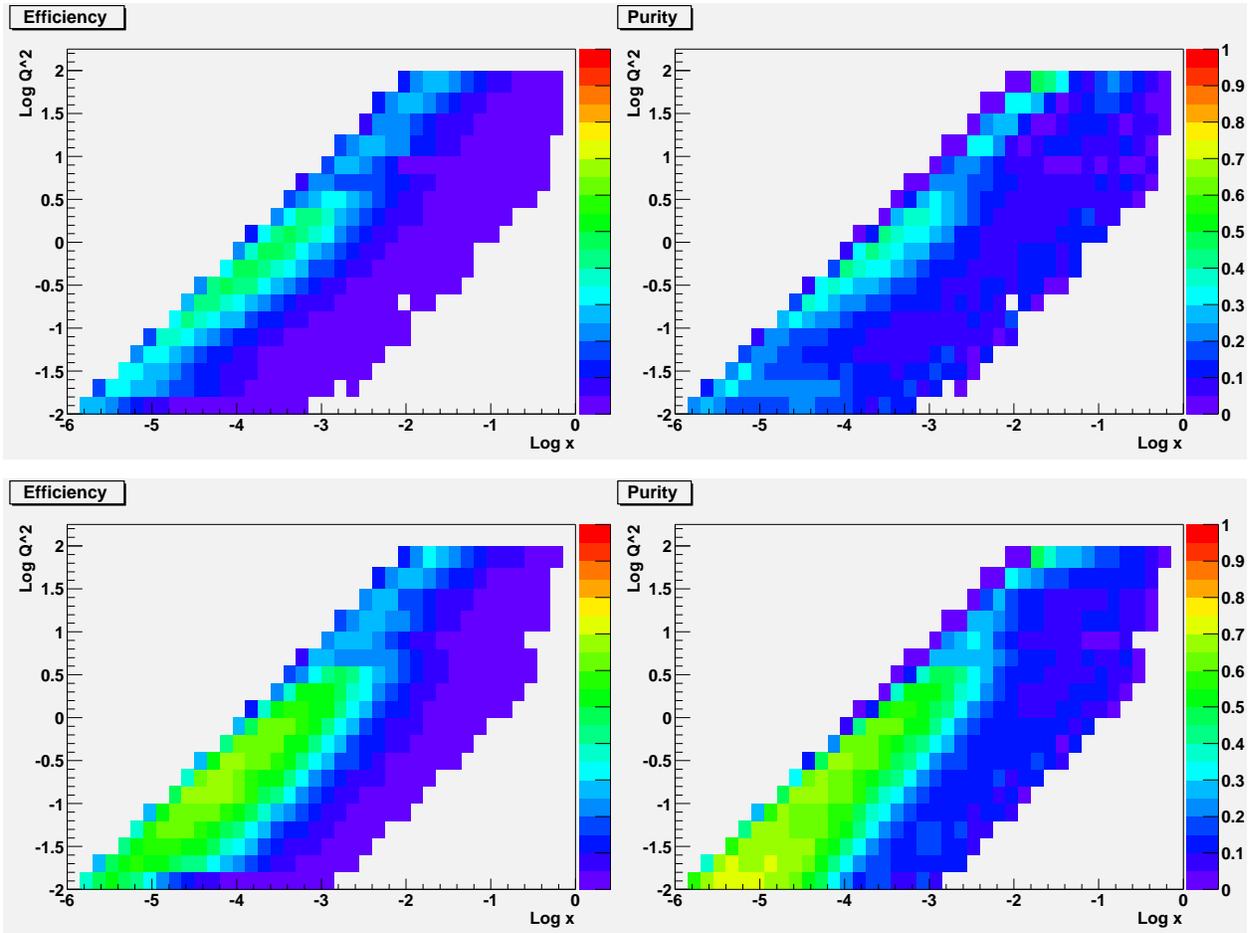
 
\begin{center}
\hbox to \linewidth{\hss
\put(0.,0.){\epsfig{file=elight200_h2d_effi.epsi,width=7.6cm,angle=-90}}
\put(230.,0.){\epsfig{file=elight200_h2d_puri.epsi,width=7.6cm,angle=-90}}
\put(0.,180.){\epsfig{file=std200_h2d_effi.epsi,width=7.6cm,angle=-90}}
\put(230.,180.){\epsfig{file=std200_h2d_puri.epsi,width=7.6cm,angle=-90}}
\hss}
\caption{
The efficiencies and purities are given for the standard~[top] and
extra-light~[bottom] detector.
    } 
  \label{fig:phys:efpu}
\end{center}
\end{figure}

Fig.~\ref{fig:phys:efpu} gives the results for two different material budgets
in the detector in the chosen bins. The standard detector has 
tracker support structures and a beam-pipe,
the extra-light detector does not  (see also sect.~\ref{sec:MCR}
table~\ref{tab:mc:mat}). It is obvious that 
the material in the standard detector reduces the range of good efficiency
and purity significantly.  
The better momentum resolution of the extra-light detector is reflected
in better efficiency and purity at higher x and down to lower $Q^2$.
At high $Q^2$ the electron is frequently only seen in the calorimeter.
As the energy resolution in the calorimeter is worse then the momentum
resolution in the tracker and e/$\pi$ separation less effective without
a track, the range in x is reduced in this region.

\subsubsection{Systematic errors}
\label{sec:phys:f2low:err}

The goal is to measure $F_2$ to better than 2~\%.  The statistical errors
will be negligible for all the bins considered here for the 
expected luminosities.
Thus all systematic errors should be controlled to 1~\%.
Sources of systematic error are studied by shifting the relevant
variables in the reconstruction and monitoring the shift in $F_2$.
The result is that it is required to control
\begin{itemize}
\item the energy scale to  1~\%,
\item the momentum scale  0.1~\%,
\item the hit efficiency to  1~\%,
\item the tracking efficiency to 1~\% .
\end{itemize}
These requirements are well within achievable bounds.  For the measurement
of $F_L$, tighter requirements must be set as described below.

The energy scale of the calorimeter acts through the e/$\pi$ separation
and affects mostly the low~$x$ region.
The momentum scale affects, via error propagation,
mostly the high~$x$ region.
Systematic uncertainties in the hit and tracking 
efficiency shift $F_2$ as a whole.

\subsubsection{Results}

\begin{figure} 
\begin{center}
\epsfig{file=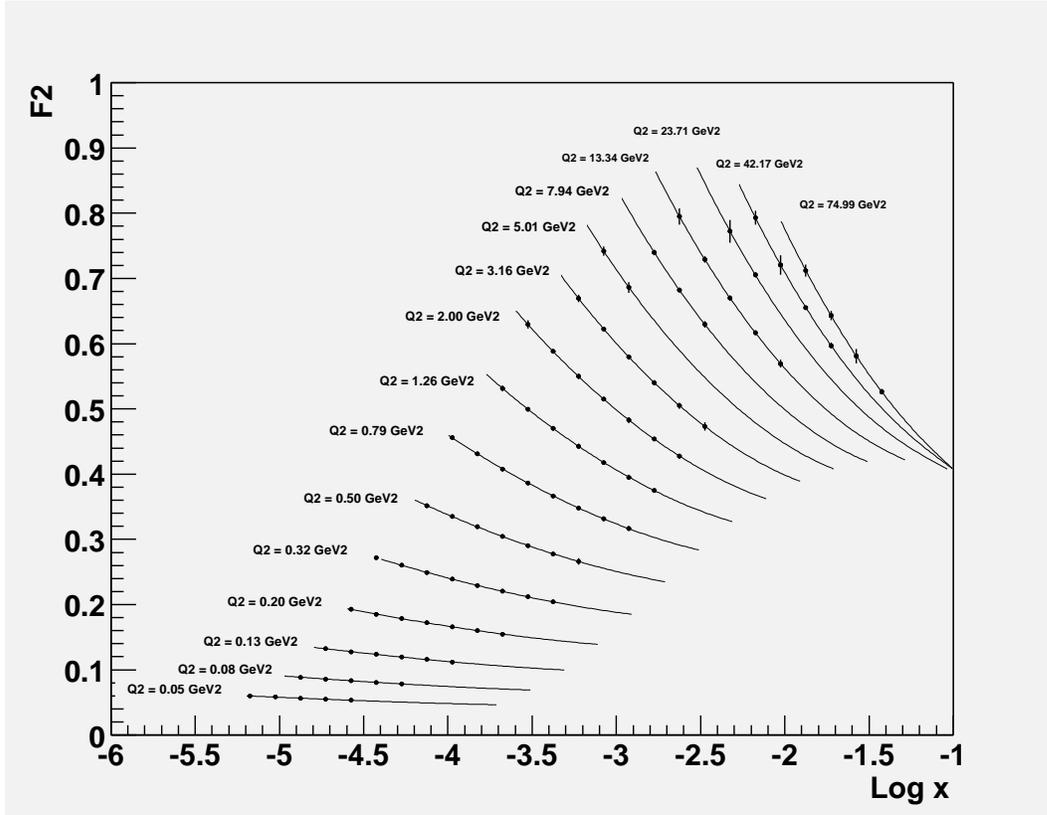,width=14cm}
\caption{
$F_2$ extracted for the standard detector.  The data points are shown for
the bins passing the purity and efficiency cuts as described in the text.
Statistical errors are shown as the inner error bar (they are smaller than
the symbols), while the quadratic sum of the statistical and systematic errors
are shown as the outer error bar. 
    } 
  \label{fig:phys:f2-std}
\end{center}
\end{figure}

\begin{figure} 
\begin{center}
\epsfig{file=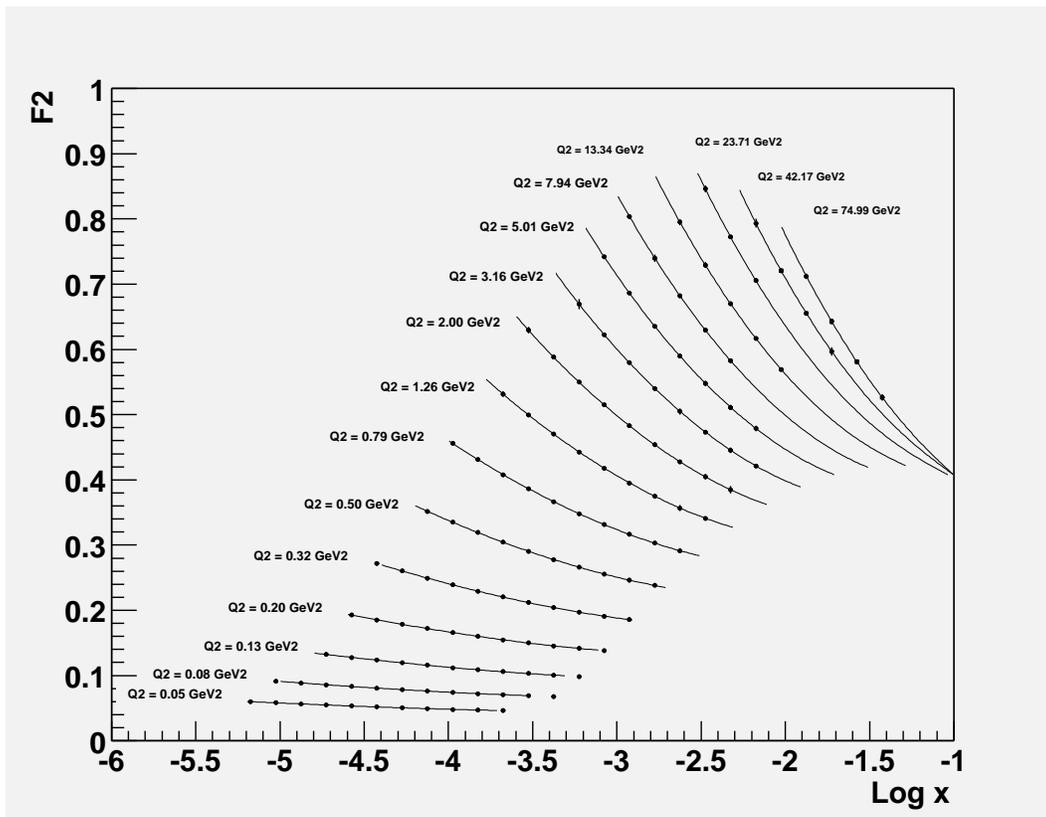,width=14cm}
\caption{
$F_2$ extracted for the extra-light detector (see caption of 
Fig.~\ref{fig:phys:f2-std}).
    } 
  \label{fig:phys:f2-elight}
\end{center}
\end{figure}

Bins where the efficiency and the purity are above 20~\% are considered
useful for the physics analysis. This is a lower cut than normally used
in structure function analyses.  It was chosen this way in order to have
the same binning for the different material budgets.  As the bins were
set using the higher resolutions of the extra-light scenario, we use a
low cut to allow also a significant number of measured bins with the standard
material budget.  The binning would be changed in a more complete analysis, but
this was not attempted here.
The $F_2$  extracted  for the standard and the extra-light detector
are depicted in Figs.~\ref{fig:phys:f2-std} and~\ref{fig:phys:f2-elight}.
An integrated luminosity of 100~pb$^{-1}$ is assumed.

The error bars containing statistical and systematical errors
are smaller than the size of the symbols for most
of the points. The advantage of the extra-light detector is that a larger
kinematic range is accessible. The reduction in 
errors as seen in some points is less significant.

\begin{figure} 
\begin{center}
\epsfig{file=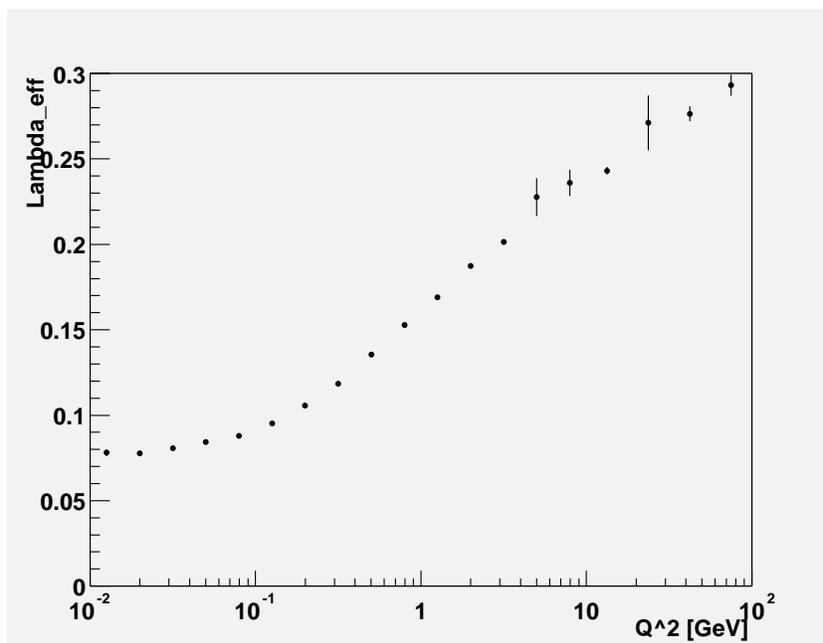,width=11cm}
\caption{
$\lambda_{\mathrm{eff}}$ as extracted for 100~pb$^{-1}$ with 
the standard etector. The error bar is the quadratic sum of the statistical
and systematic errors.  The value of $\lambda_{\mathrm{eff}}$ follows that
of the ALLM parametrization~\cite{ref:ALLM}.
    } 
  \label{fig:phys:lamda}
\end{center}
\end{figure} 

Fig.~\ref{fig:phys:lamda} depicts the extracted 
logarithmic derivative
\begin{displaymath}
\lambda_{\mathrm{eff}} = \frac{\partial ln F_2}{\partial \ln \frac{1}{x} }|_{Q^2}
\end{displaymath}
as discussed in sect.~\ref{sec:precsf}.
The errors shown come from the fit to $F_2$ with statistical and
systematical errors quadratically combined for each point.
 
In the transition region and below the 
error bars are smaller than the symbols. The precision and range in 
Fig.~\ref{fig:phys:lamda} has to be compared with Fig.~\ref{fig:lamda}.  A
very significant improvement in the precision and range is possible.  The
extra-light detector gives an even more significant improvement.

\subsection{Structure function $F_L$ at low~$x$}

The measurement of $F_L$ requires running at different center-of-mass 
energies and comparing cross sections for a given $x,Q^2$ but different
$y$, as is clear from the cross section formula
\begin{displaymath}
\frac{d^2 \sigma}{dx dQ^2} = \frac{2\pi\alpha^2}{xQ^4}
\left[Y_+ F_2(x,Q^2) - y^2F_L(x,Q^2)\right]
\end{displaymath}
where $Y_+ = 1+(1-y)^2$, and $xF_3$ has been neglected.

The reach in $F_L$ was studied by 
using the two Monte Carlo event sets NC1 and NC1L
(see sect.~\ref{sec:MCR})
using proton energies of 200~GeV and 100~GeV.
In both cases the analysis described in sect.~\ref{sec:phys:f2low} 
was applied. The usable overlap was determined by requiring both
the efficiency and the purity 
for both energies to be above 20~\% for a given bin.  The 
statistical error on $R= \frac{F_L}{F_2}$ was calculated from
the cross sections measured with the different center-of-mass
energies for a given bin.   A luminosity of 100~pb$^{-1}$
for each proton energy was assumed.
The systematic error on $R$ was then evaluated by
repeating the cross section extraction and $R$ calculation
for each systematic test.  The
systematic uncertainties clearly dominate, even for luminosities of
$100$~pb$^{-1}$.  The errors are minimized
by having the maximum y range for
a given $Q^2$ and x.  This implies the maximum possible spread in the beam
energies.  A factor 2 was assumed for this study, and a larger range would
yield improved results.
It is also important to measure as high in y as possible, i.e., as
low in electron energy as possible.  We have assumed that electron measurement
and identification down to $1$~GeV is possible. Only two beam energies 
were used
in the study; more energy points would help
in reducing some of the systematic errors.

\begin{figure} [h] 
\begin{center}
\epsfig{file=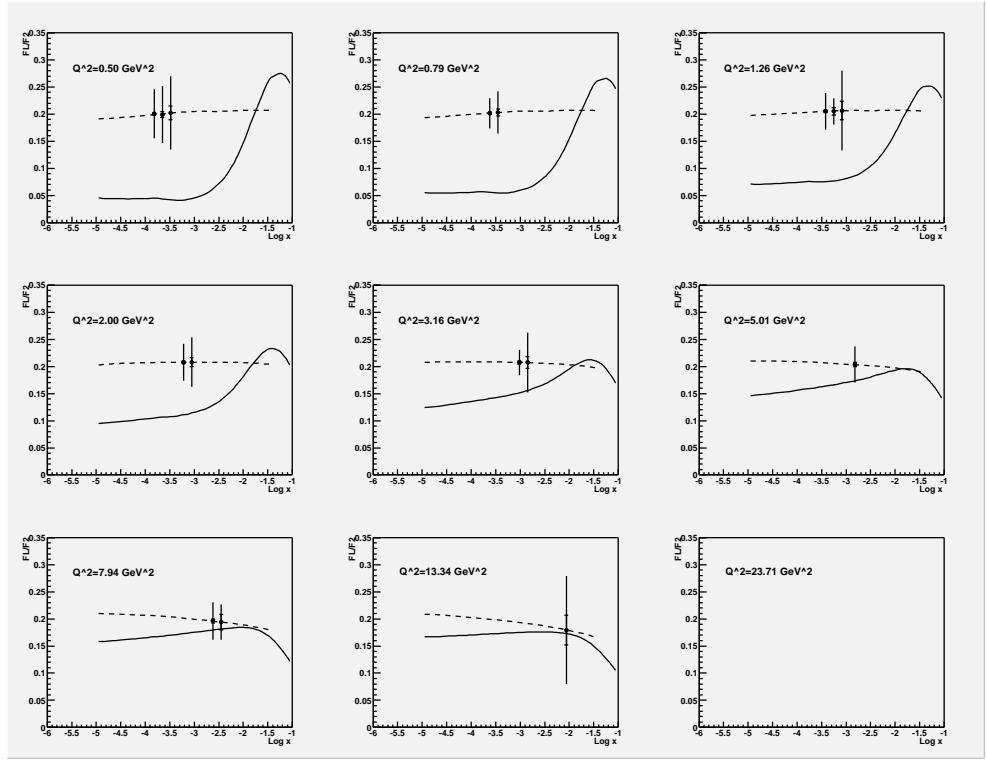,width=13cm}
\caption{The range of possible measurements of $R=F_L/F_2$ 
as attainable with the 
standard detector, 100~pb$^{-1}$ for each proton energy, 
and systematic errors according to 
sect.~\ref{sec:phys:f2low:err}. The inner error bars are statistical only,
the outer bars include systematic errors. 
The empty plot at the bottom
right reflects the inability to access this $Q^2$~region. 
The full line represents the prediction from DGLAP evolution~[MRST99], 
the broken line represents the saturation dipole model.
    } 
  \label{fig:phys:Flmeas-s-he}
\end{center}
\end{figure}

The resulting values of $R$
for the standard detector are given
in Fig.~\ref{fig:phys:Flmeas-s-he} 
along with two different sets of predictions.  The central value of $R$ was
arbitrarily chosen to lie on the curve from the saturation model.  Of 
importance here is the size of the error bars.
The inner error bars are statistical only, while the outer bars combine
statistical and systematical 
errors. Only points with errors on $R$ less than $0.1$ are shown.
The predictions come from DGLAP evolution~[full line] and the saturation 
dipole model~[dashed line].
The biggest difference is in the low~$x$ and low~$Q^2$ regime. 

\begin{figure} [h] 
\begin{center}
\hbox to \linewidth{\hss
\put(40.,300){\epsfig{file=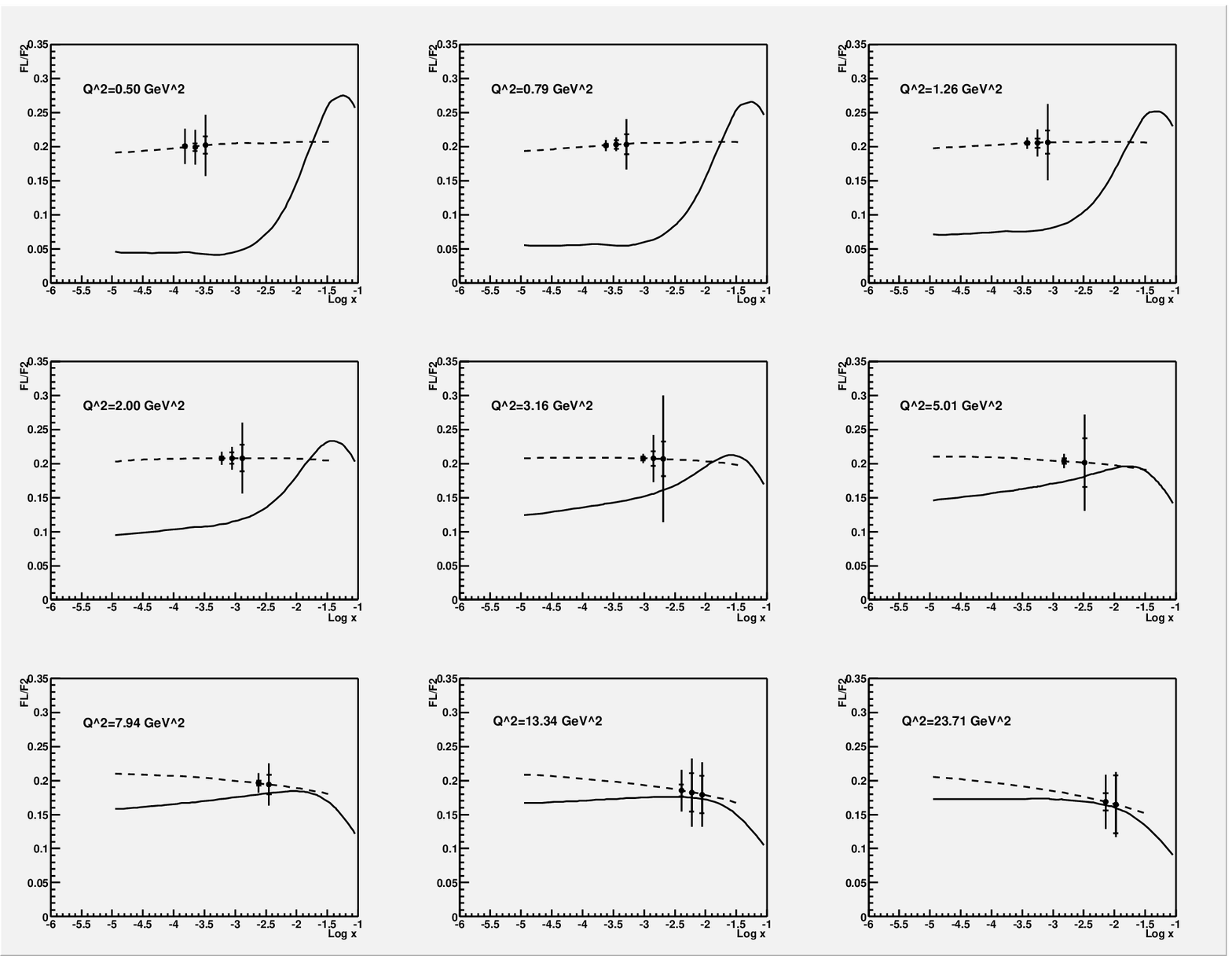,width=13cm}}
\put(40.,0.) {\epsfig{file=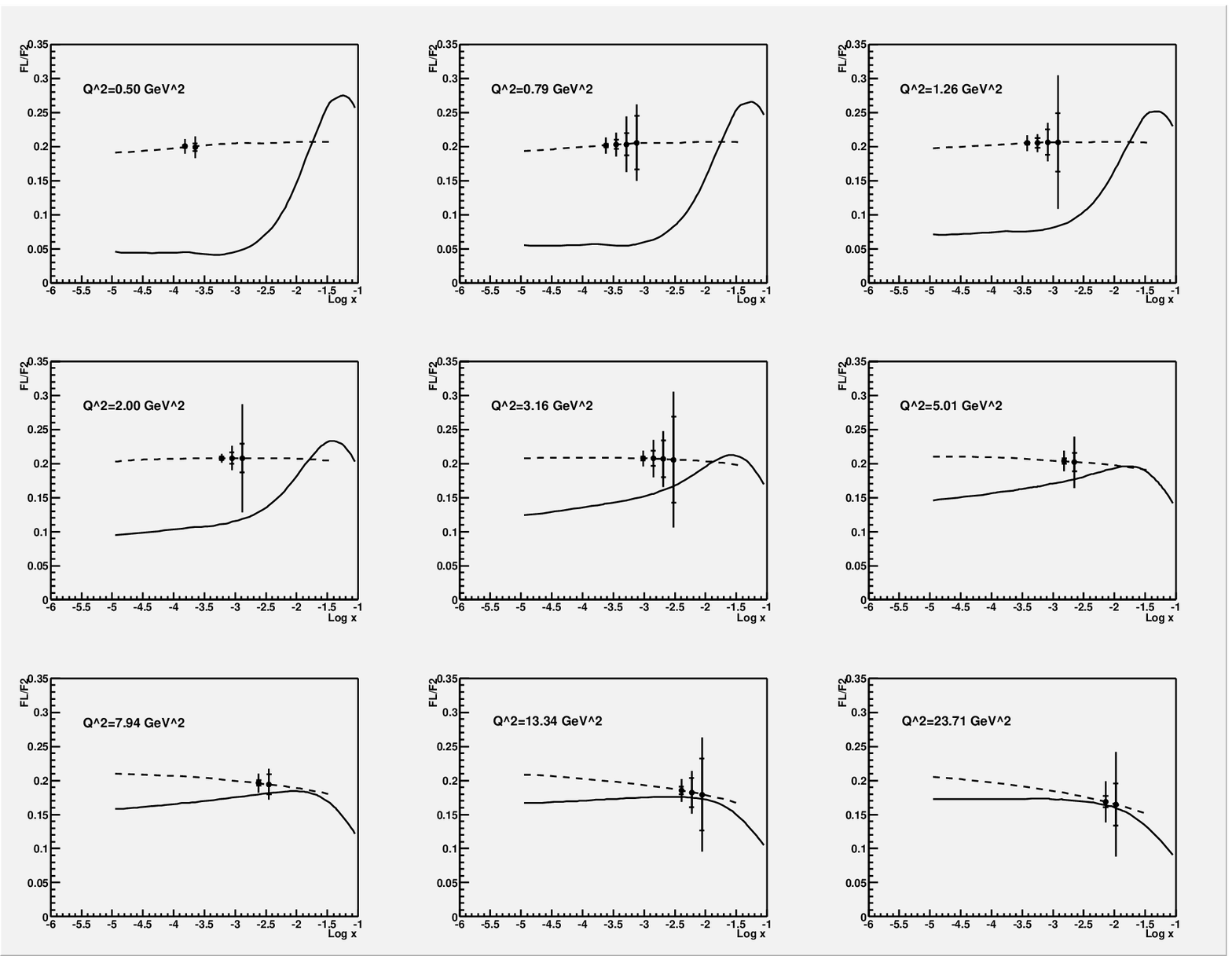,width=13cm}}
\hss}
\caption{
As in Fig.~\ref{fig:phys:Flmeas-s-he}, but [top] for reduced 
systematic errors and [bottom] for reduced errors and 
the extra-light detector. Only points with less than 10~\% error are shown.
    } 
  \label{fig:phys:Flmeas-sel-le}
\end{center}
\end{figure}


Fig.~\ref{fig:phys:Flmeas-s-he} demonstrates that the standard detector
is indeed able to provide a good measurement at low~$x$ and~$Q^2$.
Such a measurement can distinguish between models, but it is desirable
to extend the range to higher~$Q^2$. This can be done by reducing the
systematic errors. It can be realistically assumed that
\begin{itemize}
\item the energy scale is controlled to 0.3~\% ,
\item the momentum scale to 0.05~\%,
\item the hit efficiency to 0.2~\%,
\item the tracking efficiency to 0.2~\%.
\end{itemize}
This extends the range of the measurement in $Q^2$ and reduces the systematic
errors. The result is depicted in the
top plot of Fig.~\ref{fig:phys:Flmeas-sel-le}
The extra-light detector enlarges the $x$~range in the measurement
of $F_2$, and therefore also the $x$~range in~$F_L$. The result for
the extra-light detector is given
in bottom plot of Fig.~\ref{fig:phys:Flmeas-sel-le}.
As only points with a relative error of less than 10~\% are shown,
statistical fluctuations in the MC sample sometimes cause a point to
disappear. This effect can be seen for the extra-light detector 
in the lowest $Q^2$~bin.

As is clear from 
Figs.~\ref{fig:phys:Flmeas-s-he}
and~\ref{fig:phys:Flmeas-sel-le}, 
it is possible to measure $F_L$ with the required precision in
the kinematic region of highest interest. Even the standard detector
with the standard systematic errors can distinguish between the
predictions in an interesting area of phase-space.
With lower systematic errors and with an
extra-light detector the measurement is further improved.

\subsection{Forward jets and particle production}

\begin{figure} [h]
\begin{center}
\epsfig{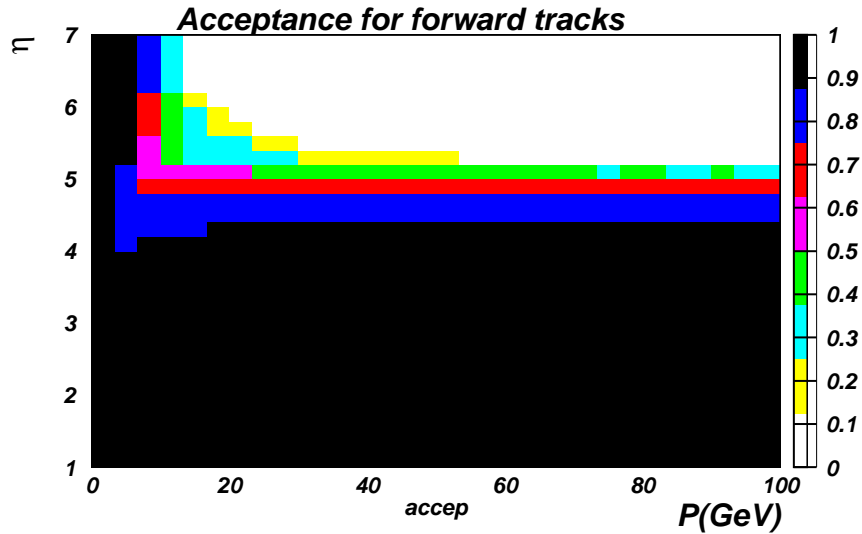}
\caption{The acceptance of the tracking system 
for positive pions in the proton direction.  
A track is considered accepted if it has crossed at least three tracking
stations. The MC sample I2 (see sect.~\ref{sec:MCR}) was used.
    } 
  \label{fig:phys:rap_p}
\end{center}
\end{figure}

One of the principal features of the detector is the possibility to track
particles over a very wide rapidity range. The tracking acceptance in the
proton direction is shown in Fig.~\ref{fig:phys:rap_p}. 
It extends to $\eta=5$ for
all momenta.  
This acceptance is to be
contrasted with the maximum pseudo-rapidity coverage of 
the currently operating e~p detectors H1 and ZEUS of
$|\eta| \leq 1.75$.  The vastly increased rapidity coverage allows the
measurement of particle production and particle correlations from the
current jet region into the target fragmentation region, thereby allowing
detailed studies of radiation patterns in QCD.

The situation for forward
jet production is also extremely advantageous, since the calorimetric
coverage extends also to very high rapidities.  
Jet reconstruction should
be possible out to $\eta=5$, again allowing for sensitive tests of
parton splitting processes.  
At this time it was impossible to find a Monte Carlo generator which
can reliably produce events at such high $\eta$ for the eRHIC kinematics
(see also next section).
The jets in Monte Carlo sample NC2 fill the pseudo-rapidity space up to
approximately $\eta$=4.

\begin{figure} 
\begin{center}
\epsfig{file=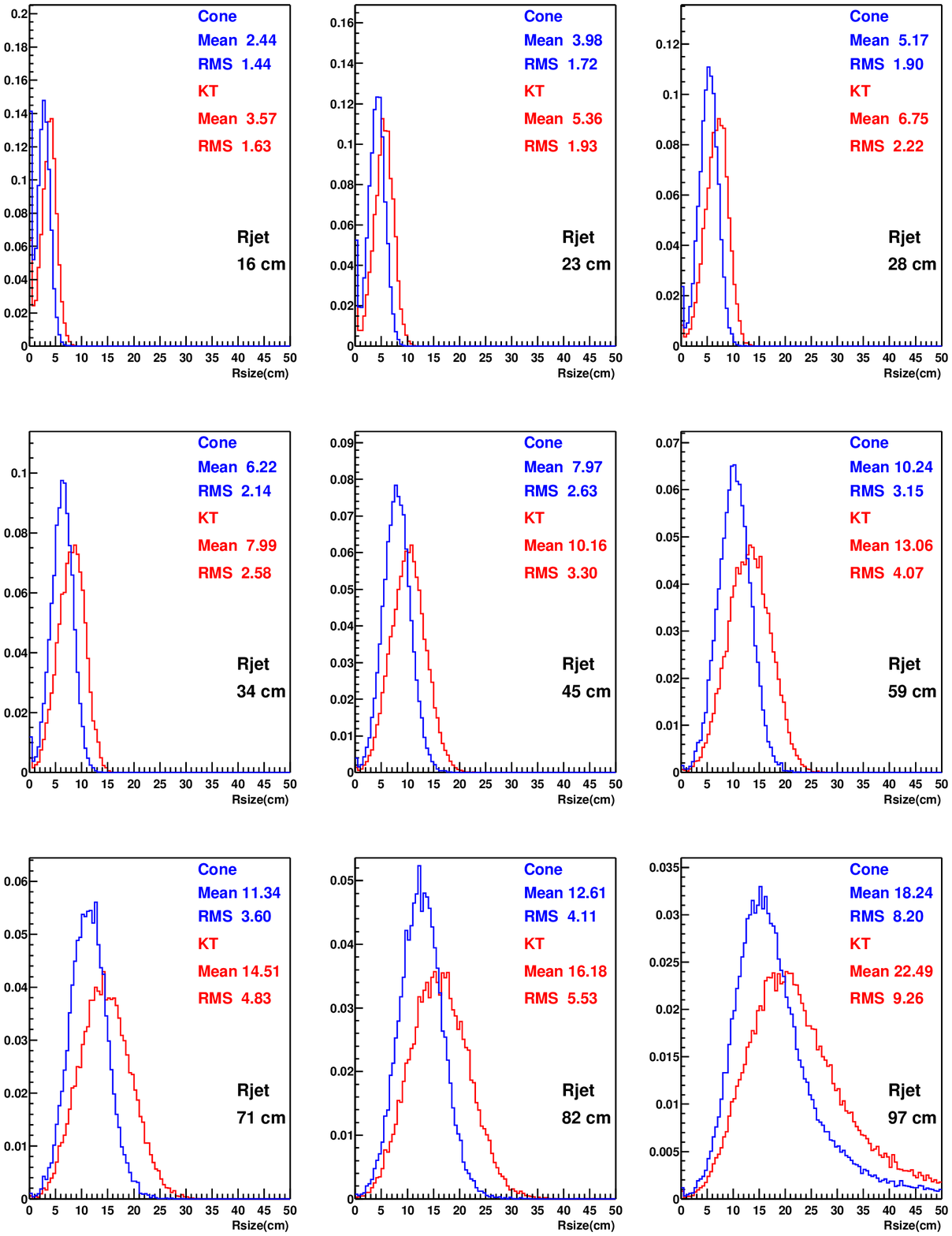,width=12.5cm}
\caption{
Jet positions and sizes 
as determined by the KT and cone jet-finding algorithms.
Rjet gives the radial distance of the jet axis from the proton beam
axis at the face of the calorimeter.
The cone algorithm is less affected by the proton remnant jet and produces
narrower jets with less contamination.
    } 
  \label{fig:phys:highx-jetsize}
\end{center}
\end{figure}

The jet sizes were studied using the KT and the cone 
jet-finding algorithms.  The jet size is defined as

$$R_{size} = \sqrt{\frac{\sum_i E_i |R_i - R_{jet}|^2}{\sum_i E_i}}$$

\noindent where the sum runs over the cells in the jet, $E_i$ is the cell energy,
$R_i$ is the cell position, and $R_{jet}$ is the jet centroid.
The KT algorithm was used in the standard 3212 mode.
The cone algorithm was used 
with a radius cut of 0.7 in $\eta$~$\phi$~phase-space.

Fig.~\ref{fig:phys:highx-jetsize} gives the distribution of $R_{size}$
for the two different algorithms for different radial distances of the
jet axis to the proton beam axis at the face of the calorimeter [z=360~cm].
The KT algorithm results in wider jets which are, as further studies
showed, pulled towards the remnant jet.  They also contain in average 
3~\% too much energy. The cone algorithm is much better in separating the
jet from the proton remnant.

The actual ability to measure the properties of these very forward jets
will certainly have to be complemented by the development of
appropriate MC generators which extend jet production to larger pseudo-rapidities.

\subsection{Structure function $F_2$ at high~$x$}

\begin{figure} 
\begin{center}
\epsfig{file=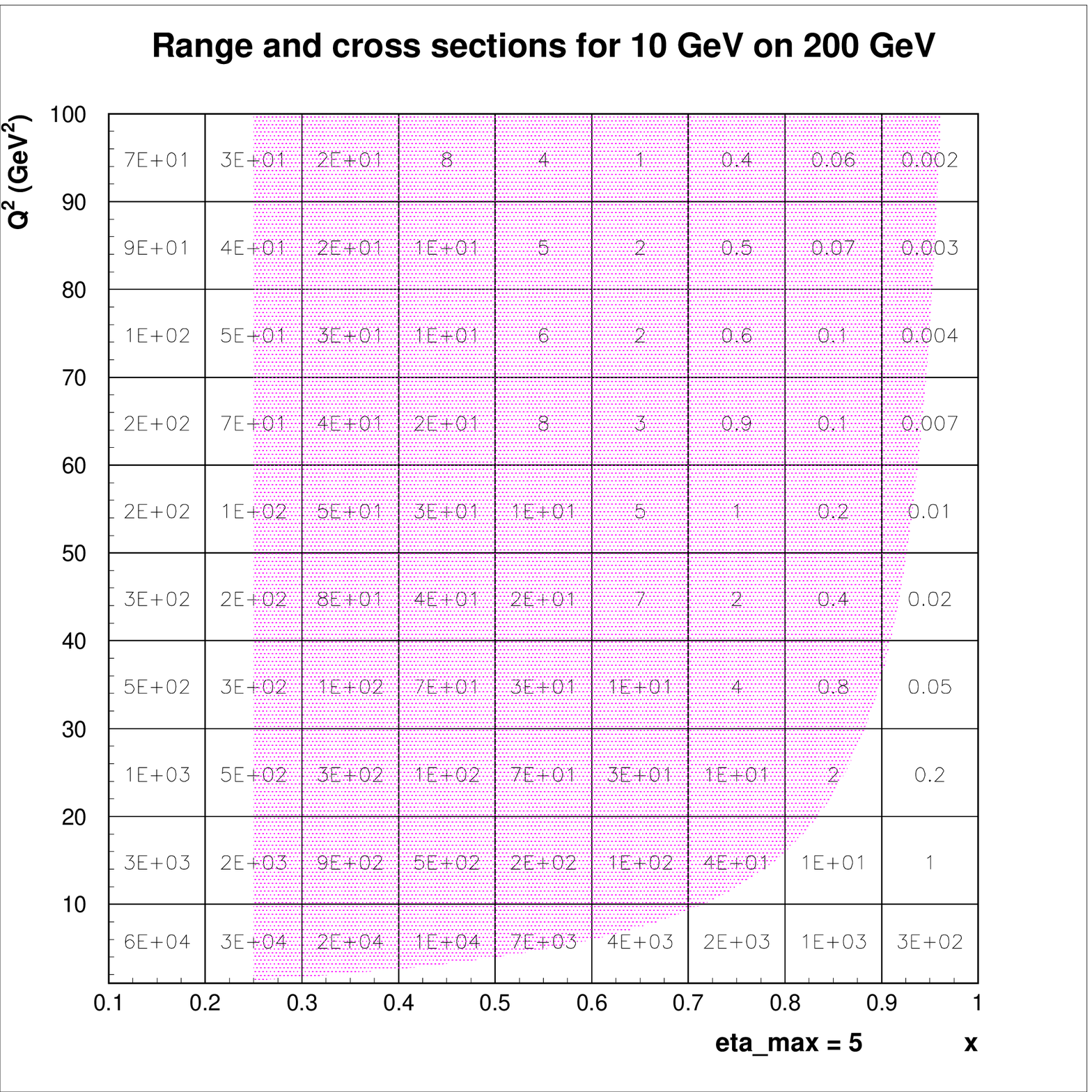,width=10cm}
\caption{ 
The cross sections at high~$x$ and the  region where the
electron+jet method can kinematically be used.  
The cross sections are estimated using the
ALLM parameterization and are in units of pb. 
    } 
  \label{fig:phys:highx-xs}
\end{center}
\end{figure}

An 
expanded view of the high-$x$ coverage, along with expected cross sections
calculated with the ALLM parameterization~\cite{ref:ALLM}, is shown in 
Fig.~\ref{fig:phys:highx-xs}.  
The large rapidity coverage of this 
detector yields 
full jet acceptance at high-$x$ already for moderate $Q^2$ values.
As the scattered 
electrons are seen in the detector up to $x=1$, an integrated
measurement of
the cross section is possible beyond the limit where jets can be used to 
measure $x$.  This allows the measurement of the integral of $F_2$ from
the edge of the phase space coverage shown in 
Fig.~\ref{fig:phys:highx-xs} to $x=1$.

\begin{figure} 
\begin{center}
\epsfig{file=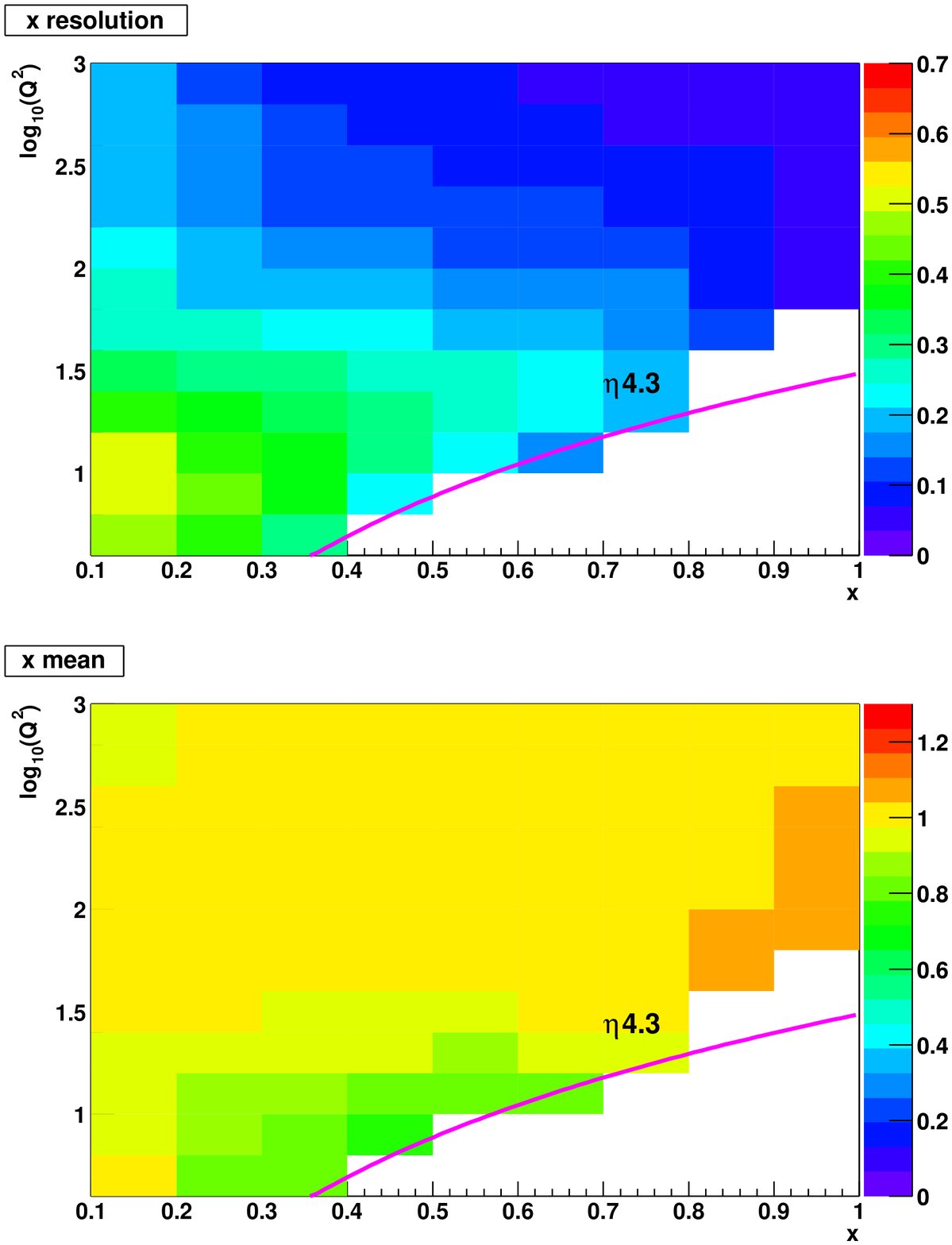,width=12cm}
\caption{
The top plot gives the $x$~resolution in the $x$ and $Q^2$~plane.
The bottom plot gives the mean ratio of reconstructed and
generated~$x$ for each bin.  
    } 
  \label{fig:phys:highx-xres}
\end{center}
\end{figure}

The Monte Carlo sample NC2 is used to study the high~$x$ regime
in more detail.
Unfortunately DjangoH cannot produce events with sufficiently high~$x$
and low $Q^2$ to fill the complete phase space indicated in 
Fig.~\ref{fig:phys:highx-xs}.
Fig.~\ref{fig:phys:highx-xres} shows the MC accessible part of the
phase-space while depicting the $x$~resolution and the accuracy of the
reconstructed~$x$. 
As was pointed out in the previous section, the KT algorithms suffers
from substantial contamination from the proton remnant jet. 
Therefore the cone algorithm is used to find the jet in this
analysis.

The $x$~resolution from the jet energy improves for high~$x$ and~$Q^2$.
As described in sect.~\ref{sec:MCR} the jet response is not fully
simulated, but approximated assuming perfect pattern recognition and
a calorimetric resolution of 35\%/$\sqrt{E}$. 
As the main factor is the finding of the jet, assuming a worse
calorimetric resolution of 50\%/$\sqrt{E}$ does not alter the results
much at this point. However, this needs further study before
the requirements  on the calorimeter can be relaxed.

The mean of the ratio between reconstructed and generated~$x$ as given
in the lower plot of Fig.~\ref{fig:phys:highx-xres} shows the effectiveness
of the cone jet algorithm over a wide range in $x$ and~$Q^2$.
At large~$\eta$ and $x\approx 0.4$ the jet energy and thus $x$ is 
underestimated. However, the effect 
is small compared to the overestimation 
that would result from the use of the KT jet-finding algorithm.
The latter would destroy a possible measurement without systematic
corrections and in any case limits the range of a possible measurement
in a much more stringent way.

The resolutions in $Q^2$ and x were used to define bins for the extraction
of $F_2$.  $Q^2$ was reconstructed using the electron method, and the $Q^2$
binning was therefore chosen to be the same as for the small-x analysis
given earlier.  The binning in x was fixed to be $\Delta x=0.1$.  This was 
possible since the resolution in x is approximately constant at large $x$.

\begin{figure} 
\begin{center}
\epsfig{file=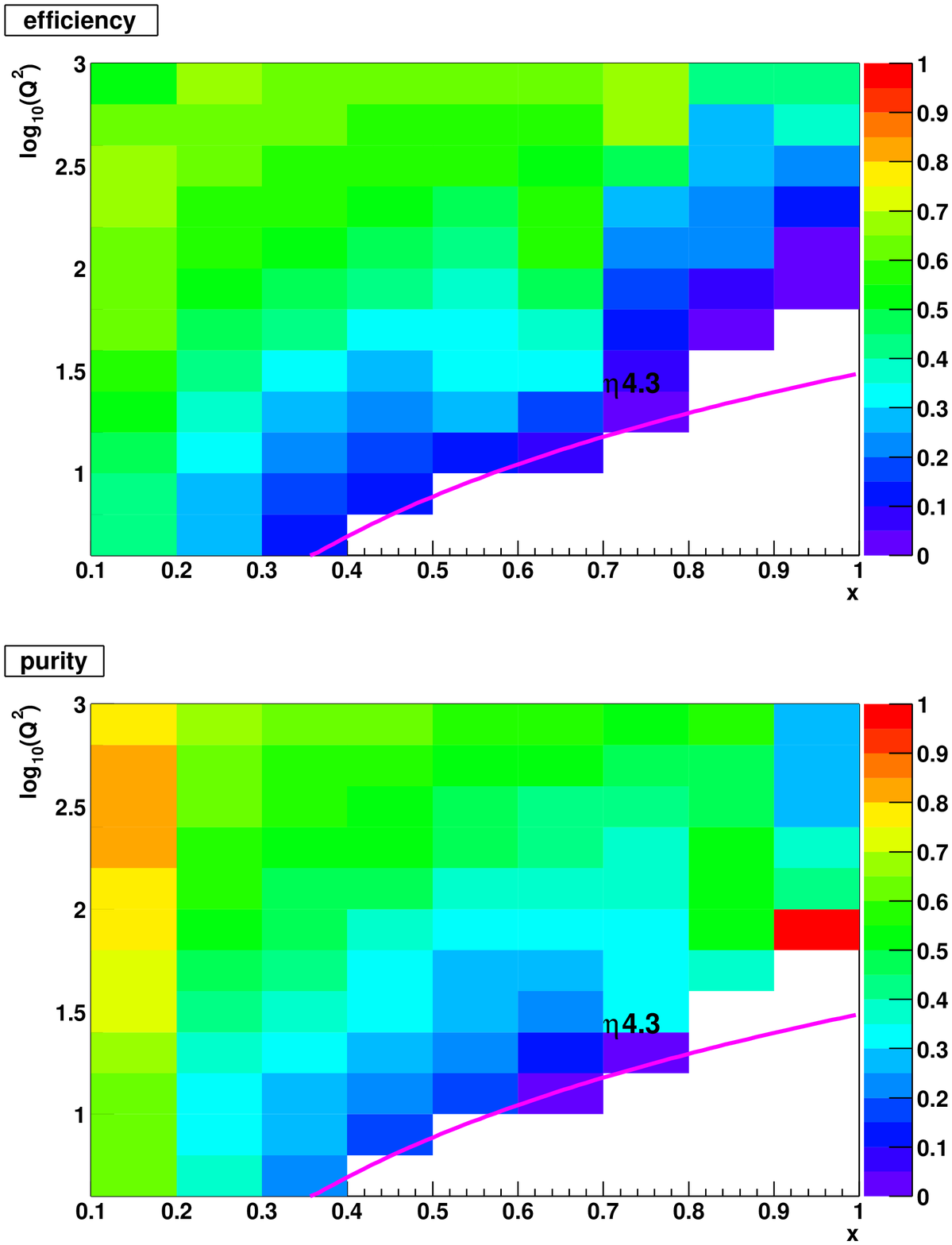,width=11.5cm}
\caption{
The top plot gives the efficiency, the bottom plot the purity
of the events in the $x$ and $Q^2$~plane. The purity of~1 in the
highest~$x$ and lowest~$Q^2$ bin is an artefact due to the inability to
generate MC events with lower $Q^2$ at the same $x$.
    } 
  \label{fig:phys:highx-efpu}
\end{center}
\end{figure}

\begin{figure} 
\begin{center}
\epsfig{file=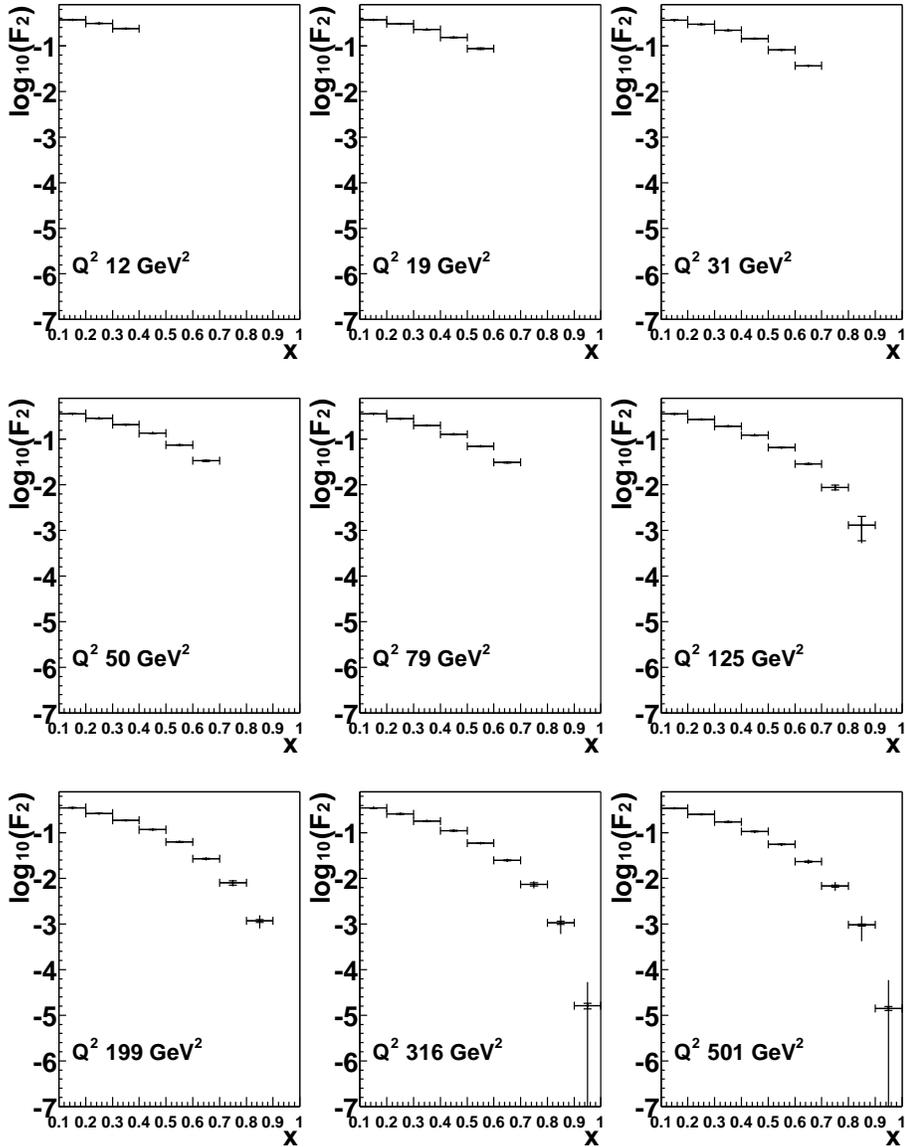,width=12cm}
\caption{
$F_2$ as extracted from the MC sample NC2. For this plot, the inner error bars
are systematical only, the outer bars include statistical errors.
These become dominant for the high~$Q^2$ and high~$x$ regime.
    } 
  \label{fig:phys:highx-f2}
\end{center}
\end{figure}

\begin{figure} 
\begin{center}
\epsfig{file=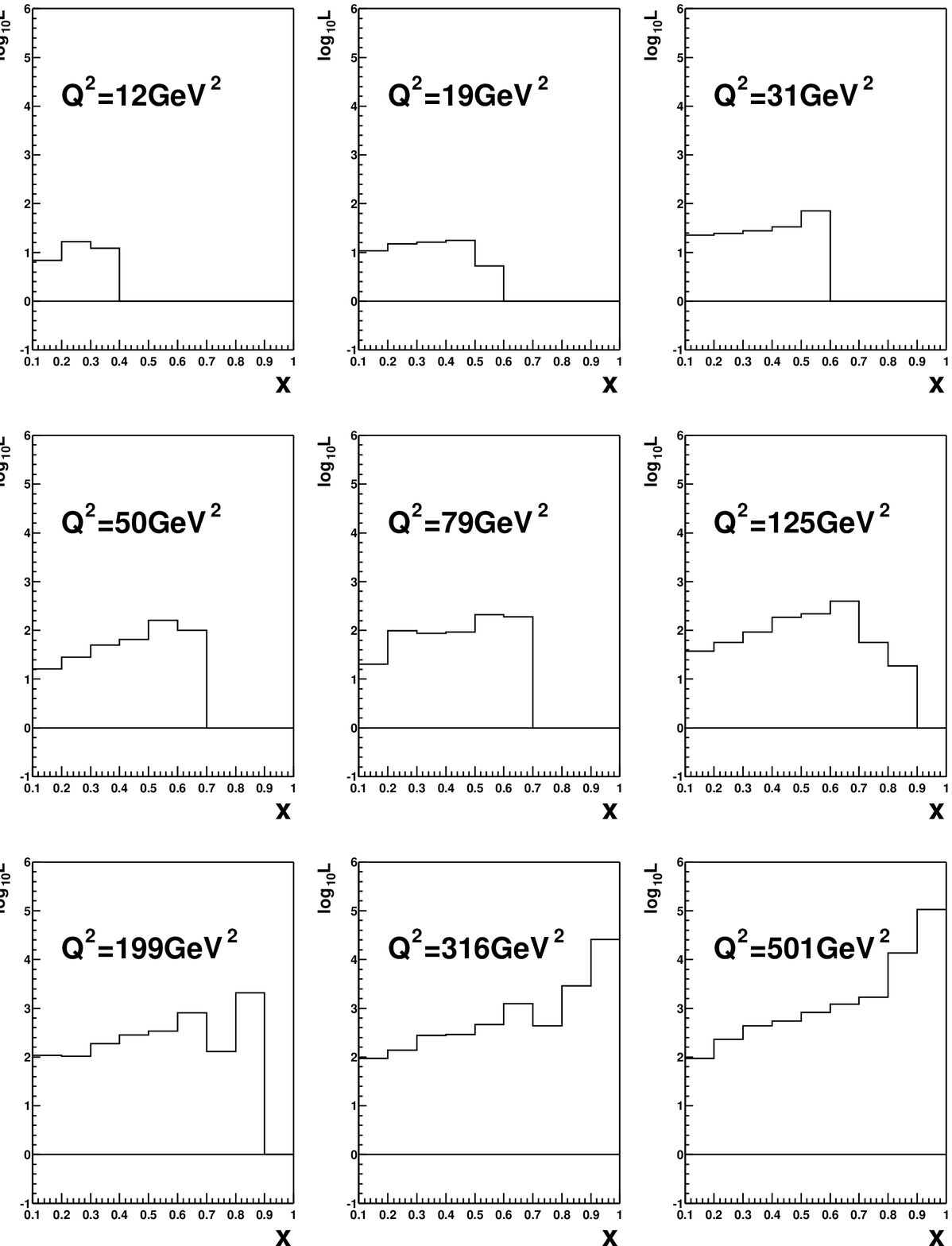,width=11.5cm}
\caption{
Shown is the luminosity for each $x$~bin at which the statistical error
becomes as small as the systematical error. The scale is logarithmic and
in units of pb$^{-1}$. For the highest $Q^2$~bin and $x>0.9$ 100~fb$^{-1}$
are needed before the statistical and systematical errors become equal.
    } 
  \label{fig:phys:highx-lumi}
\end{center}
\end{figure}

The efficiencies and purities
are computed and depicted in Fig.~\ref{fig:phys:highx-efpu}.
We consider bins with efficiency and purity above 20~\% as usable.
The systematic errors considered are the same ones as used at low~$x$
in sect.~\ref{sec:phys:f2low:err} plus an additional uncertainty
in the transverse alignment of the forward calorimeter of 1~cm.
The extracted $F_2$ is given in Fig.~\ref{fig:phys:highx-f2}.
A total luminosity of 100~pb$^{-1}$ is assumed.
The extension  in range to larger~$x$ as $Q^2$ increases
follows the limitations due to the
lack of available MC~events at $\eta$ above 4.
The detector is able to use the full MC accessible phase-space.
At high~$x$ and $Q^2$ the measurement is limited by the luminosity, not by
systematic errors.
Fig.~\ref{fig:phys:highx-lumi} shows the luminosity required to bring down
the statistical errors to level of the systematic ones.
For medium~$Q^2$ [middle plots] the 100~pb$^{-1}$ assumed in this study
are quite sufficient.
For higher~~$Q^2$ [bottom plots] and $x>0.5$ a tenfold luminosity of 
1~fb$^{-1}$ would be required. For $x>0.9$ an additional
one or two orders of magnitude increase in
 luminosity are needed depending on $Q^2$.

\subsection{Summary}

Beam energies of 10~GeV for the  electrons and 200~GeV for the protons
allow an interesting measurement of the structure function $F_2$ in
the transition region of $Q^2$ around 1~GeV$^2$ and down to very low
$Q^2$ and $x$. An integrated luminosity of 100~pb$^{-1}$ is
sufficient. The systematic errors can easily be controlled so that
a 2~\% measurement of $F_2$ is possible. 
The material budget in the detector is of utmost importance in this.
It seems advisable to consider a roman pot system for the very 
backward part~[electron hemisphere]
 of the tracker in order to enlarge the accessible
part of the $x$ and $Q^2$~plane.

Assuming an additional 100~pb$^{-1}$ of integrated luminosity with
a proton beam energy of 100~GeV a measurement of $F_L$ at low~$x$
and low~$Q^2$ becomes possible. The precision is such that different
models can be tested. The measurement can profit from a further reduction
of the systematic errors and additional beam energies.

The detector has acceptance for jets with a pseudo-rapidity up to $\eta=5$ which
is beyond the area currently accessible with Monte Carlo generators.
The phase-space currently covered by Monte Carlo 
[$\eta \le $4]
yields  measurements of $F_2$ at $x$ up to 0.6 for $Q^2$ down to
15~GeV$^2$. For $Q^2$ of several hundred GeV$^2$ it is possible to
reach $x$ above 0.9.  At these high x and $Q^2$ values, the measurements
will be limited by statistics up to about 100~fb$^{-1}$. However, 1~fb$^{-1}$
opens a wide range up to $x$ of 0.9 and $Q^2<250$~GeV$^2$.

\clearpage
\newpage
\section{Summary, scope and time scale}

\paragraph{Physics summary}

Building an electron accelerator 
to collide electrons with protons [ions] from
the existing RHIC accelerator would yield valuable information on the
high energy limit of strong interactions.
The highlights of the program are seen as precision measurements
of the structure functions $F_2$ and $F_L$, measurements of
forward jet and particle
production in small~$x$ events in an extended rapidity range and
measurements of exclusive processes, in particular at high~$t$.  These
measurements should be performed for
protons, deuterons, and at least one heavy nucleus.
  A new detector optimized for this physics program would greatly enhance
the kinematic range covered today by H1 and ZEUS.  Precision measurements
of $F_2$ would be possible in a continuous way across the parton-to-hadron
transition region, where current data are sparse but very interesting.
$F_2$ would also be measureable with good statistics up to
$x \approx 1$ for $Q^2 >200$~GeV$^2$, which to date is 
a completely unexplored region.  $F_L$
would be measured directly 
in the region of greatest interest, $Q^2\le 10$~GeV$^2$, where we currently 
have large theoretical uncertainties.  Exclusive vector meson production
and DVCS could be studied across the full $W$ range and up to high 
$|t|$, thereby allowing a much higher precision determination
of the energy dependence of the cross sections, and a three-dimensional
mapping of hadrons down to small impact parameters.  The study of forward
jet production and particle production over a wide range of pseudorapidities
(up to $\eta=5$, compared to $\eta=2.7$ at H1, ZEUS) would
allow stringent tests of our understanding of radiation processes in QCD,
as well as show us where collective phenomena become important.  Performing
these measurements for the first time on deuterons would allow us to test
the universality of parton distributions at small~$x$, as well as allow an
extraction of individual 
parton densities at high~$x$.  Scattering on a heavy nuclei
would test the onset of saturation effects, with the
possible discovery of the color glass condensate.  The data would also be
invaluable for the interpretation of heavy ion scattering experiments.
It is clear to us that this program is 
critical for the further development of our 
understanding of strong interaction physics.

\paragraph{New detector summary}

The major subcomponents of the proposed detector are
\begin{itemize}
\item the precision tracking system;
\item the $4\pi$ electromagnetic calorimeter;
\item the forward hadronic calorimeter;
\item the dipole magnet and beampipe.
\end{itemize}
In addition, we foresee the need for a photon calorimeter downstream in the
electron direction, and several small detectors in the proton beam direction:
\begin{itemize}
\item a proton remnant tagger at $z \approx 20$~m;
\item a forward neutron calorimeter;
\item a forward proton calorimeter.
\end{itemize}
The scale of the experiment is intermediate between
a typical fixed target experiment and a typical collider experiment.  A first,
crude, estimate of the cost is $50$~MEuro, approximately evenly divided 
between the tracking system, the electromagnetic calorimeter, and the beamline 
(including the dipole).  This estimate assumes the hadron calorimeter would
come from other sources.

We would aim to complete the design and construction of the detector
within six years.  A 
possible schedule would therefore be as follows:
\begin{description}
\item[T$_0$+1 year] Letter of intent and technical proposal;
\item[T$_0$+3 years] Detailed detector design and R\&D;
\item[T$_0$+5 years] Detector construction;
\item[T$_0$+6 years] Detector commissioning and operation.
\end{description}

The small~$x$ (or high energy) physics opened up by HERA should
now be pursued with a new round of experiments.  Many key measurements could
not be carried out with H1 and ZEUS, as these
detectors were designed for a different physics.  Other important measurements
require the addition of nuclei to be carried out.  A dedicated experiment
optimized for this physics would significantly enhance our understanding of
QCD in its high energy limit, and possibly open the way to a complete
revision of our ideas on the strong interactions.
The eRHIC facility would be well suited for the 
measurement program described in this document.

\newpage

%
\end{document}